\documentclass[twocolumn,showpacs,aps,floatfix,prd,nofootinbib]{revtex4}
\usepackage{graphicx}
\usepackage{dcolumn}
\usepackage{array, longtable}
\usepackage{epsfig}
\usepackage{amsmath}
\usepackage{colordvi}
\usepackage{hhline}

\newcommand{\BaBarNumber}{PUB-15/005}
\newcommand{\SLACPubNumber}{SLAC-PUB-16940}

\def\Ecm       {\ensuremath {E_{\rm c.m.}}\xspace}

\def\K         {\ensuremath{K}}

\def\Kst       {\ensuremath{\K^*(892)}\xspace}
\def\kst       {\ensuremath{\K^*(892)}\xspace}
\def\kstz      {\ensuremath{\K^{*}(892)^{0}}\xspace}
\def\kstc      {\ensuremath{\K^{*}(892)}\xspace}
\def\kstp      {\ensuremath{\K^{*}(892)^{+}}\xspace}
\def\kstm      {\ensuremath{\K^{*}(892)^{-}}\xspace}
\def\kstpm     {\ensuremath{\K^{*}(892)^{\pm}}\xspace}
\def\kstmp     {\ensuremath{\K^{*}(892)^{\mp}}\xspace}
\def\ktwo      {\ensuremath{\K^*_2(1430)}\xspace}
\def\Kstzero   {\ensuremath{\K^*_0(1430)}\xspace}

\def\Ecm       {\ensuremath{E_{\rm c.m.}}\xspace}

\input babarsym

\begin{document}

\begin{flushleft}
{\babar}-\BaBarNumber \\
\SLACPubNumber \\
\end{flushleft}

% Title of the paper
\title{\large \bf
\boldmath Measurement of the $\epem\to \KS\Kpm\pimp\piz$ and $\KS\Kpm\pimp\eta$ cross 
sections using initial-state radiation
} % end title

% Input author list file ?
% NOTES
% 22-Jan-2016 Changes requested by Babar Spokesperson (M. Roney)       J.W. Gary
%    Remove V. Luth; add H. Lacker and R. Sobie
%    Add Institute of Particle Physics label for C. Hearty, S. Robertson,
%    R. Sobie and ``a'' ``b'' superscripts for all Canadian authors
%    Move R. Cheaib from McGill to Mississippi
% 20-Jan-2016 Add ``deceased'' footnote for Erwin Gabathuler           J.W. Gary
% 16-Dec-2016 Move Miriam Fritsch from Mainz to Bochum                 J.W. Gary
% 04-AUG-2016 Add ``deceased'' footnote for Giancarlo Piredda          J.W. Gary
% 02-JUN-2016 Move Marcello Rotondo from Padova to Frascati            J.W. Gary
% 20-FEB-2016 Add footnote for Liang Sun                               J.W. Gary
% 21-DEC-2015 Add Bologna alternative address for Claudia Patrignani   J.W. Gary
% 
\author{J.~P.~Lees}
\author{V.~Poireau}
\author{V.~Tisserand}
\affiliation{Laboratoire d'Annecy-le-Vieux de Physique des Particules (LAPP), Universit\'e de Savoie, CNRS/IN2P3,  F-74941 Annecy-Le-Vieux, France}
\author{E.~Grauges}
\affiliation{Universitat de Barcelona, Facultat de Fisica, Departament ECM, E-08028 Barcelona, Spain }
\author{A.~Palano}
\affiliation{INFN Sezione di Bari and Dipartimento di Fisica, Universit\`a di Bari, I-70126 Bari, Italy }
\author{G.~Eigen}
%\author{B.~Stugu}
\affiliation{University of Bergen, Institute of Physics, N-5007 Bergen, Norway }
\author{D.~N.~Brown}
%\author{L.~T.~Kerth}
\author{Yu.~G.~Kolomensky}
%\author{M.~J.~Lee}
%\author{G.~Lynch}
\affiliation{Lawrence Berkeley National Laboratory and University of California, Berkeley, California 94720, USA }
\author{M.~Fritsch}
\author{H.~Koch}
\author{T.~Schroeder}
\affiliation{Ruhr Universit\"at Bochum, Institut f\"ur Experimentalphysik 1, D-44780 Bochum, Germany }
% ...
\collaboration{The \babar\ Collaboration}
\noaffiliation

\begin{abstract}
The processes $\epem\to\KS\Kpm\pimp\piz$ and $\epem\to\KS\Kpm\pimp\eta$
  are studied over a continuum of energies from threshold to 4 GeV with
  the initial-state photon radiation method.  Using 454 \invfb of data
  collected with the \babar\ detector at the SLAC PEP-II storage ring,
  the first measurements of the cross sections for these processes
  are obtained. The intermediate resonance structures from $\Kstarz(\K\pi)^0$,
  $\kstc^{\pm}(\K\pi)^{\mp}$ and $\KS\Kpm\rho^{\mp}$ are studied. The \jpsi is observed in
  all of these channels, and corresponding branching fractions are measured.
\end{abstract}

% PACS

\vfill
\maketitle

% reset footnote counter
\setcounter{footnote}{0}

% The body of the paper starts here
\section{Introduction}
\label{sec:Introduction}

Measurements of low-energy $\epem$ hadronic cross sections are important
ingredients for the standard model prediction of the muon anomalous
magnetic moment~\cite{g-2} and provide a wealth of spectroscopic information.
At an $\epem$ collider, a continuous spectrum of collision energies
below the nominal $\epem$ center-of-mass (c.m.) energy can be attained
by selecting events with initial-state radiation (ISR), as proposed
in Ref.~\cite{baier} and discussed in Refs.~\cite{arbus,kuehn,ivanch}.

At energies below a few $\gev$, individual exclusive final states must
be studied in order to understand the experimental acceptance.
The cross section $\sigma_{\gamma f}$ for an incoming $\epem$ pair colliding at a
c.m. energy $\sqrt{s}$ to radiate a photon of energy E$_{\gamma}$ and then
annihilate into a specific final state $f$ is related to the
corresponding direct $\epem\to f$ cross section $\sigma_f$ by:
\begin{equation}
\frac{d\sigma_{\gamma f}(s,x)}{dx} = W(s,x)\sigma_f(\Ecm)\ ,
\label{eq1}
\end{equation}
where $x=2E_{\gamma}/\sqrt{s}$ and $\Ecm = \sqrt{s(1-x)}$ is the effective center-of-mass energy at which the state $f$ is produced. 
The radiator function $W(s,x)$, or probability density for
 photon emission, can be evaluated to better than 1\% accuracy~\cite{phokara}.  

Previously, we presented measurements of low-energy
cross sections for many exclusive hadronic reactions
using the ISR method, including a number of final states
with two kaons in the final state, such as $f = \Kp\Km$~\cite{kk},
$\Kp\Km\pip\pim$~\cite{kkpipi}, $\KS\KL$, $\KS\KL\pip\pim$,
$\KS\KS\pip\pim$ and $\KS\KS\Kp\Km$~\cite{ksklpipi},  
$\KS\Kpm\pimp$~\cite{kskpi}, $\KS\KL\piz$ and $\KS\KL\piz\piz$~\cite{ksklpi0s}.
Here, we extend our program and report measurements of
the $\epem\to\KS\Kpm\pimp\piz$ and $\KS\Kpm\pimp\eta$ channels,
including studies of the intermediate
resonant substructure.

\section{\boldmath The \babar\ detector and data set}
\label{sec:babar}

The results presented in this analysis are based
on a sample of $\epem$ annihilation data
collected at \Ecm = 10.58 GeV with the
\babar\ detector~\cite{babar_det} at the SLAC \pep2 storage ring,
and correspond to an integrated luminosity 
of $454\;\fb^{-1}$~\cite{babar_lum}.

Charged-particle momenta are measured in a tracking
system consisting of a five-layer double-sided silicon
vertex tracker (SVT) and a 40-layer central drift chamber
(DCH), immersed in a 1.5  T axial magnetic field.
An internally reflecting ring-imaging Cherenkov detector
(DIRC) with fused silica radiators provides charged-particle
identification (PID).
A CsI electromagnetic calorimeter (EMC) is used to detect
and identify photons and electrons.
Muons are identified in the instrumented magnetic
flux-return system.

Charged pion and kaon  candidates are selected using
a likelihood function based on the specific ionization
in the DCH and SVT, and the Cherenkov angle measured
in the DIRC.
Photon candidates are defined as  clusters
in the EMC that have a shape consistent with an
electromagnetic shower and no associated
charged track.

To study the signal efficiency as well as backgrounds from other ISR 
processes, a special package of Monte Carlo (MC) simulation programs for 
radiative processes has been developed.
Algorithms for generating hadronic final
states via ISR are derived from Ref.~\cite{czyz}.
Multiple soft-photon emission from initial-state
charged particles is implemented by means of the
structure-function technique~\cite{kuraev,strfun}, while
extra photon radiation from
final-state particles is simulated with the PHOTOS~\cite{PHOTOS} package.

Large samples of signal $\epem\to\KS\Kpm\pimp\piz\gamma$ and $\KS\Kpm\pimp\eta\gamma$ 
events are generated with this program, as well as samples of events from
the principal ISR background sources, $\epem\to\KS\Kpm\pimp\gamma$
and $\epem\to\KS\Kpm\pimp\piz\piz\gamma$. The $\KS\Kpm\pimp\gamma$ generator is 
tuned to reproduce our measured~\cite{kskpi} \Ecm dependence and resonant
substructure.  The other modes use smooth \Ecm dependences and phase
space for the final state hadrons.  The signal and $\KS\Kpm\pimp$
generators reproduce the kaon and pion kinematic distributions
observed in the data, and we study the effect of resonances on the
efficiency in each case below. In addition to the ISR sources,
background arises from the non-ISR processes $\epem\to\qqbar$
and $\tau^+\tau^-$.  These events are simulated with the
JETSET~\cite{jetset} and KORALB~\cite{koralb} event generators, respectively.
All simulated events are processed through a detector
simulation based on the GEANT4~\cite{GEANT4} package and are analyzed
in the same manner as the data.

\section{Event selection and kinematics}
\label{sec:selection}

We require events to contain at least three
photon candidates and at least four charged tracks, including at     
least one $\KS\to\pip\pim$ candidate.  

Photon candidates must lie within the acceptance of
the EMC, defined by $0.35 < \theta < 2.4$ radians, where
$\theta$ is the polar angle relative to the $e^-$ beam
direction. The photon candidate with highest energy
is assumed to be the ISR photon, and is required to
have energy $E^*>3$~GeV, where the asterisk indicates a quantity evaluated
in the $\epem$ c.m. frame.  To reduce
background from machine-induced soft photons, at
least one additional photon candidate must have
$E^*>100$~MeV and another $E^*>60$~MeV.
We calculate the invariant mass $m_{\gaga}$ of each pair of photon
candidates, and consider a pair to be a $\piz$ candidate if
$0.09<m_{\gaga}<0.18$~\gevcc and an $\eta$ candidate if 
$0.47<m_{\gaga}<0.62$~\gevcc. Events with at least one $\piz$ or $\eta$ 
candidate are retained.

We require at least two charged tracks in an event, of
opposite charge, one identified as a kaon and one as a pion,
that appear in the polar angle range $0.45<\theta<2.40$ radians.
Each track must extrapolate to within 0.25~cm of the nominal
$\epem$ collision point in the plane perpendicular to the beam axis
and to within 3~cm along the axis.

The \KS candidates are reconstructed in the $\pip\pim$ decay mode from pairs  
of oppositely charged tracks not identified as electrons.  They must
have an invariant mass within 15 \mevcc of the nominal \KS mass, and a
well reconstructed vertex at least 2 mm away from the beam axis.  The
angle $\theta_{\KS}$ between their reconstructed total momentum and the line
joining their vertex with the primary vertex position must satisfy   
$\cos(\theta_{\KS})>0.99$.

Each of these events is subjected to a set of 5-constraint (5C)
kinematic fits, in which the four-momentum of the
$\KS\Kpm\pimp\gamma_{ISR}\gaga$ system is required to equal that of the 
initial \epem system and the invariant mass of the two non-ISR photon
candidates is constrained to the nominal $\piz$ or $\eta$ mass.  The fits
employ the full covariance matrices and provide      
$\chi^2$ values and improved determinations of the particle   
momenta and angles, which are used in the subsequent analysis.  
Fits are performed for every $\piz$ and $\eta$
candidate in the event, and we retain the combinations giving the
lowest values of $\chi^2_{\KS\Kpm\pimp\piz}$ and $\chi^2_{\KS\Kpm\pimp\eta}$.

\begin{figure}[h]
\includegraphics[width=0.8\columnwidth]{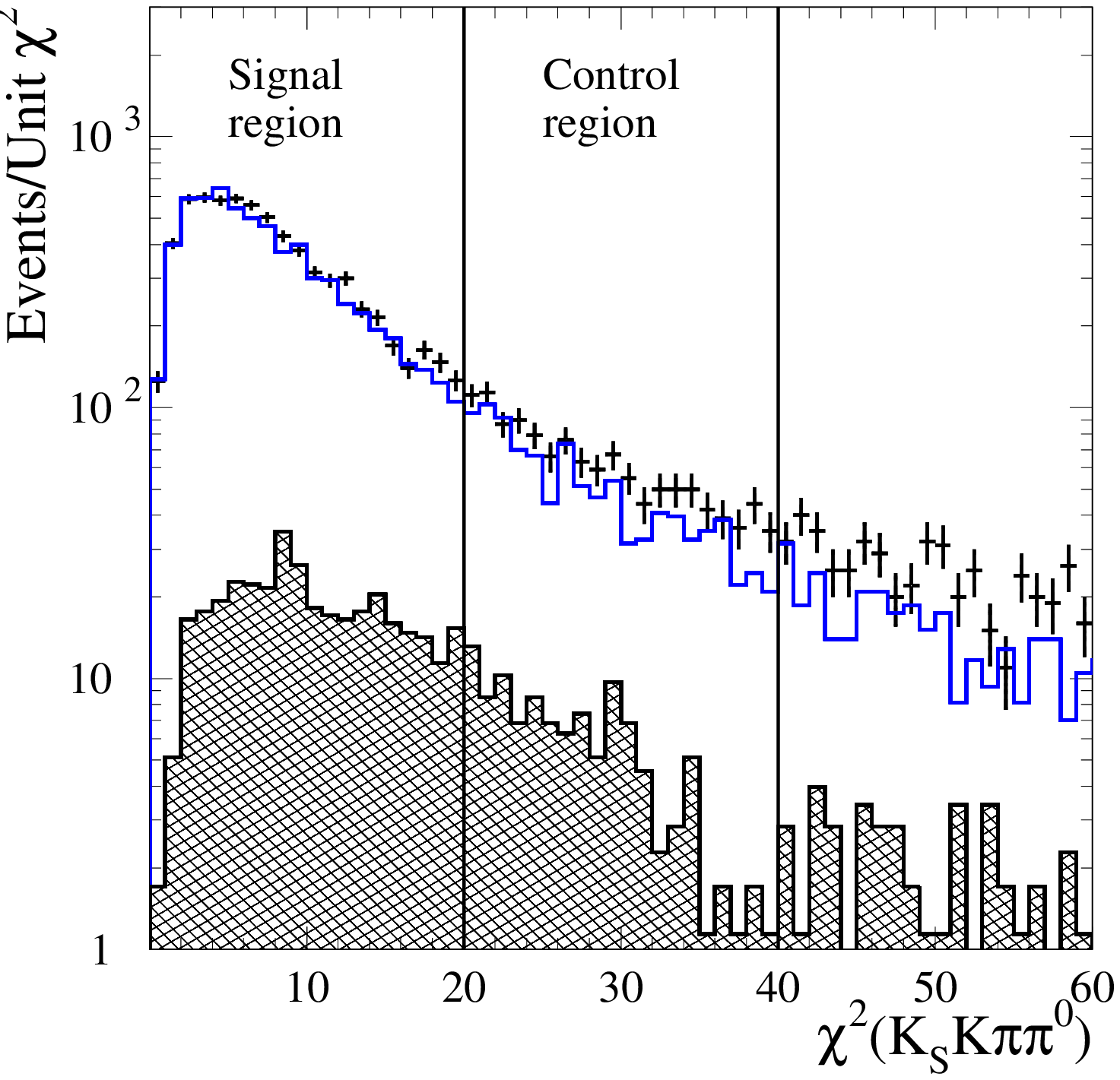}
\caption{Distribution of $\chi^2$ from the 5-constraint fit for $\KS\Kpm\pimp\piz\gamma$ candidates in the 
data (points). The open and cross-hatched histograms are the distributions for
 simulated signal and \qqbar background events, respectively, 
 normalized as described in the text.  The signal
and control regions are indicated. }
\label{fig:chi2-isr}
\end{figure}

\section{The $\KS\Kpm\pimp\piz$ final state}
\subsection{Event selection}\label{sec:kskpipi0}

The $\chisq_{\KS\Kpm\pimp\piz}$ distribution for the selected $\epem\to\KS\Kpm\pimp\piz\gamma$ events is
shown in Fig.~\ref{fig:chi2-isr}, after subtraction of the small background 
from \qqbar events, which is discussed below and shown in the figure as the
cross-hatched histogram.  The corresponding distribution for simulated,
selected signal events is shown as the open histogram.  It is
normalized to the data integrated over the first five bins, where the
lowest ISR background contributions are expected.
These distributions are broader than a
typical 5C \chisq distribution because of multiple soft-photon emission
from the initial state, which is not taken into account in the fit 
but is present in both the data and simulation. 
Previous studies have found these effect to be well simulated, and we
assign a systematic uncertainty in Section~\ref{sec:efficiency}.  The remaining
differences can be explained by ISR backgrounds, which we discuss in this 
subsection.

Signal event candidates are selected by requiring $\chi_{\KS\Kpm\pimp\piz}^2<20$.
Events with $20<\chi_{\KS\Kpm\pimp\piz}^2<40$ are used as a control sample to evaluate
background. The signal and control samples contain 6859 (5656) and
1257(870) experimental (simulation) events, respectively.
 
\begin{figure}[tbh]  
\begin{tabular}{cc}
\includegraphics[width=.5\columnwidth]{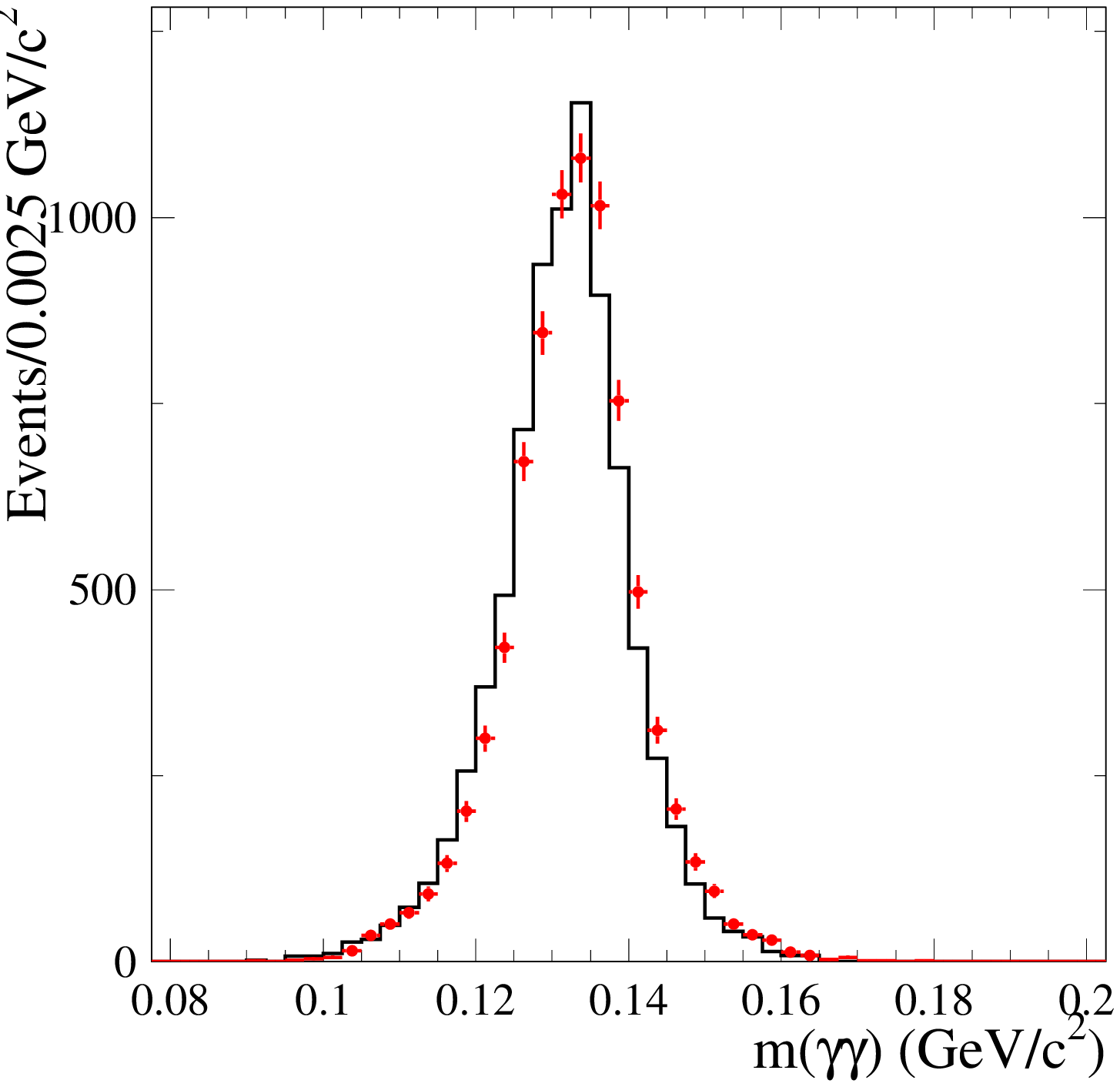}
\put(-25,105){(a)}
\includegraphics[width=.5\columnwidth]{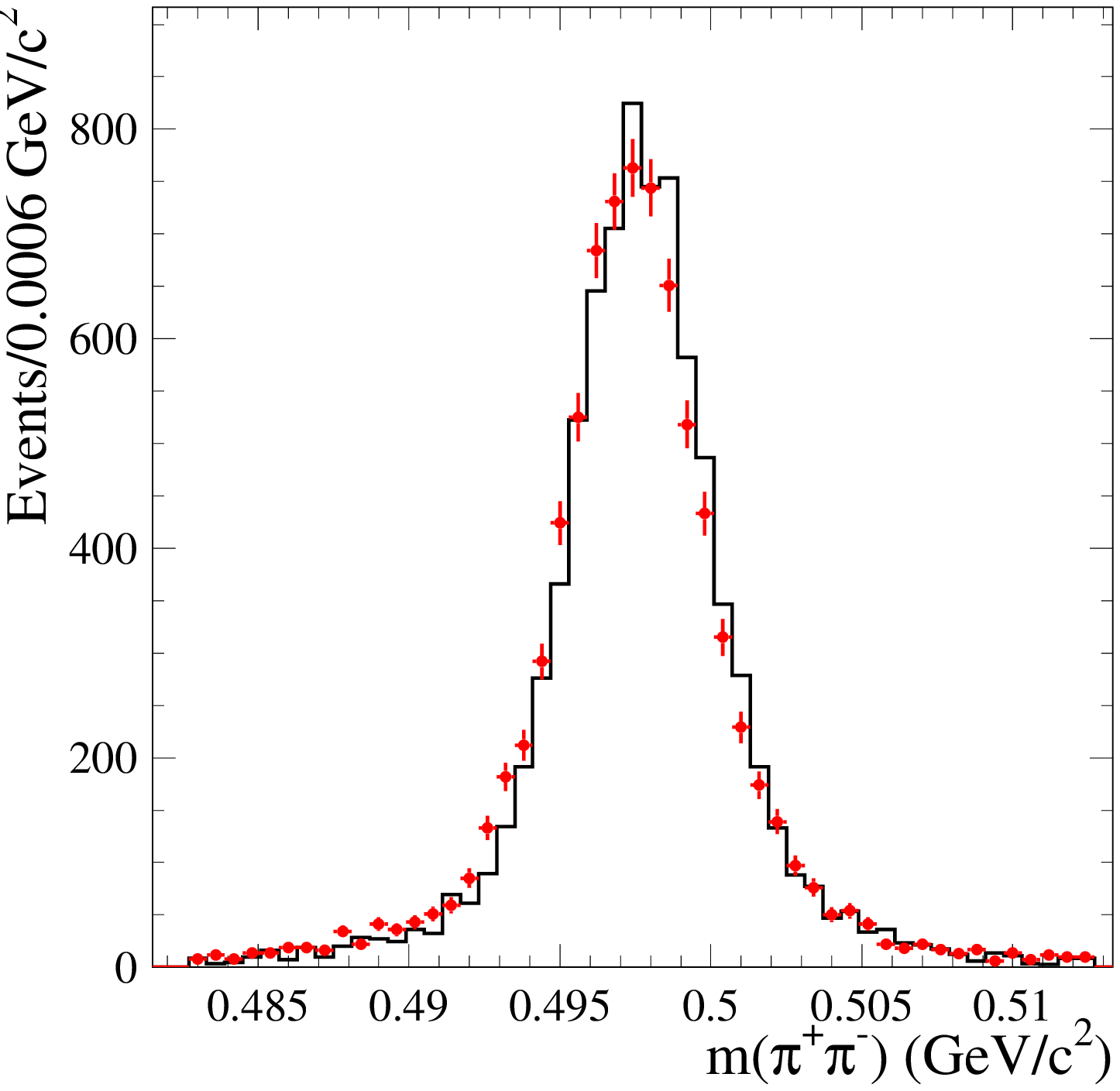}
\put(-25,105){(b)}
\end{tabular}
\caption{The (a) \gaga and (b) $\pip\pim$ invariant-mass distributions of
the $\piz$ and $\KS$ candidates, respectively, in $\KS\Kpm\pimp\piz$ events
in the $\chi^2_{\KS\Kpm\pimp\piz}$ signal region, for the selected data (points) and
the signal simulation (histograms).}
\label{fig:ks_pi0}
\end{figure}

Figure~\ref{fig:ks_pi0}(a) compares the $\gaga$ invariant-mass distribution of
the \piz candidate for data events in the signal region with the
prediction of the signal-event simulation.
The $\pi^0$ peak in the simulation is shifted with respect to the data  by 
$-0.6\pm$0.2 MeV/c$^2$, while the 
standard deviations are consistent with each other ($\sigma_{\rm DATA} = 6.65\pm0.14$ MeV/c$^2$ and 
$\sigma_{\rm MC} = 6.70\pm 0.12$ MeV/c$^2$). 

The corresponding distributions of the $\pip\pim$ invariant mass of the
 \KS candidate are shown in Fig.~\ref{fig:ks_pi0}(b).
In this
case, a shift in the peak values of $0.23\pm0.05$~MeV/c$^2$ is observed between
data and simulation.  The widths are found to be somewhat different:
$\sigma_{DATA}=2.40\pm0.03$~MeV/c$^2$ and $\sigma_{MC}=2.30\pm0.03$~MeV/c$^2$.
Our selection 
criteria on the $\piz$ and $\KS$ masses are unrestrictive enough to ensure the shifts do not affect the result. 

\begin{figure}[tbh]
\includegraphics[width=0.8\columnwidth]{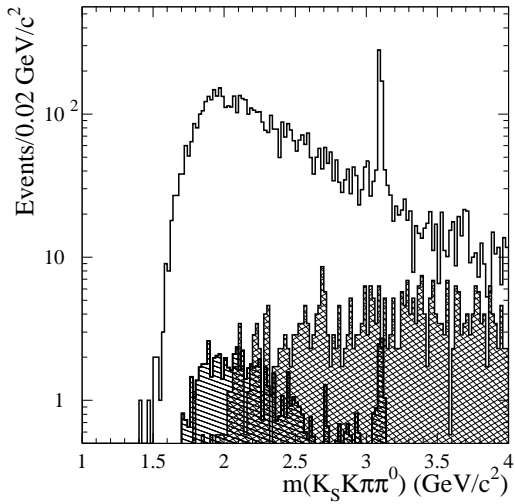}
\caption{Distribution of the fitted $\KS\Kpm\pimp\piz$ invariant mass for data
 events in the $\KS\Kpm\pimp\piz$ signal region.  The hatched and cross-hatched distributions show the estimated backgrounds 
evaluated from ISR and $\qqbar$ events, respectively.}
\label{fig:mkskpipi0_sig_bkg}
\end{figure}

The distribution of the invariant mass of the final-state hadronic system for all data events in the signal region is 
shown as the open histogram  in Fig.~\ref{fig:mkskpipi0_sig_bkg}. A narrow peak due to $\jpsi\to\KS\Kpm\pimp\piz$ decays is clearly visible.

Cross sections for backgrounds from $\qqbar$ processes are
poorly known.  In simulation, the dominant such process is
$\epem\to\KS\Kpm\pimp\piz\piz$, in which an energetic photon from one of the \piz decays 
is erroneously taken as the ISR photon. These events have kinematic properties similar to signal events and yield a \chisq 
distribution peaked 
at low values.  
This component can be evaluated from the data, since such events produce a peak at the
\piz invariant mass when the photon erroneously identified as the ISR photon is combined with another photon in the 
event. 
Following the procedure described in Ref.~\cite{kskpi}, we use the MC mass distribution, and normalize it to the
data in the region $2<m<4$~\gev, where the \piz peak is
prominent. A consistent normalization factor is obtained from the 4--6~\gevcc
region. 
For lower masses, we see no significant \piz peak in the data, and we use
the very small MC prediction with the same normalization.
The normalized contribution of the $\qqbar$ background to the distributions of Figs.~\ref{fig:chi2-isr} 
and~\ref{fig:mkskpipi0_sig_bkg} is shown by the cross-hatched histograms.  For subsequent distributions, the $\qqbar$ 
background is subtracted.

\begin{figure}[tbh]
\includegraphics[width=0.8\columnwidth]{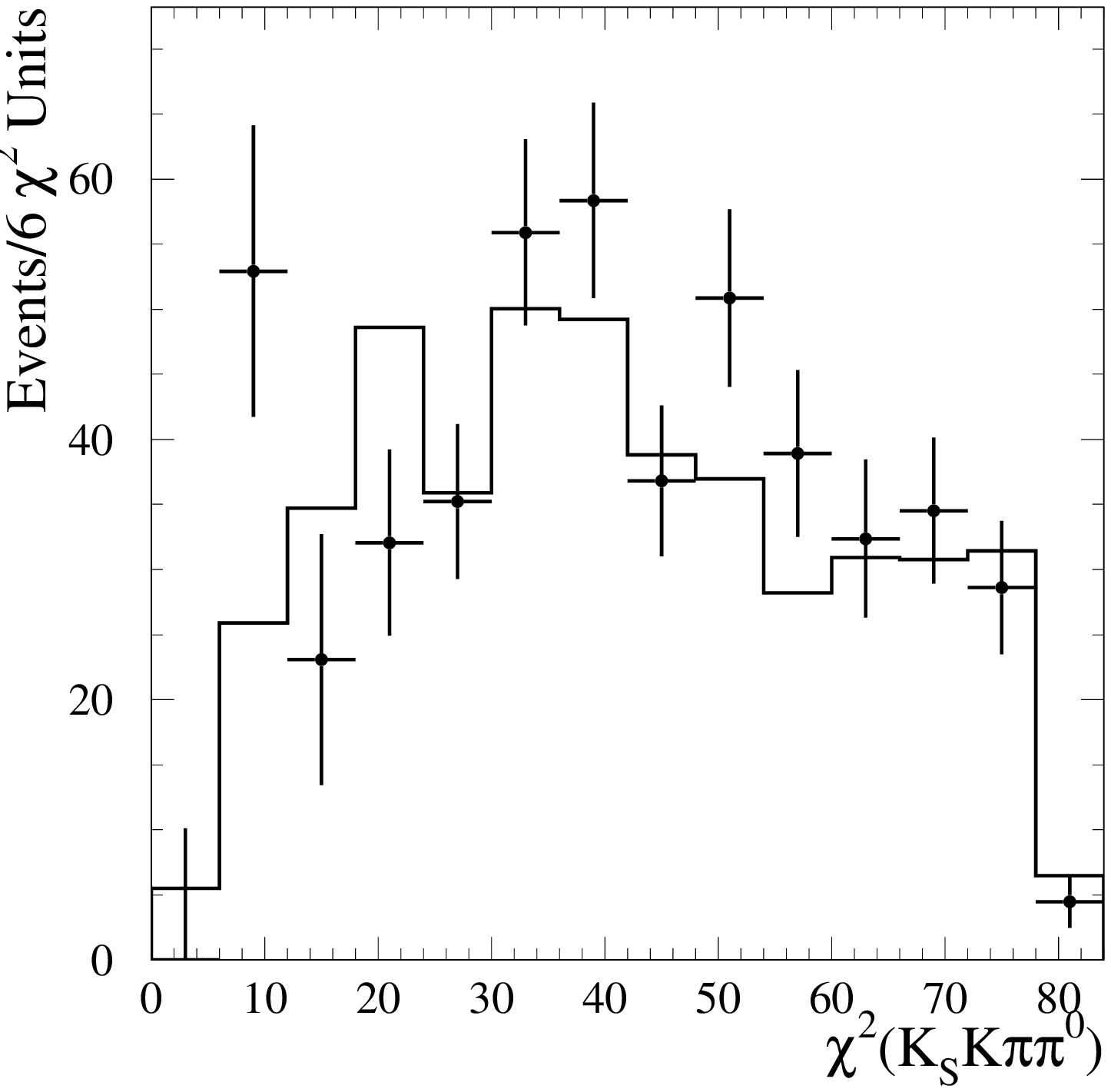}
\caption{The $\chi^2_{\KS\Kpm\pimp\piz}$ distributions of the ISR background determined from the data (points with errors) and   
the sum of MC simulations for the processes $\epem\to\KS\Kpm\pimp\gamma$ and $\epem\to\KS\Kpm\pimp\piz\piz\gamma$ 
(open histogram) described in the text.} 
\label{fig:isrbkg}
\end{figure}

The remaining background arises from ISR processes, dominated
by $\epem\to\KS\Kpm\pimp\gamma$ events combined with random photons, and by $\epem\to\KS\Kpm\pimp\piz\piz\gamma$ 
events. 
These have broad distributions in $\chisq$, and can be estimated from
the control region of the \chisq distribution.  The points with errors in Fig.~\ref{fig:isrbkg} show the
difference between the data and the normalized simulated
$\chisq_{\KS\Kpm\pimp\piz}$ distributions of Fig.~\ref{fig:chi2-isr}.
Assuming good signal simulation and low ISR backround at low
$\chisq$, this gives an estimate of the shape of the distribution for
the total remaining background.  The simulation of the ISR
$\KS\Kpm\pimp$ background shows a consistent shape and, when normalized
to our previous measurement~\cite{kskpi}, accounts for about 10\% of the
entries.  The simulated ISR $\KS\Kpm\pimp\piz\piz$ background also has
a consistent shape, and is expected to be much larger.  Normalizing
to a cross section nine times larger and adding the ISR $\KS\Kpm\pimp$
prediction, we obtain the simulated distribution shown as the
histogram in Fig.~\ref{fig:isrbkg}.  This demonstrates sufficient understanding of
the shape of the background distribution, and we assume that all
remaining background has the simulated shape.
The genuine signal and the ISR background in any distribution other than
the \chisq are estimated bin-by-bin using the numbers of selected events
in that bin in the signal and control regions, $N_1$ and $N_2$, after
subtraction of the respective \qqbar backgrounds.  We take $N_1$ ($N_2$) to
be the sum of the numbers of genuine signal $N_{1S}$ ($N_{2S}$) and ISR background
events $N_{1B}$ ($N_{2B}$) in the signal (control) region.  From the signal
simulation, we obtain $N_{1S}/N_{2S}=\alpha=6.59\pm0.24$, and from the ISR
background simulation $N_{1B}/N_{2B}=\beta=0.49\pm0.07$.  The observed values of $N_1$ and $N_2$ are $6509\pm81$ and 
$1146\pm34$, 
respectively. We then solve for
\begin{equation}
N_{1S} = \alpha\cdot\frac{N_1-\beta\cdot N_2}{\alpha-\beta},
\end{equation}
and $N_{1B}$ in that bin.

The ISR background evaluated in this manner is shown by the
hatched histogram in Fig.~\ref{fig:mkskpipi0_sig_bkg}. 

We find $N_{1S}=6430\pm90$, where the uncertainty is
statistical.  
The systematic uncertainty in the $\qqbar$ background estimate
is taken to be 50\%,
to account for the limited knowledge of the $\qqbar$ cross section.
The systematic uncertainty in the ISR background estimate is,
more conservatively, taken to be 100\%.
The total systematic uncertainty is evaluated in
three regions of \Ecm.
This yields relative uncertainties in $N_{1S}$ of 2.5\% for $\Ecm<2$~\gev,
6.25\% for $2<\Ecm<3$~\gev, and 10\% for $\Ecm>3$~\gev.

\subsection{Detection efficiency}
\label{sec:efficiency}

\begin{figure}[tbh]
\includegraphics[width=0.8\columnwidth]{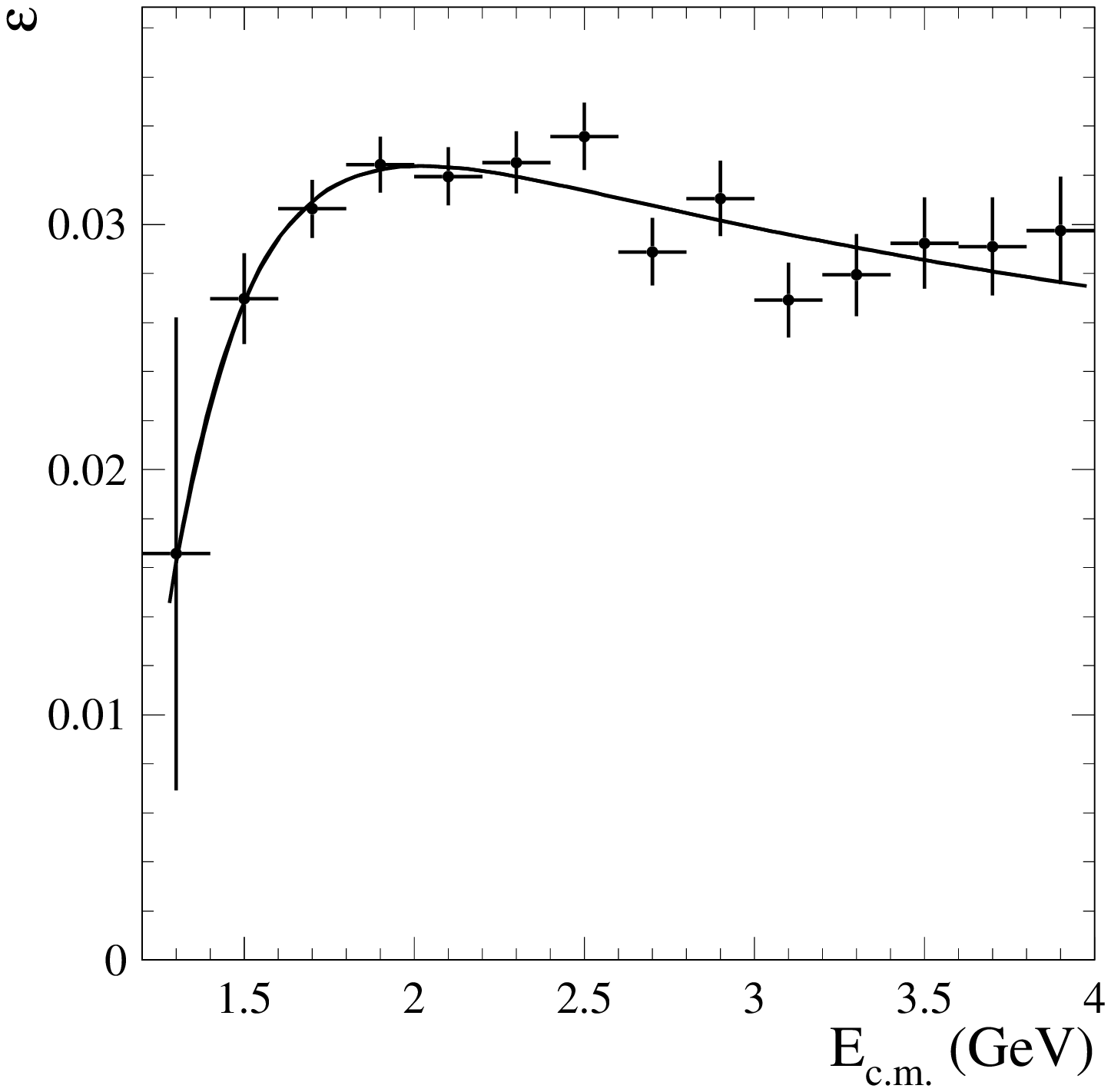}
\caption{ Detection efficiency for $\epem\to\KS\Kpm\pimp\piz$ events
as a function of the hadronic invariant mass $\Ecm=m(\KS\Kpm\pimp\piz)$.
The solid curve shows a fitted parametrization.}
\label{fig:effreg_kskpipi0}
\end{figure}

The reconstruction and selection efficiency for signal events is
determined from the  signal simulation, corrected for known
differences with respect to data.  The
efficiencies for charged-track, photon, and \KS reconstruction depend
on the momentum and polar angle of the particle.  The distributions
of these variables are well described by the simulation for all
relevant particles.  The total event detection efficiency from the
simulation, including the \KS\to\pip\pim branching fraction of 0.692~\cite{pdg}
is shown as a function of \Ecm in Fig.~\ref{fig:effreg_kskpipi0}.  A smooth
parametrization, shown by the solid line, is used.

The $\piz$ detection efficiency was studied in our previous analysis~\cite{pi0rec}
of $\epem\to\omega\gamma\to\pip\pim\piz\gamma$ events, yielding corrections
to the simulation as a function of the $\piz$ momentum and polar angle.
Applying these event-by-event to the signal simulation yields an
overall correction of +2$\pm$1\%, independent of \Ecm.
Similarly, we incorporate corrections to the charged-track and \KS
reconstruction efficiencies making use of the results found in our
previous studies of $\epem\to\pip\pim\pip\pim\gamma$~\cite{ar06020} and
$\epem\to\KS\KL\gamma$~\cite{ksklpipi} events, respectively, where the latter corrections
also depend on the flight length of the \KS meson transverse to
the beam direction.  Corrections of $+0.8\pm1.0$\% for each of the \pipm and
\Kpm, and $+1.1\pm1.0$\% for the \KS, are derived, again independent of $\Ecm$.
Similar corrections to the pion and kaon identification efficiencies amount to 0$\pm$2\%.

We study a possible data-MC difference in the shape of the $\chi^2$ distribution using the
\jpsi signal, which has negligible non-ISR background.  The increase
in the \jpsi yield when loosening the $\chi^2$ requirement from 20 to 200 
is consistent with the expectation from simulation, and
we estimate a correction of $+3.7\pm4.6$\%.

As a cross-check, using a fast simulation of the detector
response for computational simplicity, we compare the results
obtained for signal events generated with a phase-space model 
to those obtained for signal events generated with intermediate 
$\KS\pimp$ resonances, specifically $\epem\to\kstpm\KS\pimp$  and $\Kstarz\Kpm\pimp$.
No difference in efficiency
larger than 0.5\% is seen, and we assign a systematic uncertainty of 0.5\%
to account both for possible model dependence and for the choice
of parametrization of the efficiency as a function of \Ecm.  These
corrections and uncertainties are listed in Table~\ref{tab:kskpipi0_syst}.  The total
correction is +8.6\%
  
\subsection{The cross section for $\epem\to \KS\Kpm\pimp\piz$}
\label{sec:xsec}

The $\epem\to \KS\Kpm\pimp\piz$ cross section is obtained from:
\begin{equation}
\sigma(\Ecm) = 
\frac{dN_{\KS\Kpm\pimp\piz}(\Ecm)}{d\mathcal{L}(\Ecm)\varepsilon(\Ecm)R(\Ecm)},
\label{eq:xsec}
\end{equation}
where $\Ecm$ is the invariant mass of the $\KS\Kpm\pimp\piz$ system, 
$dN_{\KS\Kpm\pimp\piz}$  is the number of signal
$\KS\Kpm\pimp\piz$ events in the interval $d\Ecm$, $d\mathcal{L}(\Ecm)$ is the differential luminosity,
$\varepsilon(\Ecm)$ is the corrected efficiency
  discussed in Section~\ref{sec:efficiency}, 
and $R(\Ecm)$ is the correction to account for additional soft radiative photon emission from the initial state. 

The differential luminosity  $d\mathcal{L}(m)$ is calculated using the total PEP-II integrated 
luminosity $\mathcal{L} = 454\invfb$ and the probability density function for ISR photon emission. To first order it 
can be written as:
\begin{equation}
\frac{d\mathcal{L}}{dm} = \frac{\alpha}{\pi x}\left((2-2 x + x^2) \log\frac{1+C}{1-C}-x^2 
C\right)\frac{2 m}{s}\mathcal{L}.
\label{eq:lum}
\end{equation}
Here $m = m(\KS\Kpm\pimp\piz)$, $x = 1-m^2/s$, $C = \cos\theta^*_0$, and $\theta^*_0$ 
defines the acceptance of the analysis in the polar angle of the ISR
photon in the $\epem$  c.m. frame,
$\theta^*_0 < \theta^*_{\gamma} < 180^o - \theta^*_0$. Here, $\theta^*_0 = $ 20$^o$.

The radiative correction $R(\Ecm)$ 
is determined using generator-level MC (without simulation of the detector response) 
as the ratio of the $\KS\Kpm\pimp\piz$ spectrum with soft photon
emission to that at the Born level. 
We determine $R=1.0029\pm0.0065$, independent of $\Ecm$. 
The combined systematic uncertainty in the luminosity 
and radiative correction is estimated to be 1.4\%.

\begin{figure}[tbh]
\includegraphics[width=0.8\columnwidth]{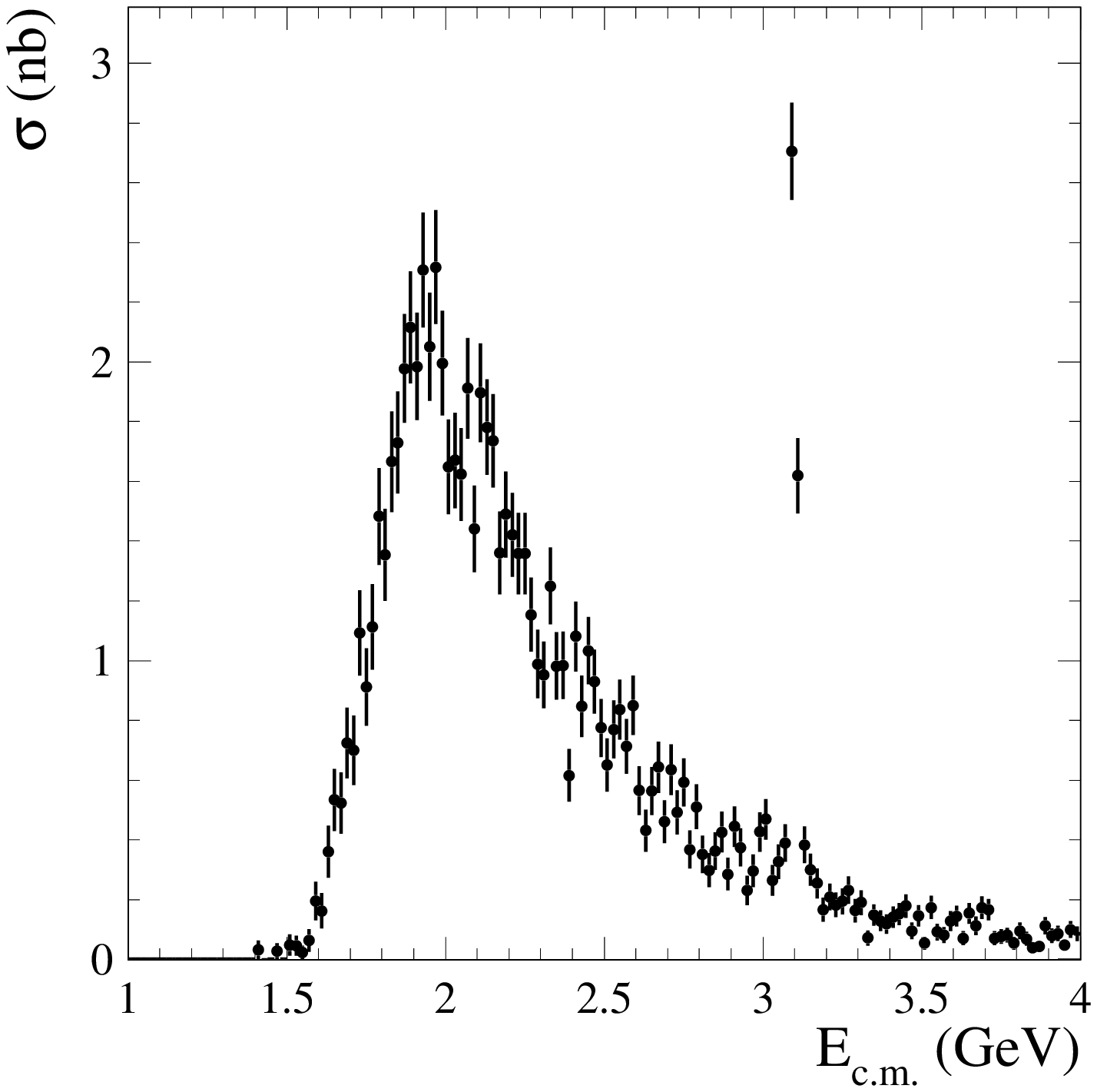}
\caption{Cross section for the process $\epem\to\KS\Kpm\pimp\piz$. The uncertainties are statistical.}
\label{fig:cs_kskpipi0}
\end{figure}

The fully corrected $\epem\to\KS\Kpm\pimp\piz$ cross section is shown
in Fig.~\ref{fig:cs_kskpipi0} and listed in Table~\ref{tab:cs_kskpipi0}, with statistical uncertainties.
The relative systematic uncertainties are summarized in Table~\ref{tab:kskpipi0_syst};
their total ranges from 6.2\% for $\Ecm<2$~\gev to 11.6\%
for $\Ecm>3$~\gev.
  
\begin{table}[tbh]
\caption{Summary of the corrections to, and systematic uncertainties in the
 \epem\to\KS\Kpm\pimp\piz cross section.}
\label{tab:kskpipi0_syst}
\begin{ruledtabular}
\begin{tabular}{l c l}
Source                                    &  Correction & Systematic  \\
                                          &   (\%)       & uncertainty (\%)   \\
\hline
$\piz$ reconstruction                     &     +2.0       &  \, 1.0          \\
$\Kpm$, $\pipm$  reconstruction           &     +1.6       &  \, 2.0          \\
$\KS$ reconstruction                      &     +1.1       &  \, 1.0          \\
PID efficiency                            &   \, 0.0       &  \, 2.0          \\
\chisq selection                          &     +3.7       &  \, 4.6          \\
Background subtraction                    &     ---        &  \, 2.5, $<$ 2.0 GeV          \\
                                          &                &  \, 4.2, 2.0-3.0 GeV    \\
                                          &                &   10.0, $>$ 3.0 GeV \\
Model acceptance                          &     ---        &  \, 0.5          \\
Luminosity and Rad.Corr.                  &     ---        &  \, 1.4          \\
\hline
 Total                                    &     +8.6       &  \, 6.3, $<$ 2.0 GeV         \\
                                          &                &  \, 7.1, 2.0-3.0 GeV   \\
                                          &                &   11.5, $>$ 3.0 GeV \\
\end{tabular}
\end{ruledtabular}
\end{table}

%\begin{linenomath}
\begin{table*}[tbh]
\caption{Measurements of the $\epem\to\KS\Kpm\pimp\piz$ cross section versus $\Ecm = m(\KS\Kpm\pimp\piz)$. The 
uncertainties are statistical only; systematic uncertainties are given in Table~\ref{tab:kskpipi0_syst}.}
\label{tab:cs_kskpipi0}
\begin{ruledtabular}
\begin{tabular}{c c c c c c c c c c}
 $E_{c.m.}$           & $\sigma$       &  $E_{c.m.}$ & $\sigma$ &  $E_{c.m.}$ & $\sigma$ & $E_{c.m.}$ & $\sigma$ & $E_{c.m.}$ & $\sigma$ \\
 (GeV)                 & (nb)           &  (GeV)      & (nb)     &  (GeV)      & (nb)     & (GeV)      & (nb)     & (GeV)  & (nb)     \\
\hline
 1.51 & 0.05 $\pm$ 0.03 & 2.01 & 1.65 $\pm$ 0.16 & 2.51 & 0.65 $\pm$ 0.09 & 3.01 & 0.47 $\pm$ 0.07 & 3.61 & 0.14 $\pm$ 0.03 \\
 1.53 & 0.05 $\pm$ 0.03 & 2.03 & 1.67 $\pm$ 0.16 & 2.53 & 0.77 $\pm$ 0.10 & 3.03 & 0.26 $\pm$ 0.05 & 3.63 & 0.07 $\pm$ 0.02 \\
 1.55 & 0.02 $\pm$ 0.02 & 2.05 & 1.62 $\pm$ 0.16 & 2.55 & 0.83 $\pm$ 0.10 & 3.05 & 0.33 $\pm$ 0.06 & 3.65 & 0.15 $\pm$ 0.04 \\
 1.57 & 0.06 $\pm$ 0.04 & 2.07 & 1.91 $\pm$ 0.17 & 2.57 & 0.71 $\pm$ 0.09 & 3.07 & 0.39 $\pm$ 0.06 & 3.67 & 0.11 $\pm$ 0.03 \\
 1.59 & 0.19 $\pm$ 0.06 & 2.09 & 1.44 $\pm$ 0.15 & 2.59 & 0.85 $\pm$ 0.10 & 3.09 & 2.69 $\pm$ 0.16 & 3.69 & 0.17 $\pm$ 0.04 \\
 1.61 & 0.16 $\pm$ 0.06 & 2.11 & 1.90 $\pm$ 0.17 & 2.61 & 0.56 $\pm$ 0.08 & 3.11 & 1.61 $\pm$ 0.13 & 3.71 & 0.16 $\pm$ 0.04 \\
 1.63 & 0.36 $\pm$ 0.09 & 2.13 & 1.78 $\pm$ 0.16 & 2.63 & 0.43 $\pm$ 0.07 & 3.13 & 0.38 $\pm$ 0.06 & 3.73 & 0.07 $\pm$ 0.02 \\
 1.65 & 0.53 $\pm$ 0.10 & 2.15 & 1.73 $\pm$ 0.16 & 2.65 & 0.56 $\pm$ 0.08 & 3.15 & 0.30 $\pm$ 0.05 & 3.75 & 0.08 $\pm$ 0.02 \\
 1.67 & 0.52 $\pm$ 0.10 & 2.17 & 1.36 $\pm$ 0.14 & 2.67 & 0.64 $\pm$ 0.09 & 3.17 & 0.25 $\pm$ 0.05 & 3.77 & 0.08 $\pm$ 0.03 \\
 1.69 & 0.72 $\pm$ 0.12 & 2.19 & 1.49 $\pm$ 0.14 & 2.69 & 0.46 $\pm$ 0.07 & 3.19 & 0.16 $\pm$ 0.04 & 3.79 & 0.05 $\pm$ 0.02 \\
 1.71 & 0.70 $\pm$ 0.12 & 2.21 & 1.42 $\pm$ 0.14 & 2.71 & 0.63 $\pm$ 0.08 & 3.21 & 0.21 $\pm$ 0.04 & 3.81 & 0.09 $\pm$ 0.03 \\
 1.73 & 1.09 $\pm$ 0.14 & 2.23 & 1.36 $\pm$ 0.14 & 2.73 & 0.49 $\pm$ 0.07 & 3.23 & 0.18 $\pm$ 0.04 & 3.83 & 0.07 $\pm$ 0.02 \\
 1.75 & 0.91 $\pm$ 0.13 & 2.25 & 1.36 $\pm$ 0.14 & 2.75 & 0.59 $\pm$ 0.08 & 3.25 & 0.19 $\pm$ 0.04 & 3.85 & 0.04 $\pm$ 0.02 \\
 1.77 & 1.11 $\pm$ 0.14 & 2.27 & 1.15 $\pm$ 0.12 & 2.77 & 0.37 $\pm$ 0.06 & 3.27 & 0.23 $\pm$ 0.05 & 3.87 & 0.04 $\pm$ 0.02 \\
 1.79 & 1.48 $\pm$ 0.16 & 2.29 & 0.99 $\pm$ 0.12 & 2.79 & 0.51 $\pm$ 0.07 & 3.29 & 0.16 $\pm$ 0.04 & 3.89 & 0.11 $\pm$ 0.03 \\
 1.81 & 1.35 $\pm$ 0.15 & 2.31 & 0.95 $\pm$ 0.11 & 2.81 & 0.35 $\pm$ 0.06 & 3.31 & 0.19 $\pm$ 0.04 & 3.51 & 0.05 $\pm$ 0.02 \\
 1.83 & 1.67 $\pm$ 0.17 & 2.33 & 1.25 $\pm$ 0.13 & 2.83 & 0.30 $\pm$ 0.06 & 3.33 & 0.07 $\pm$ 0.03 & 3.53 & 0.17 $\pm$ 0.04 \\
 1.85 & 1.73 $\pm$ 0.17 & 2.35 & 0.98 $\pm$ 0.11 & 2.85 & 0.36 $\pm$ 0.06 & 3.35 & 0.15 $\pm$ 0.04 & 3.55 & 0.09 $\pm$ 0.03 \\
 1.87 & 1.98 $\pm$ 0.18 & 2.37 & 0.98 $\pm$ 0.11 & 2.87 & 0.42 $\pm$ 0.07 & 3.37 & 0.13 $\pm$ 0.03 & 3.57 & 0.08 $\pm$ 0.03 \\
 1.89 & 2.12 $\pm$ 0.19 & 2.39 & 0.61 $\pm$ 0.09 & 2.89 & 0.28 $\pm$ 0.05 & 3.39 & 0.12 $\pm$ 0.03 & 3.59 & 0.13 $\pm$ 0.03 \\
 1.91 & 1.99 $\pm$ 0.18 & 2.41 & 1.08 $\pm$ 0.12 & 2.91 & 0.44 $\pm$ 0.07 & 3.41 & 0.14 $\pm$ 0.03 & 3.91 & 0.08 $\pm$ 0.02 \\
 1.93 & 2.31 $\pm$ 0.19 & 2.43 & 0.84 $\pm$ 0.10 & 2.93 & 0.37 $\pm$ 0.06 & 3.43 & 0.15 $\pm$ 0.04 & 3.93 & 0.08 $\pm$ 0.03 \\
 1.95 & 2.05 $\pm$ 0.18 & 2.45 & 1.03 $\pm$ 0.11 & 2.95 & 0.23 $\pm$ 0.05 & 3.45 & 0.18 $\pm$ 0.04 & 3.95 & 0.05 $\pm$ 0.02 \\
 1.97 & 2.32 $\pm$ 0.19 & 2.47 & 0.93 $\pm$ 0.11 & 2.97 & 0.29 $\pm$ 0.06 & 3.47 & 0.09 $\pm$ 0.03 & 3.97 & 0.10 $\pm$ 0.03 \\
 1.99 & 2.00 $\pm$ 0.18 & 2.49 & 0.77 $\pm$ 0.10 & 2.99 & 0.42 $\pm$ 0.07 & 3.49 & 0.14 $\pm$ 0.04 & 3.99 & 0.08 $\pm$ 0.02 \\

\end{tabular}
\end{ruledtabular}
\end{table*}
%\end{linenomath}

\subsection{Substructure in the $\KS\Kpm\pimp\piz$ final state}
\label{sec:kskpipi0_substr}
Previously, we studied single \kst production in the processes
$\epem\to\KS\Kpm\pimp$ and $\Kp\Km\piz$~\cite{kskpi},  and double
\kst production, as well as $\phi$, $\rho$, and $f_0$ production, in
$\epem\to\Kp\Km\pip\pim$, $\Kp\Km\piz\piz$~\cite{kkpipi} and $\KS\KL\pip\pim$~\cite{ksklpipi}.
Here, we expect single \kst, double \kst, $\rho$, and possibly
other resonance contributions, but the statistical precision of the data sample
is insufficient for competitive
measurements of such processes.  Since it is important to confirm, as
far as possible, resonant cross sections measured in different final
states, and to verify expected isospin relations, we perform a simple
study of those resonant subprocesses accessible with our data.

\begin{figure}[tbh]  
\begin{tabular}{cc}
\includegraphics[width=.5\columnwidth]{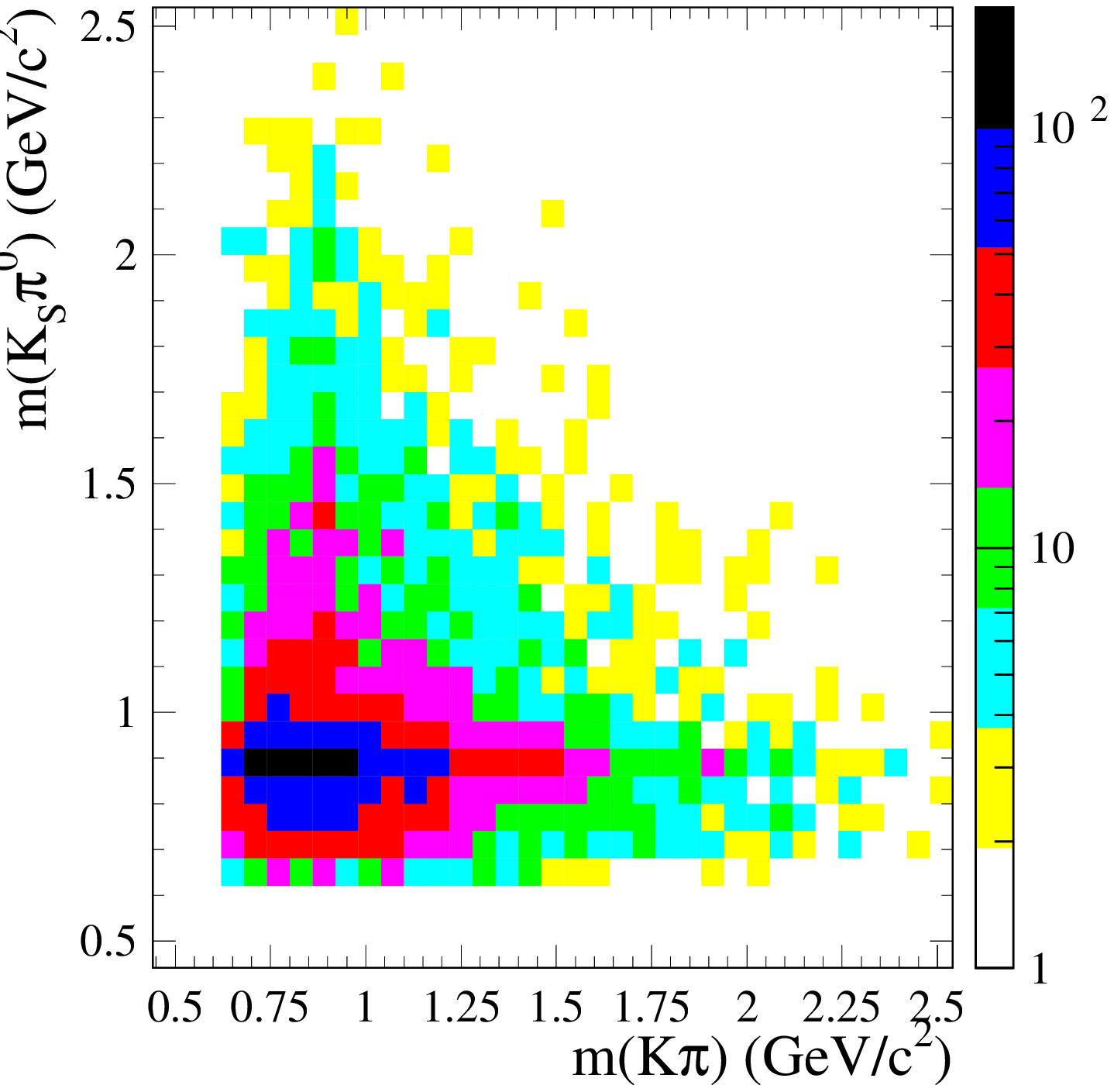}
\put(-55,105){(a)} 
\includegraphics[width=.5\columnwidth]{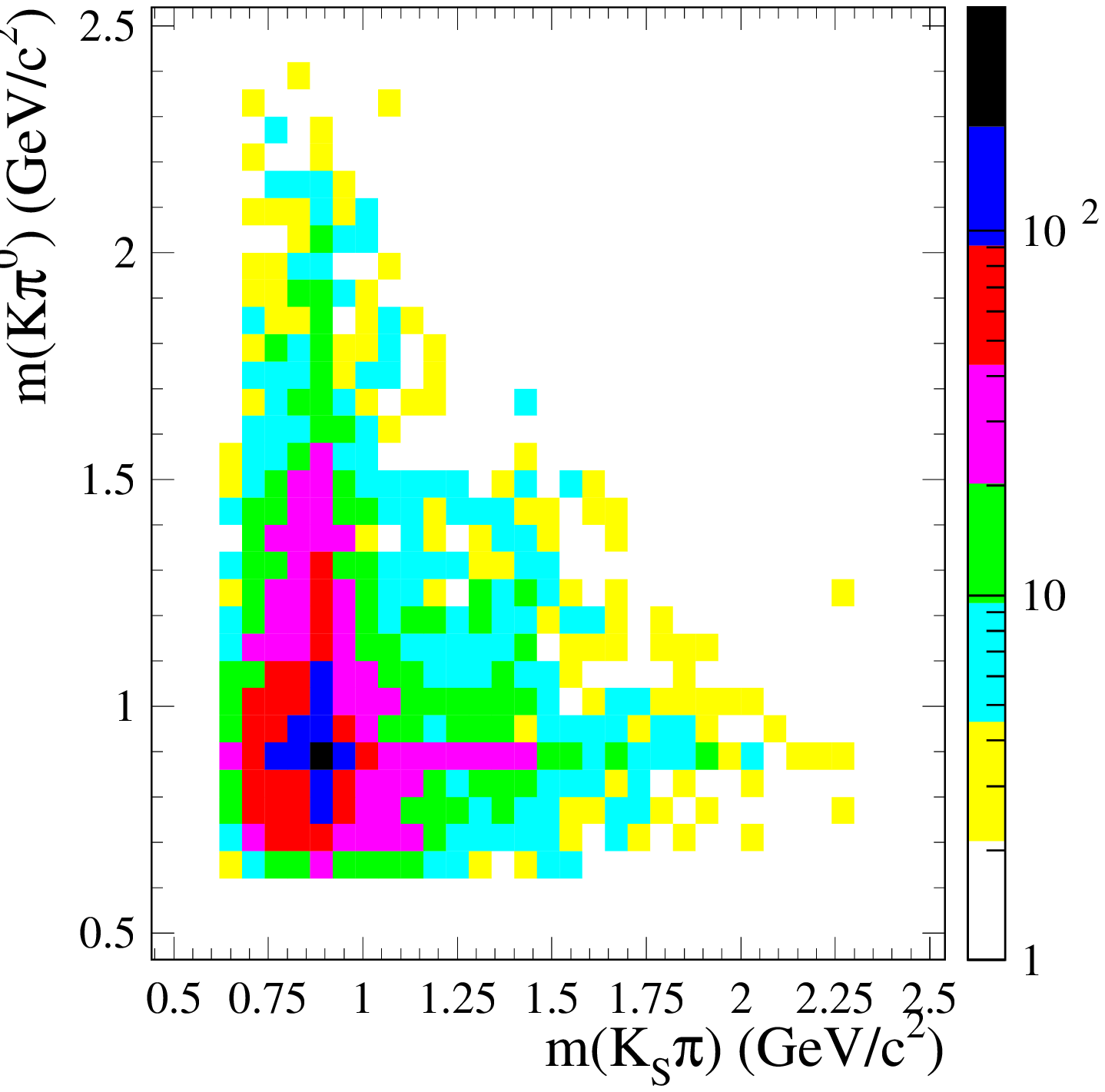}
\put(-55,105){(b)} 
\end{tabular}
\caption{Scatter plots of (a) the $\KS\piz$ vs.\ $\Kpm\pimp$ and (b) $\Kpm\piz$ vs.\
$\KS\pimp$ invariant masses in $\epem\to\KS\Kpm\pimp\piz$ events.}
\label{fig:kstar}
\end{figure}

\begin{figure}[t]  
\begin{tabular}{cc}
\includegraphics[width=.5\columnwidth]{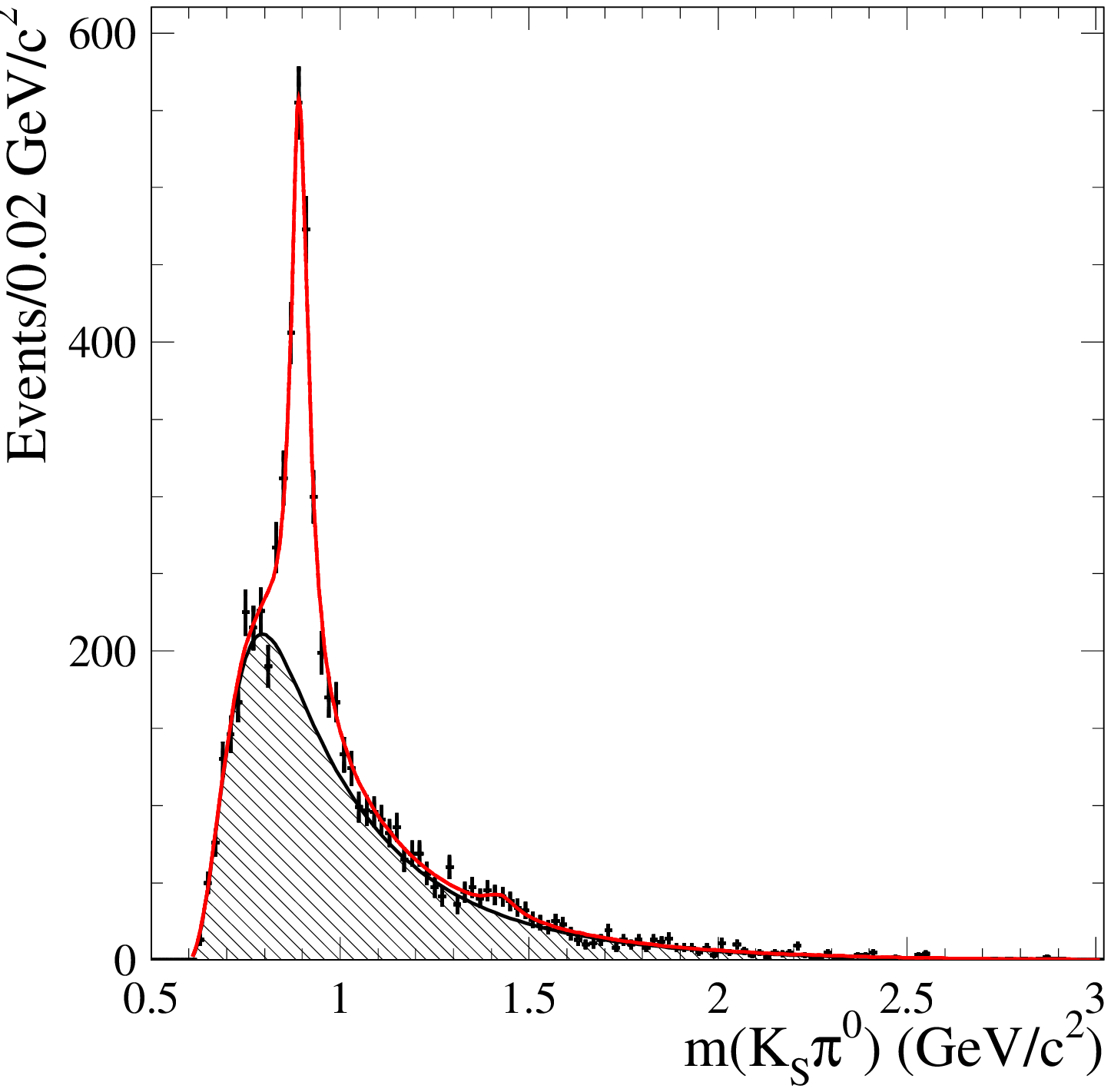}
\put(-25,105){(a)} 
\includegraphics[width=.5\columnwidth]{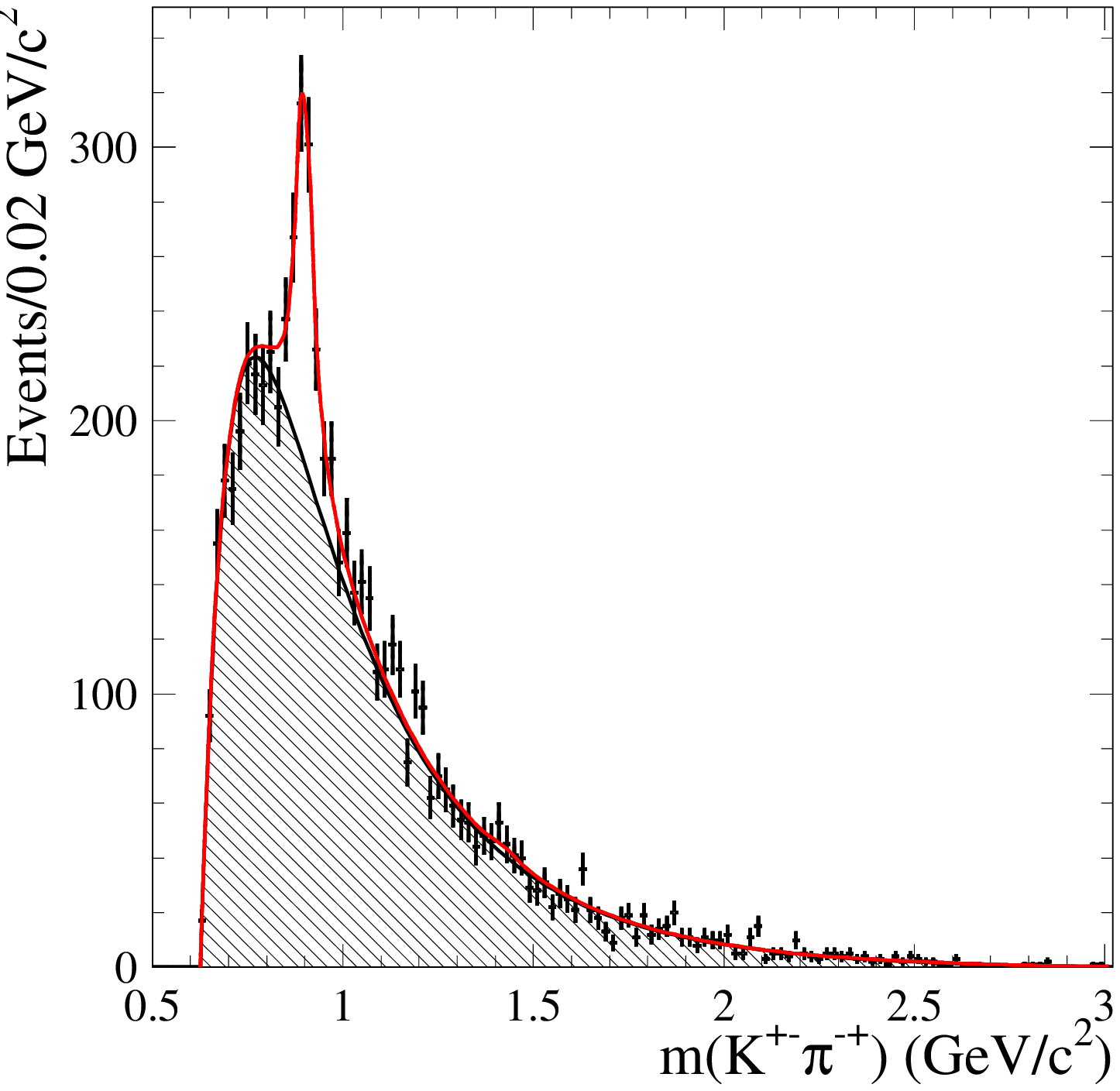}
\put(-25,105){(b)} 
\\
\includegraphics[width=.5\columnwidth]{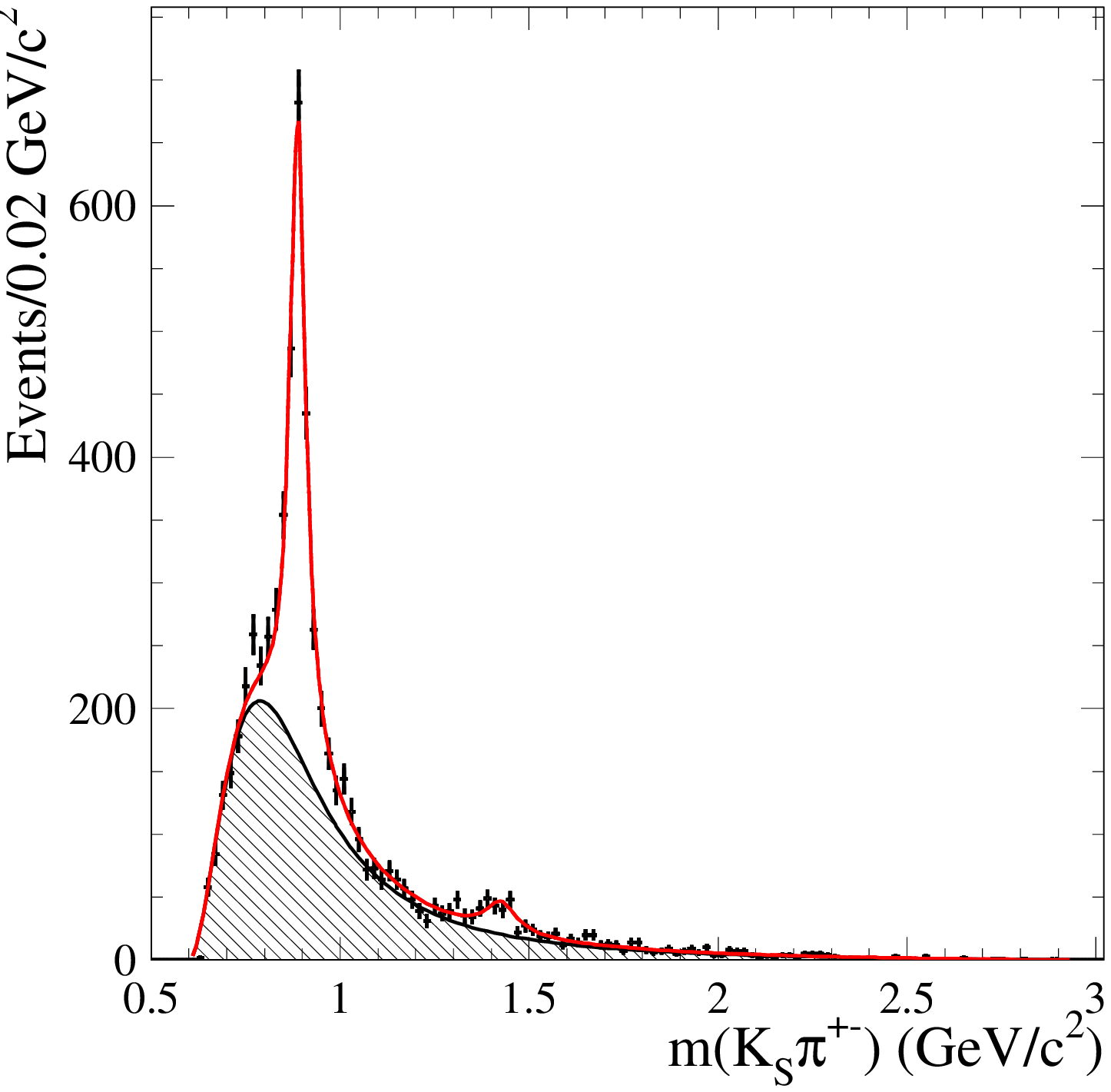}
\put(-25,105){(c)} 
\includegraphics[width=.5\columnwidth]{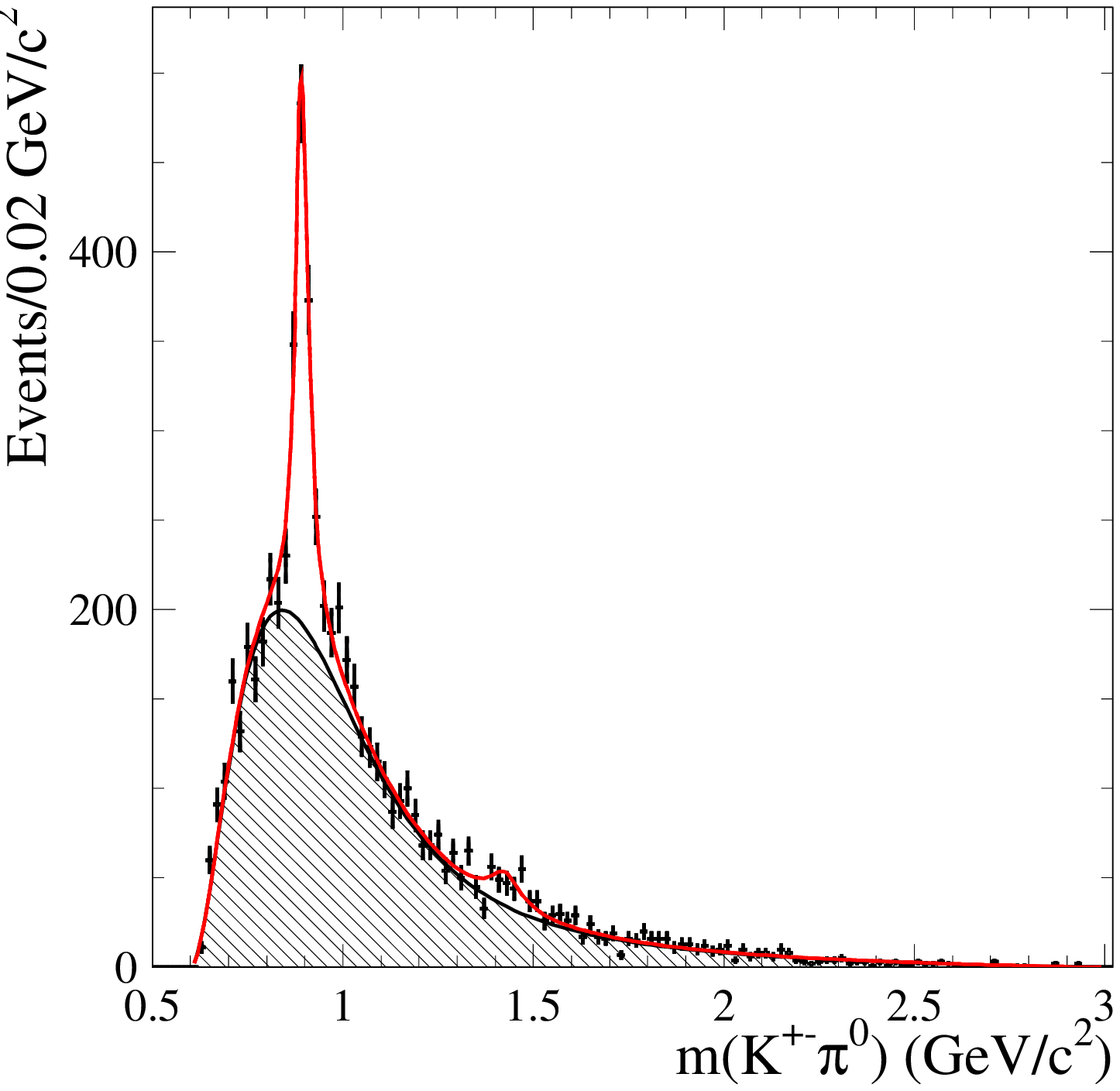}
\put(-25,105){(d)} 
\end{tabular}
\caption{The (a) $\KS\piz$ and (b) $\Kpm\pimp$ invariant mass projections of
Fig.~\ref{fig:kstar}(a), and the (c) $\KS\pipm$ and (d) $\Kpm\piz$ invariant mass
projections of Fig.~\ref{fig:kstar}(b).  The lines represent the results of the fits
described in the text, with the hatched areas denoting their non-resonant components.}
\label{fig:kst0kst0}
\end{figure}

Decays of the \jpsi are discussed below (Sec.~\ref{sec:jpsi}), and for the 
study presented in this section
we exclude the region $3.0<\Ecm<3.2$~\gev.  Figure~\ref{fig:kstar}(a) shows a scatter plot of
the $\KS\piz$ vs.\ $\Kpm\pimp$ invariant masses in the selected data
sample, corrected for backgrounds as described above, while Fig.~\ref{fig:kstar}(b) 
shows the $\Kpm\piz$ vs.\ \KS\pimp masses.  Clear signals for charged
and neutral \kstz states are seen.  Figure~\ref{fig:kst0kst0}(a) is the projection of
Fig.~\ref{fig:kstar}(a) onto the vertical axis, and shows a large \kstz peak as well
as possible structure near 1.43~\gevcc.  This could arise from the
\ktwo or \Kstzero resonances, or any combination.  We cannot study
this structure in detail, but must take it into account in any fit.

We fit this distribution with a sum of two incoherent resonances 
and a non-resonant (NR) component.  The \kstz is described by a
relativistic P-wave Breit-Wigner (BW) function with a threshold term,
with mass and width fixed to the world-average values~\cite{pdg}.  The NR
function is the product of a fifth-order polynomial in the inverse of
the mass and an exponential cutoff at threshold.  The second peak is
described by a relativistic D- or S-wave BW with parameters fixed to
the nominal values~\cite{pdg} for \ktwo or \Kstzero. The narrower
\ktwo gives better fits here and in most cases below, so we use it
everywhere.  The result of the fit is shown as the line in Fig.~\ref{fig:kst0kst0}(a),
with the NR component indicated by the hatched area.

The fit yields $1671\pm 60$ $\kstz\Kpm\pimp$ events and $85\pm24$
 $\ktwo^{\star}\Kpm\pimp$ events, where the uncertainties are statistical only.  We 
 do not claim observation of any particular state near 1.43~\gevcc, but  
we quote a generic number of events from this fit and those below for completeness.  
Some of the $\Kstarz(892)\Kpm\pimp$ events are
produced through the $\Kstarz(892)\Kstarzb$ channel, which we study below.  In
order to avoid double counting, we subtract the latter yield to obtain
$1533\pm 60$ quasi-three-body $\kstz\Kpm\pimp$ events.  

The projection of Fig.~\ref{fig:kstar}(a) onto the horizontal axis is 
shown in Fig.~\ref{fig:kst0kst0}(b), along with the results of a corresponding fit, which, after $\Kstarz(892)\Kstarzb(892)$ 
subtraction, yields 454$\pm$60 $\kstz\KS\piz$ and 20$\pm$25 $\ktwo\KS\piz$ events, respectively.  

Corresponding fits to the projections of Fig.~\ref{fig:kstar}(b), shown in
Figs. ~\ref{fig:kst0kst0}(c) and ~\ref{fig:kst0kst0}(d), followed by $\kstp\kstm$ subtraction, yield
 1173$\pm$64 $\kstpm\Kmp\piz$ events, 157$\pm$50 $\kstpm\KS\pimp$ events,  187$\pm$25 $\ktwo\Kmp\piz$ events, and
141$\pm$27 $\ktwo\KS\pimp$ events.  The uncertainties are statistical only; systematic uncertainties are discussed below.

\begin{figure}[h]
\begin{tabular}{cc}
\includegraphics[width=0.5\columnwidth]{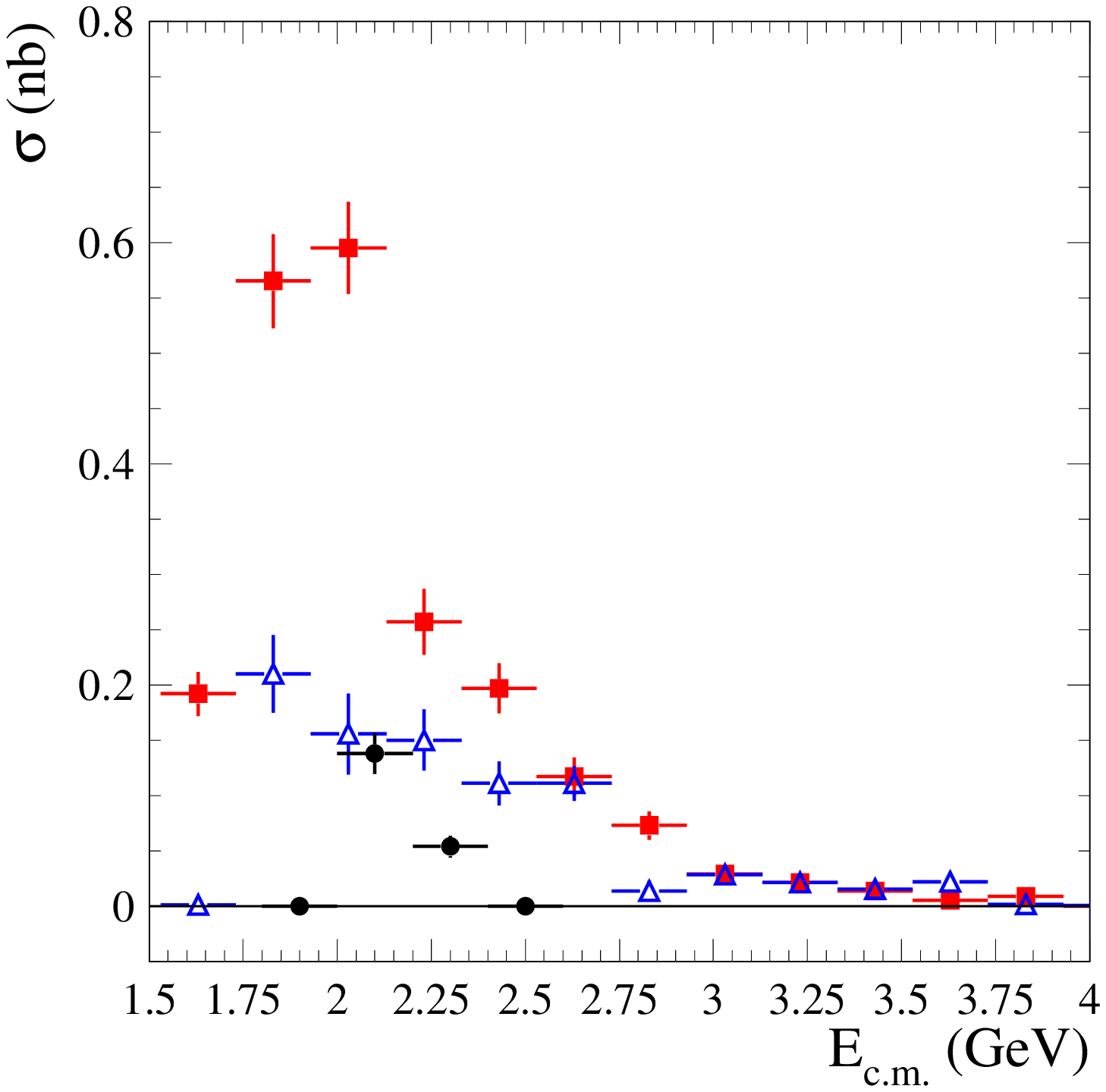}
\put(-25,105){(a)}
\includegraphics[width=0.5\columnwidth]{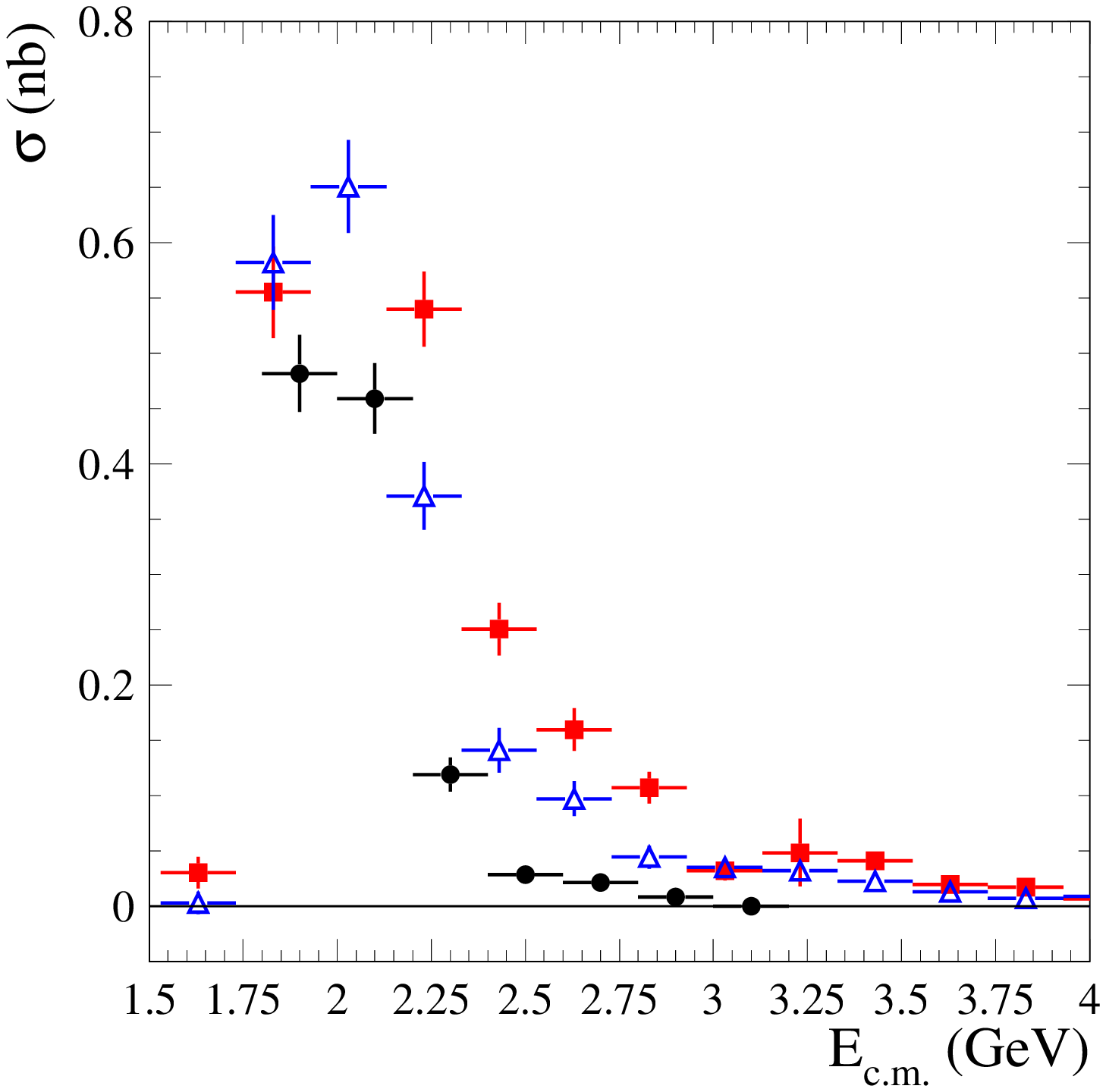}
\put(-25,105){(b)}
\end{tabular}
\caption{Cross sections for (a) the processes $\epem\to\Kstarz\Kpm\pimp$(squares),
 $\epem\to\Kstarz\KS\piz$(triangles), and $\epem\to\Kstarz\Kstarzb$(circles), and (b) the
 processes $\epem\to\kstpm\KS\pimp$(squares), $\epem\to\kstpm\Kmp\piz$
 (triangles), and $\epem\to\kstp\kstm$ (circles).  The uncertainties are statistical
 only, and in each case the $\Kstar\Kstarb$ cross section is included in
 both of the others.}  
\label{fig:cskst0kst0}
\end{figure}

Repeating these fits in 0.2~\gev bins of \Ecm, and using Eq.~(\ref{eq:xsec}),
we extract the cross sections for the
processes $\epem\to\kstz\Kpm\pimp$, $\kstz\to\KS\piz$, and 
 $\epem\to\kstz\KS\piz$, $\kstz\to\Kpm\pimp$ shown in Fig.~\ref{fig:cskst0kst0}(a), as well
 as for the processes $\epem\to\kstpm\KS\pimp$, $\kstpm\to\Kpm\piz$ and
 $\epem\to\kstpm\Kmp\piz$, $\kstpm\to\KS\pimp$ shown in Fig.~\ref{fig:cskst0kst0}(b).  They
 are similar in size and shape, except that the $\Kstarz\KS\piz$ cross section is a factor of 
2~--~3 lower.  Accounting for the $K^*(892)$ branching
 fractions, the $\kstz\Kpm\pimp$ and $\kstpm\Kmp\piz$ cross sections are
 consistent with those we measured previously~\cite{kkpipi} in the
 $\Kp\Km\pip\pim$ and $\Kp\Km\piz\piz$ final states, respectively, and the 
 $\kstpm\KS\pimp$ cross section is consistent with our previous
 measurement~\cite{ksklpipi} in the $\KS\KS\pip\pim$ final state.

\begin{figure}[h]
\begin{tabular}{cc}
\includegraphics[width=0.5\columnwidth]{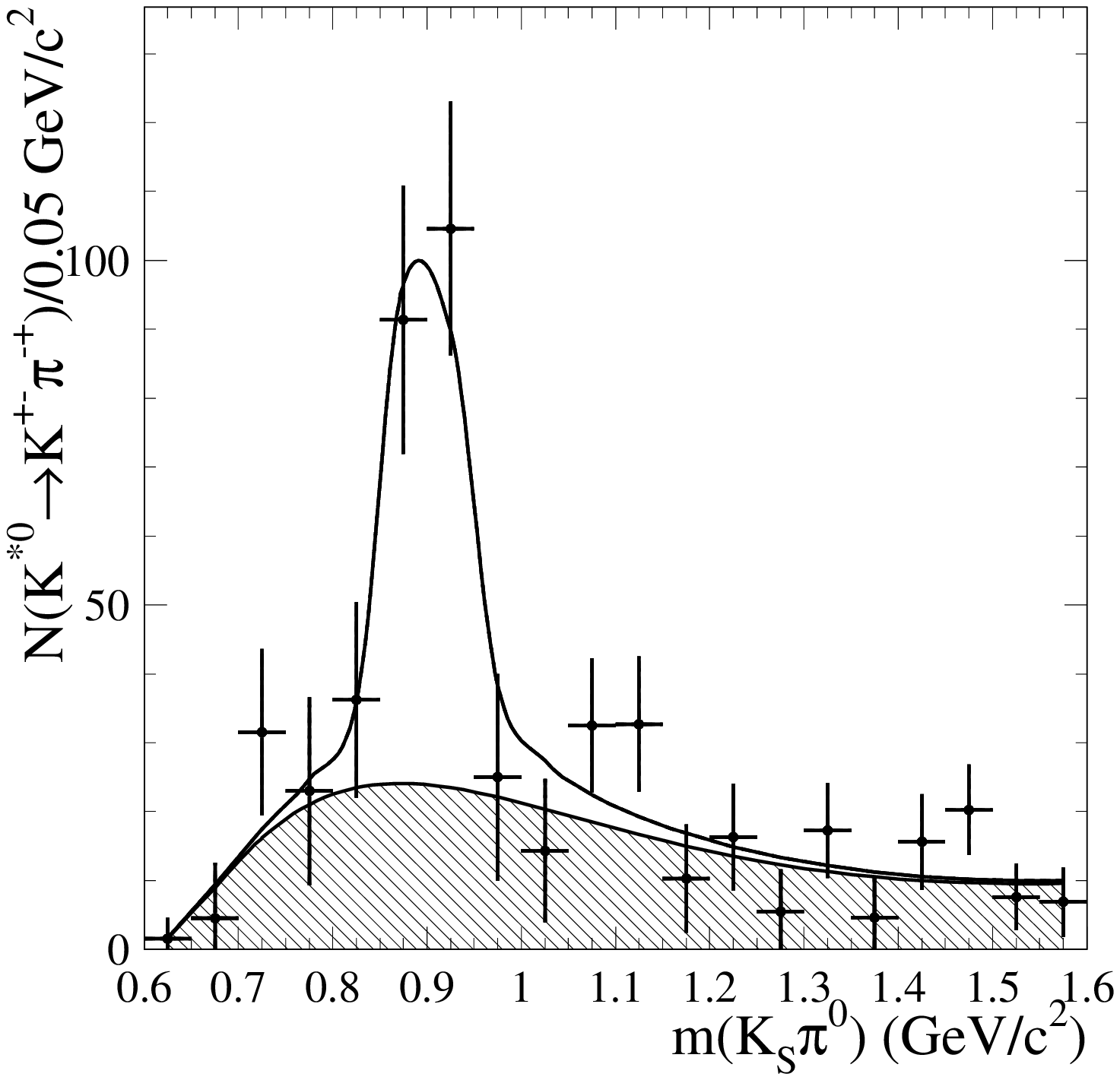}
\put(-25,105){(a)}
\includegraphics[width=0.5\columnwidth]{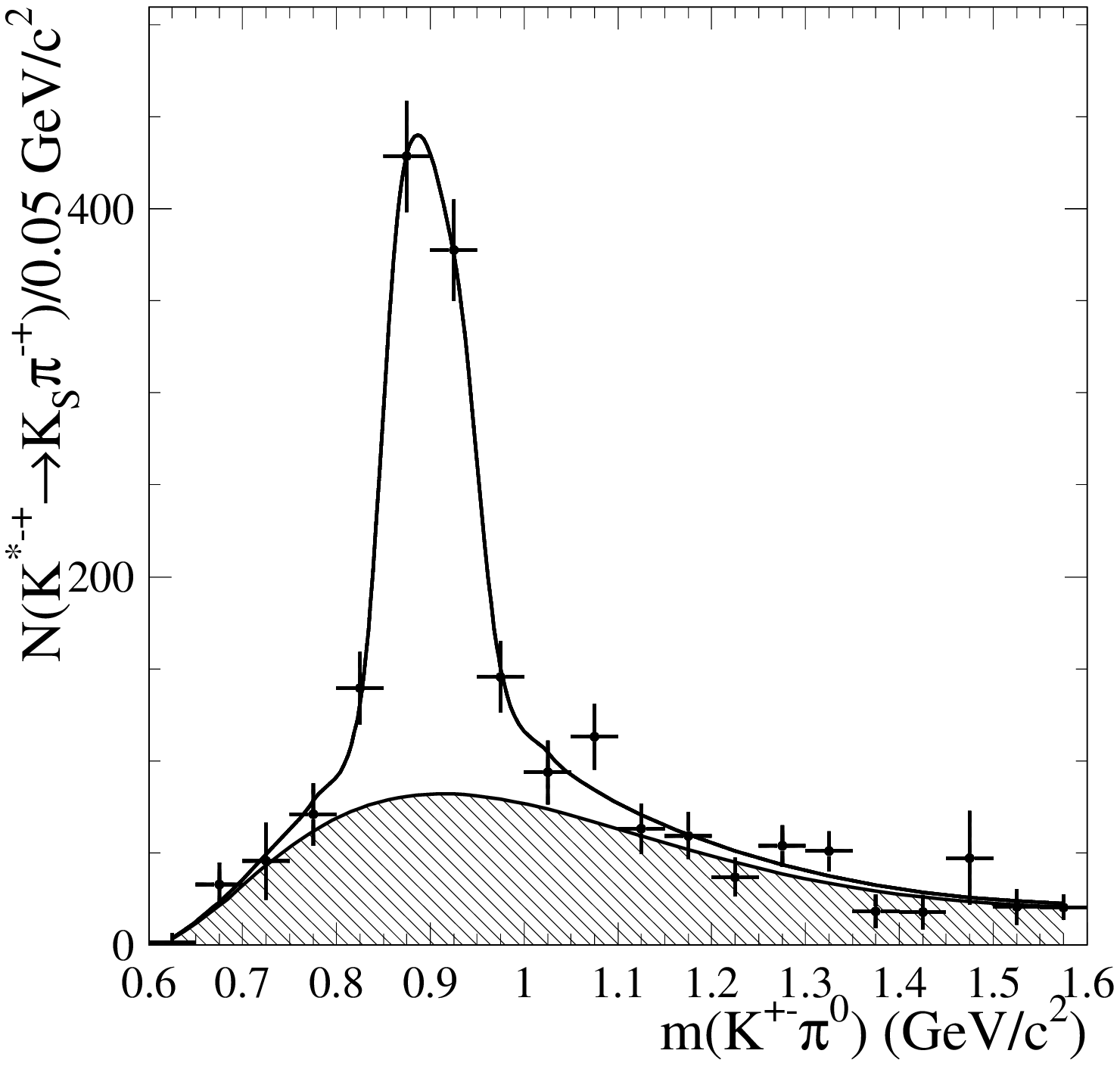}
\put(-25,105){(b)}
\end{tabular}
\caption{The number of events in the $\KS\Kpm\pimp\piz$ sample containing (a) a
$\Kstarz\to\Kpm\pimp$ decay as a function of the $\KS\piz$ invariant mass,
and (b) a $\kstpm\to\KS\pipm$ decay as a function of the $\Kpm\piz$
invariant mass.  The lines represent the result of the fits described
in the text, with the hatched areas denoting their non-resonant
components.}
\label{fig:nkst0corr}
\end{figure}

We investigate the correlated production of \Kstarz and
$\Kstarzb$ directly by repeating the fit of the $\Kpm\pimp$ invariant mass
distribution in 0.05~\gevcc bins of the $\KS\piz$ invariant mass.  The
resulting numbers of \kstz decays in each bin are shown in Fig.~\ref{fig:nkst0corr}(a),
and there is a substantial peak near 892~\mevcc.  Fitting these
points with the same NR function plus a single BW function yields
138$\pm$16 $\epem\to\Kstarz\Kstarzb$ events.  Similarly, fitting the $\KS\pipm$
invariant-mass distribution in bins of the $\Kpm\piz$ invariant mass
yields the results for \kstpm decays shown in Fig.~\ref{fig:nkst0corr}(b), and a
single-resonance plus NR fit to those results yields 814$\pm$36
$\epem\to\kstp\kstm$ events.  Repeating this procedure in 0.2~\gevcc  
bins of \Ecm, and applying Eq.~(\ref{eq:xsec}) provides the cross sections for $\epem\to\Kstarz\Kstarzb\to\KS\Kpm\pimp\piz$ 
and $\epem\to\kstpm\kst^{\mp}\to\KS\Kpm\pimp\piz$ shown in Figs.~\ref{fig:cskst0kst0}(a) and~\ref{fig:cskst0kst0}(b), 
respectively.  

The $\kstp\kstm$ intermediate state dominates both 
$\kstpm\KS\pimp$ and $\kstpm\Kmp\piz$ production, whereas the
$\Kstarz\Kstarzb$ intermediate state (Fig.~\ref{fig:cskst0kst0}(a)) provides a significant
fraction of $\kstz\K\pi$ production only near 2.1~\gev.  Accounting
for the $K^*(892)$ branching fractions, the
 $\kstp\kstm$ cross section is consistent with our previous measurement~\cite{kkpipi}
  in  the $\Kp\Km\piz\piz$ final state, where it also dominated 
 $\kstpm\Kmp\piz$ production, and the $\Kstarz\Kstarzb$ cross section is
 consistent with our previous measurement~\cite{kkpipi} in the $\Kp\Km\pip\pim$
 final state, where it also represented only a small fraction of 
 $\kstz\Kp\pim$ and $\Kstarzb\Km\pip$ production.

\begin{figure}[t]
\begin{tabular}{cc}
\includegraphics[width=0.5\columnwidth]{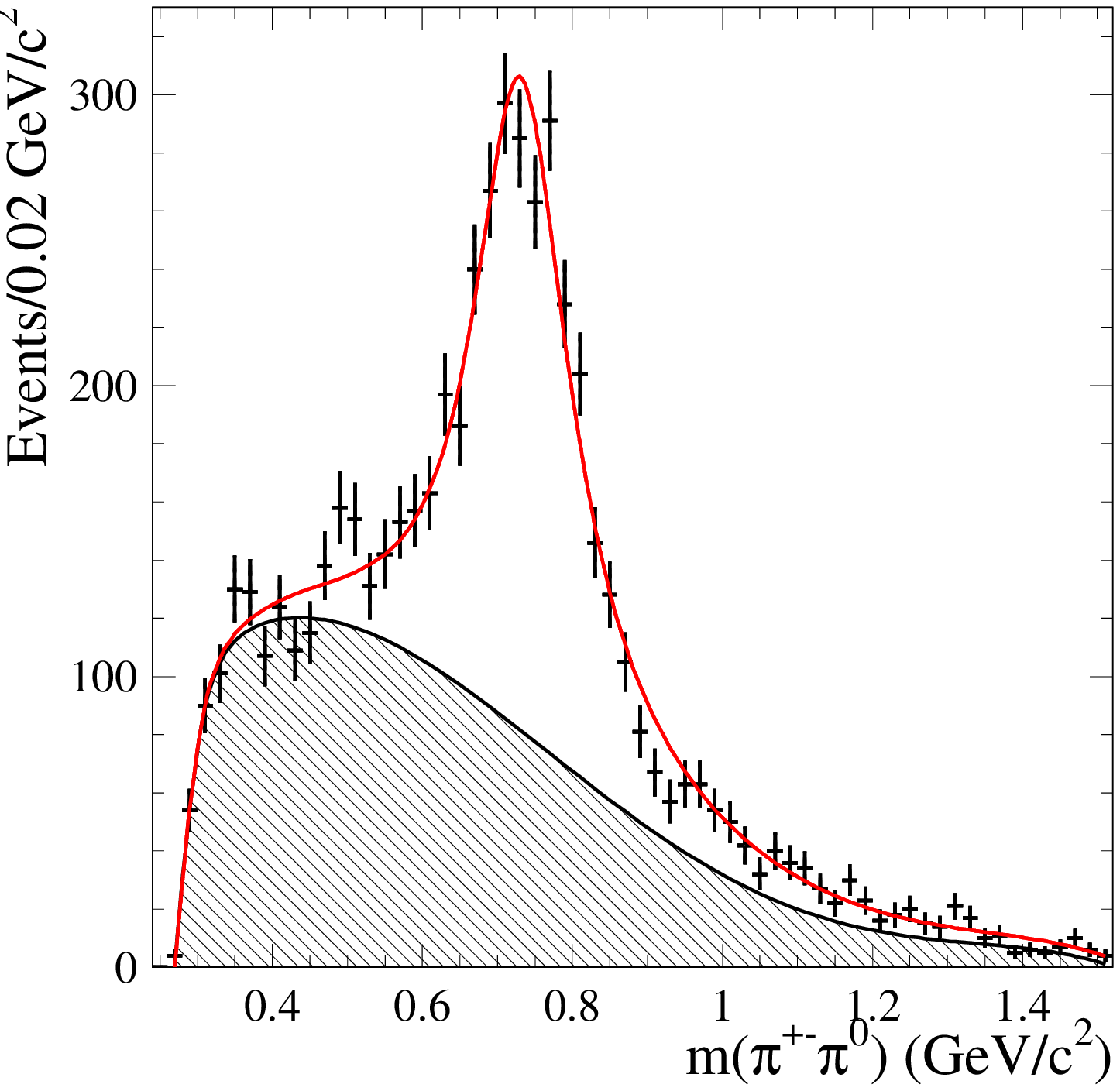} &
\put(-25,105){(a)}
\includegraphics[width=0.5\columnwidth]{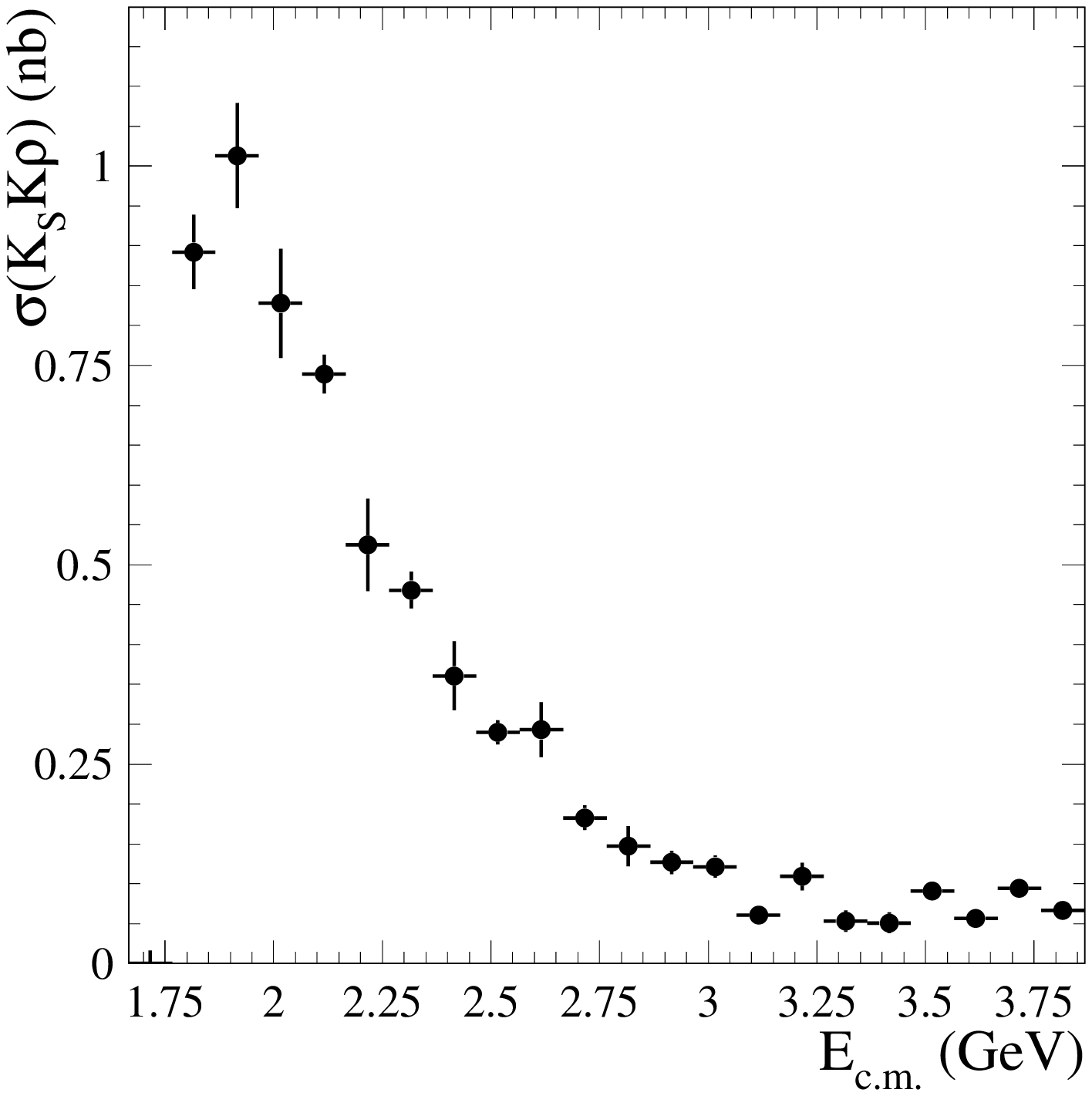} 
\put(-25,105){(b)} \\
\end{tabular}
\caption{(a) The $\pipm\piz$ invariant-mass distribution in
$\epem\to\KS\Kpm\pimp\piz$ events.  The line represents the result of
the fit described in the text, with the hatched area denoting its
non-resonant component.  (b) Cross section for the process
$\epem\to\KS\Kpm\rho^{\mp}$ as a function of \Ecm.  The uncertainties are
statistical only.}
\label{fig:rho}
\end{figure}   

Figure~\ref{fig:rho}(a) shows the distribution of the $\pipm\piz$ invariant mass in
selected, background-subtracted, $\KS\Kpm\pimp\piz$ events, which features a prominent
$\rho(770)$ peak.  The limited size of the data sample precludes
a detailed study of the $\rho$ region, and insteaad we perform a simple
fit, using the the same NR function plus a relativistic P-wave BW
with parameters fixed to those of the $\rho(770)^{\pm}$~\cite{pdg}.  The result is
shown as the line and hatched area in Fig.~\ref{fig:rho}(a). The fitted number
of $\KS\Kpm\rho^{\mp}$ events, $2498 \pm 100$, is a large fraction of the
$\KS\Kpm\pimp\piz$ signal.  Again, the uncertainty is statistical only, and systematic
uncertainties, discussed below, are large.

Repeating this fit in 0.1~\gev bins of \Ecm and using Eq.~(\ref{eq:xsec}), we
extract the cross section for the process $\epem\to\KS\Kpm\rho^{\mp}$,
shown in Fig.~\ref{fig:rho}(b).  It peaks at lower \Ecm and at approximately twice the
value of a typical $\kst K\pi$ cross section, and is consistent with
our previous measurement of the $\Kp\Km\rho^0$ cross section~\cite{kkpipi}.

\begin{figure}[t]  
\begin{tabular}{cc}
\includegraphics[width=.5\columnwidth]{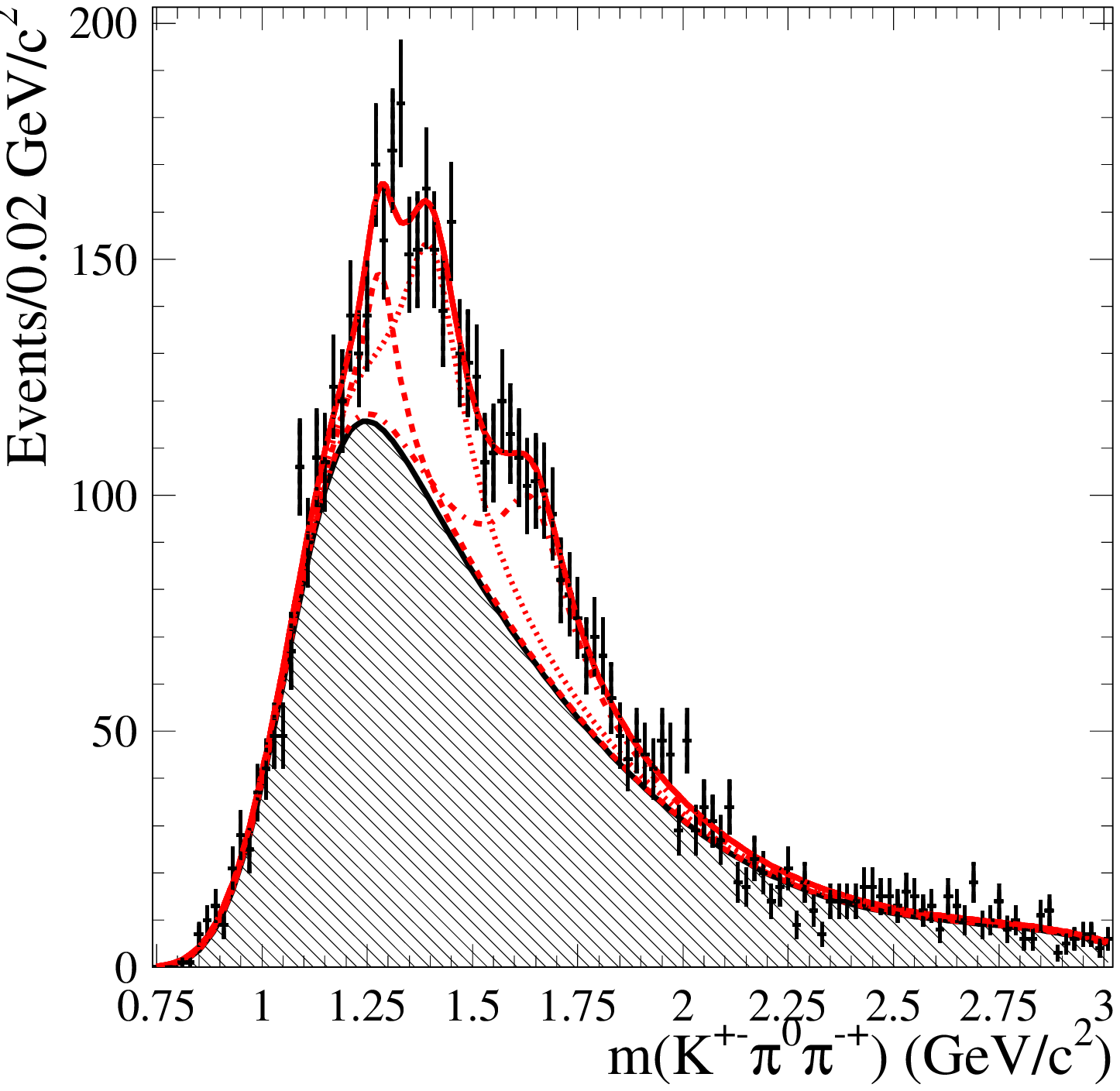}
\put(-25,105){(a)} 
\includegraphics[width=.5\columnwidth]{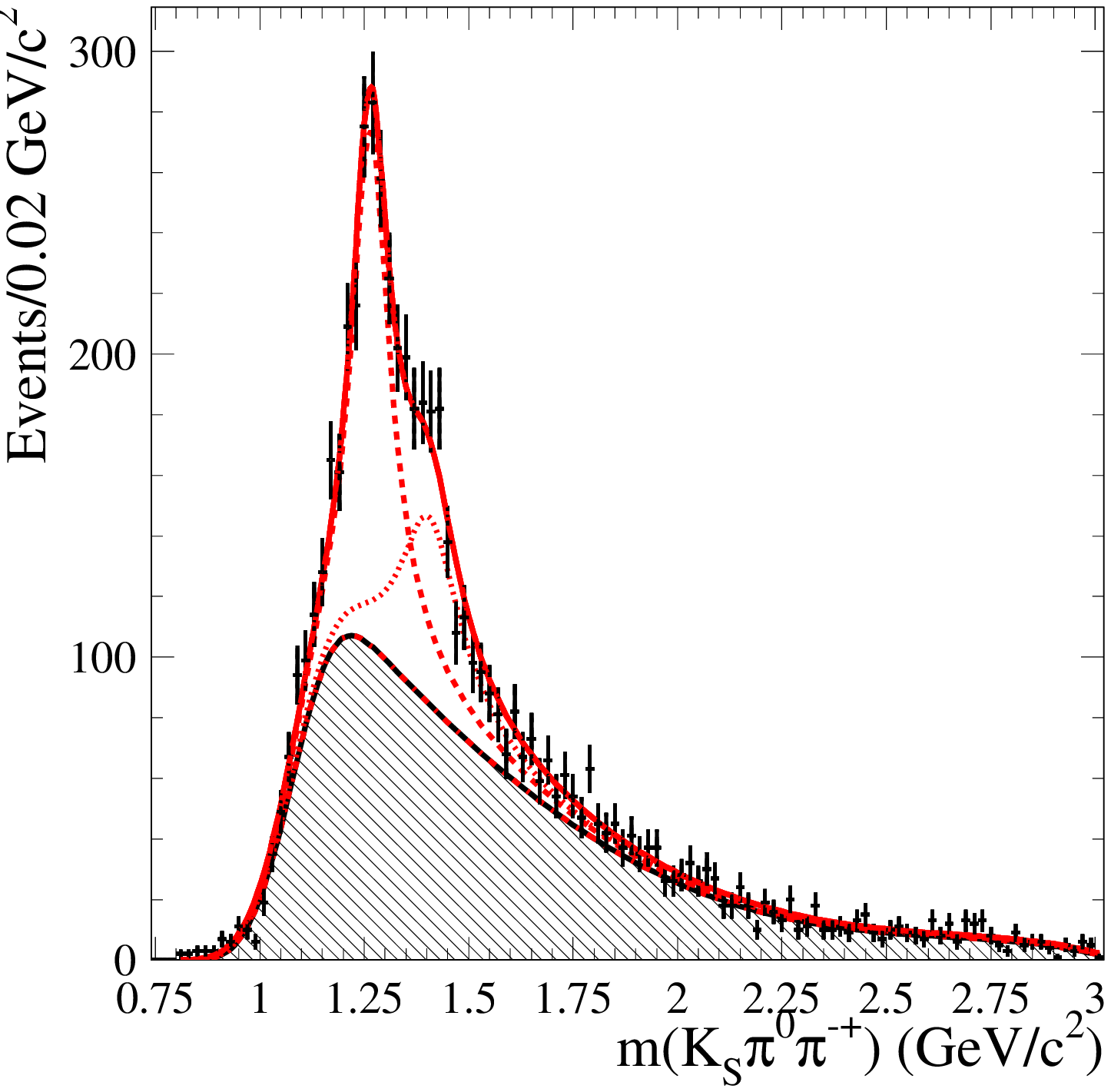}
\put(-25,105){(b)} 
\end{tabular}
\caption{The (a) $\Kpm\pimp\piz$ and (b) $\KS\pipm\piz$ invariant mass
 distributions in $\epem\to\KS\Kpm\pimp\piz$ events.  The solid lines
 represent the results of the fits described in the text;  the
 hatched areas denote their non-resonant components, while the
 dashed, dotted, and dash-dotted lines indicate the contrubutions
 from $\K K_1$(1270), $\K K_1$(1400), and $\K K_1$(1650) events, respectively.}
\label{fig:k1270k1400}
\end{figure}

Some of these events may arise from $\epem\to\K K_1$ events, with
$K_1\to\K\rho^{\pm}$, $\rho^{\pm}\to\pipm\piz$.  Figures~\ref{fig:k1270k1400}(a) 
and~(b) show the $\Kpm\pimp\piz$ and $\KS\pipm\piz$ invariant-mass distributions,
respectively.  There is some apparent structure in the peak regions
of both distributions, and, as an exercise, we perform fits to each
distribution with a sum of the same NR function and three incoherent
P-wave BW functions with parameters fixed to world-average~\cite{pdg}
values for the $K_1$(1270), $K_1$(1400), and $K_1$(1650) resonances.  We note
that other nearby resonances, such as \ktwo or $K^*$(1680), could
contribute in addition or instead.  The results are shown as the
lines in Fig.~\ref{fig:k1270k1400}, with the hatched areas denoting the NR components.
The fit to the spectrum in Fig.~\ref{fig:k1270k1400}(a) yields $230\pm70$ $\KS K_1(1270)^0$ events, 739$\pm$101
$\KS K_1(1400)^0$ events, and 537$\pm$126 $\KS K_1(1650)^0$ events, where all
uncertainties are statistical only.  The fit to Fig.~\ref{fig:k1270k1400}(b) yields 1593$\pm$76
$\Kpm K_1(1270)^\mp$ events, 547$\pm$60 $\Kpm K_1(1400)^\mp$ events,
and 0$\pm$49 $\Kpm K_1(1650)^\mp$ events.  Systematic uncertainties,
discussed below, are large, but at least three (two) neutral
(charged) $K_1$ states are required to describe the data.  Far more
charged than neutral $\K K_1$(1270), but far fewer charged than neutral
$\K K_1$(1650), are produced.

Systematic uncertainties are substantial and difficult to evaluate.
The NR function must describe a distribution complicated by
resonances in, and kinematic constraints on, the other particles in
the event, and the widths and positions of the $\rho(770)$ and $K^*(890)$
resonances do not allow strong constraints from the data.  
%multi-dimensional fit is not feasible because of the limited statistical precision
%and wide \Ecm range.  After considerable study, 
We adopt a simple,
conservative procedure, based on the largest sources of variation.
We repeat each fit with the NR function reduced to a fourth-order
polynomial, and, separately, with the parameters of each resonance
under study allowed to vary.  The two resulting differences in yield
are added in quadrature.  To this we add, linearly, a 10\% relative
uncertainty to account for possible interference between resonances,
the use of fixed vs.\ energy-dependent widths, and the choice of
parametrization for the $\rho^{\pm}$ lineshape.  This procedure is applied to the
\Ecm-integrated distributions in Figs.~\ref{fig:kst0kst0}, \ref{fig:nkst0corr}, and 
~\ref{fig:rho}(a), yielding
systematic uncertainties in the respective total yields.  In each case, the same
relative uncertainty is applied as an overall normalization
uncertainty in the cross sections (Figs.~\ref{fig:cskst0kst0} and~\ref{fig:rho}(b)).

The total yields of all measured $\Kstar\K\pi$, $\Kstarz\Kstarzb$, and $\K\K\rho$ intermediate states
and their uncertainties are listed in Table~\ref{tab:kskpipi0_dyn_sum}.  We do not quote
yields for any of the $\K K_1$ modes, as the uncertainties are very large.  Here,
we have subtracted each $\Kstar\Kstarb$ yield from both of the relevant $\Kstar\K\pi$
yields, so that the sum of all yields, 7013$\pm$683 events, can be
compared with the total number of $\KS\Kpm\pimp\piz$ events, which is
6430$\pm$90. The two numbers are consistent, leaving little room for additional resonant
contributions.

\begin{table}
\caption{Summary of intermediate processes contributing to the $\KS\Kpm\pimp\piz$ final
state. The results  for the $\Kstar\K\pi$ channels do
not include contributions from the $\Kstar\Kstarb$ channels. The first uncertainty
  is statistical and the second systematic.}
\label{tab:kskpipi0_dyn_sum}
\begin{ruledtabular}
\begin{tabular}{lrcccl}
Intermediate state &\multicolumn{5}{c}{Number of events} \\
\hline
$\Kstarz\KS\piz$        & 454&$\pm$&60&$\pm$&74 \\
$\Kstarz\Kpm\pimp$      & 1533&$\pm$&60&$\pm$&296 \\
$\ktwo^0\KS\piz$        & 20&$\pm$&25&$\pm$&4 \\
$\ktwo^0\Kpm\pimp$      & 85&$\pm$&24&$\pm$&18 \\
$\kstpm\KS\pimp$        & 157&$\pm$&50&$\pm$&117\\
$\kstpm\Kmp\piz$        & 1173&$\pm$&64&$\pm$&170\\
$\ktwo^{\pm}\KS\pimp$   & 141&$\pm$&27&$\pm$&28 \\
$\ktwo^{\pm}\Kmp\piz$   & 187&$\pm$&25&$\pm$&35 \\
$\Kstarz\Kstarzb$       & 138&$\pm$&16&$\pm$&55 \\
$\kstp\kstm$            & 814&$\pm$&36&$\pm$&229 \\
$\KS\Kpm\rho(770)^{\mp}$& 2498&$\pm$&100&$\pm$&521 \\
\hline
Total                   & 7013&$\pm$&167&$\pm$&682 \\
\end{tabular}
\end{ruledtabular}
\end{table}  

From Table~\ref{tab:kskpipi0_dyn_sum} we see that $\kstp\kstm$ events account for most of
the $\kstpm\KS\pimp$ production, but only half the $\kstpm\Kmp\piz$
production.  Neutral \kst pair production is much lower than
charged, whereas $\Kstarz\K\pi$ and $\kstpm\K\pi$ are similar.  The rate of
charged $\kstp\kstm$ production is about three times that of neutral
$\Kstarz\Kstarzb$, and these are about four and fifteen times lower than those
of the respective \kst states.  This pattern in the data after \qqbar background subtraction is consistent
with that seen in our previous study of $\epem\to\Kp\Km\pip\pim$ and
$\Kp\Km\piz\piz$~\cite{kkpipi}.

\begin{figure}[tbh]
\includegraphics[width=0.8\columnwidth]{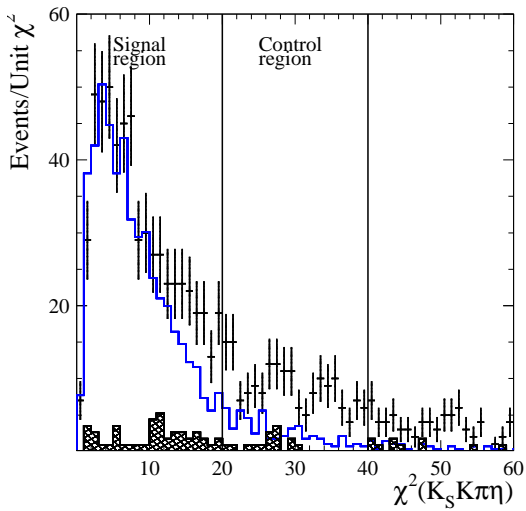}
\caption{Distribution of $\chi^2$ from the 5-constraint fit of the $\KS\Kpm\pimp\eta\gamma$ candidates in the
data (points). The open and cross-hatched histograms are the distributions for
 simulated signal and \qqbar background events, respectively,
 normalized as described in the text.  The signal
and control regions are indicated.}
\label{fig:chi2_kskpieta_exp_mc}
\end{figure}

\section{The $\KS\Kpm\pimp\eta$ final state}
\subsection{Event selection}

The $\chi^2_{\KS\Kpm\pimp\eta}$ distribution for the selected $\epem\to\KS\Kpm\pimp\eta$
 events in the data is shown in Fig.~\ref{fig:chi2_kskpieta_exp_mc}, together with the
 corresponding distributions of simulated signal and \qqbar
 background events.  Again, the \qqbar background is normalized using
 the \piz peaks in the data and simulated invariant-mass distributions
 of the ISR photon candidate combined with all other photon candidates
 in the event.  The signal simulation is normalized to have the same
 integral in the first five bins as the data minus the \qqbar
 background.  We define signal and control regions by
 $\chi^2_{\KS\Kpm\pimp\eta}<20$ and $20<\chi^2_{\KS\Kpm\pimp\eta}<40$, respectively, containing
 459 (1418) and 128 (147) data (simulated) events.

\begin{figure}[t]  
\begin{tabular}{cc}
\includegraphics[width=.5\columnwidth]{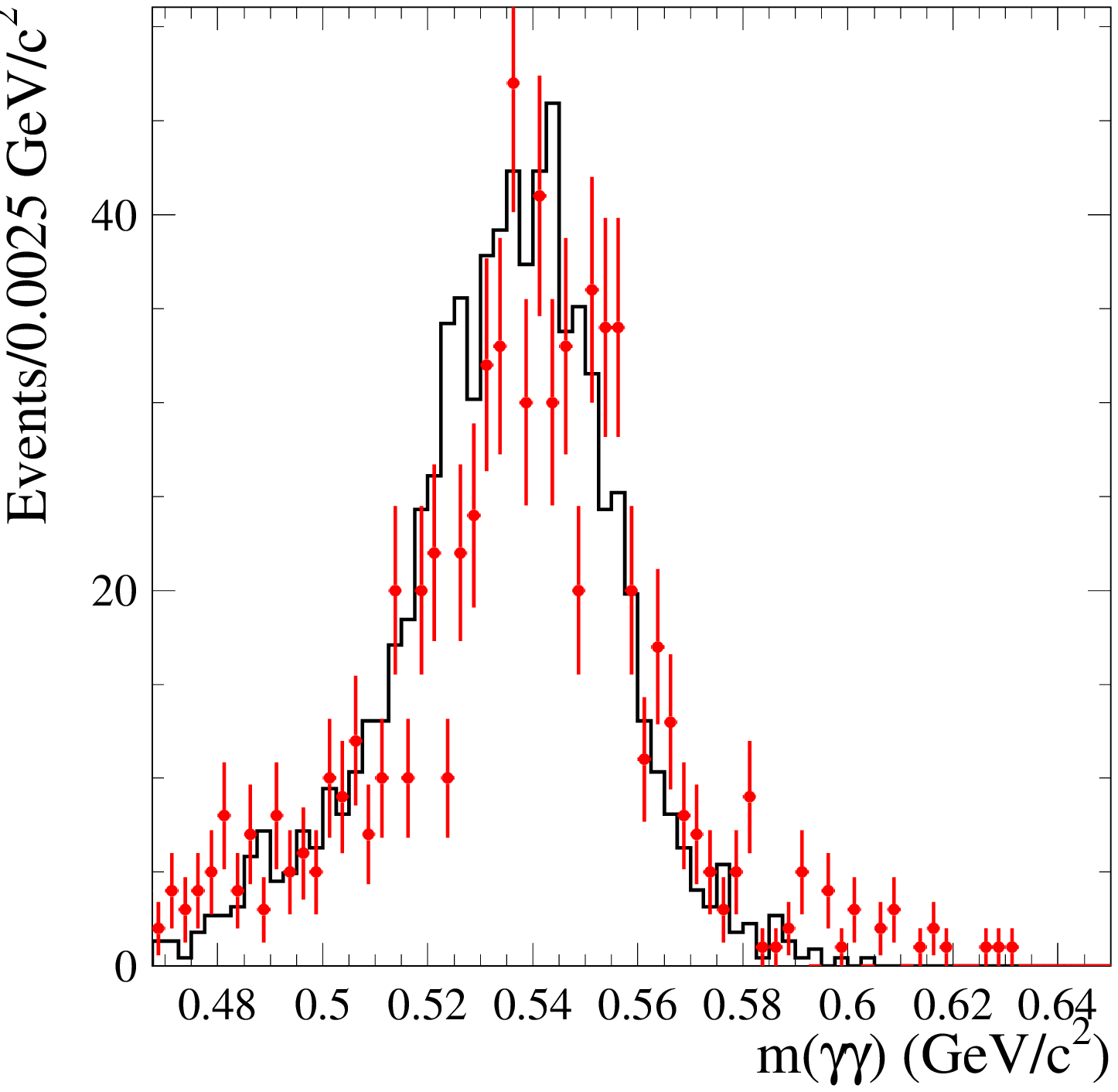}
\put(-25,105){(a)}
\includegraphics[width=.5\columnwidth]{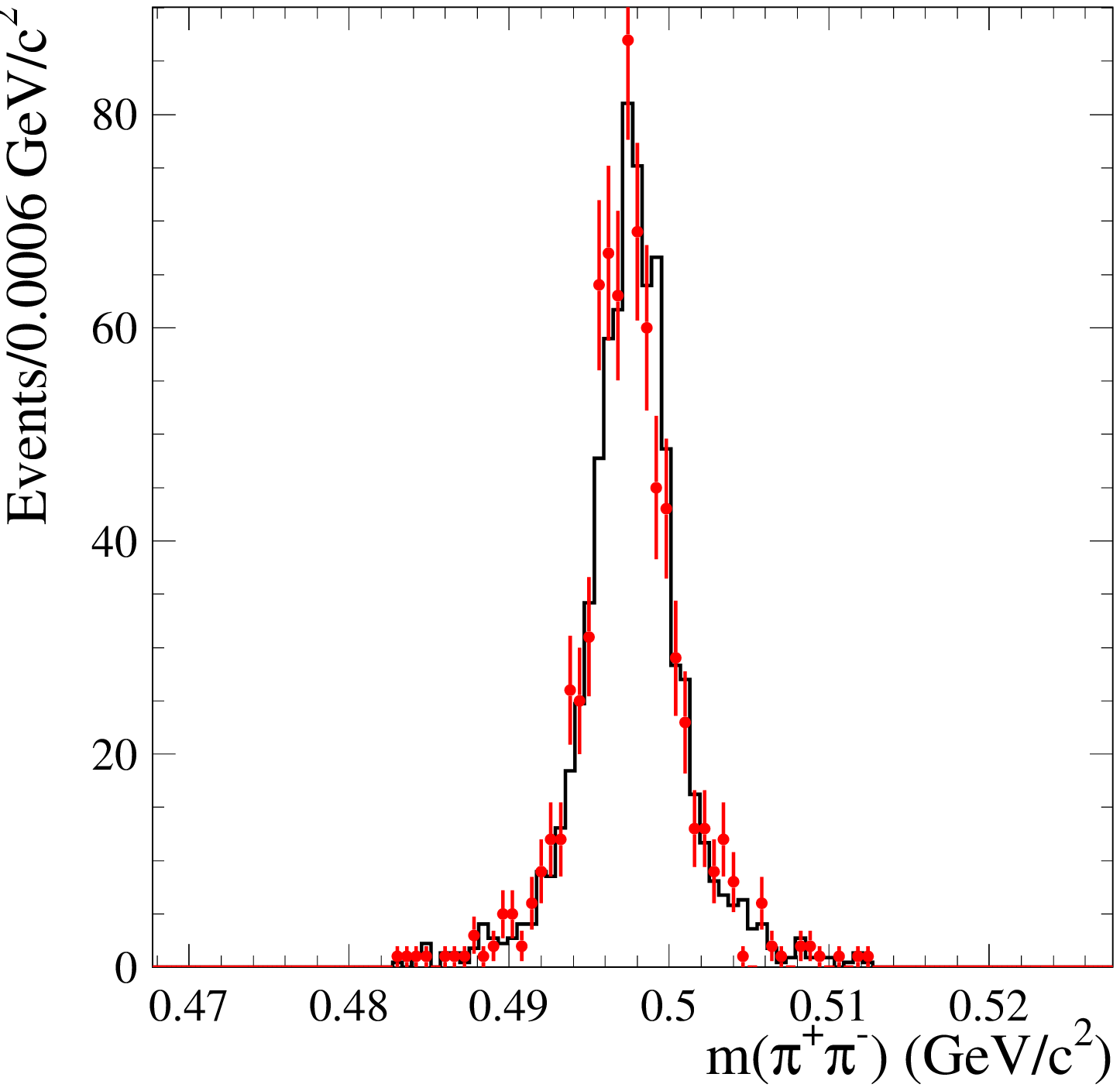}
\put(-25,105){(b)}
\end{tabular}
\caption{The (a) \gaga and (b) $\pip\pim$ invariant-mass distributions of
the $\eta$ and $\KS$ candidates, respectively, in $\KS\Kpm\pimp\eta$ events
in the $\chi^2_{\KS\Kpm\pimp\eta}$ signal region, for the selected data (points) and
the signal simulation (histograms).}
\label{fig:ks_eta}
\end{figure}

Figure~\ref{fig:ks_eta}(a) compares the \gaga invariant-mass distribution of
 the $\eta$ candidate for data events in the signal region with the
 prediction of the signal-event simulation, and Fig.~\ref{fig:ks_eta}(b) shows the
 corresponding $\pip\pim$ invariant-mass distributions of the \KS
 candidate.  The $\eta$ peak is wider and more skewed than the \piz
 peak in Fig.~\ref{fig:ks_pi0}(a), but the selection criteria are sufficiently loose
 that there is no effect on the results.

\begin{figure}[b]
\includegraphics[width=0.8\columnwidth]{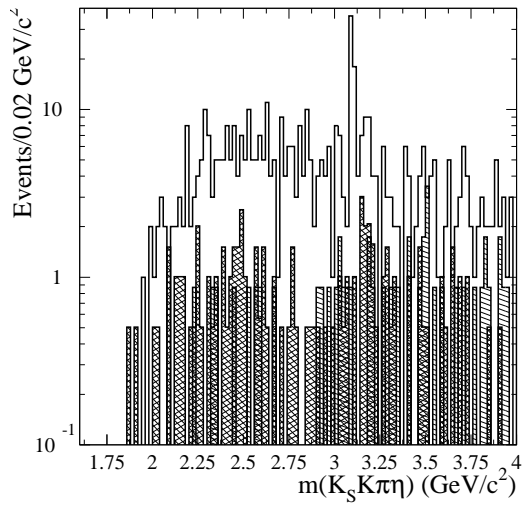}
\caption{Distribution of the fitted $\KS\Kpm\pimp\eta$ invariant mass for data
 events in the signal region (open histogram).  The hatched and cross-hatched
 distributions show the estimated backgrounds evaluated from the
 control region and from \qqbar events, respectively.}
\label{fig:bkg_kskpieta}
\end{figure}

The distribution of the invariant mass of the final-state hadronic
 system for data events in the signal region is shown in Fig.~\ref{fig:bkg_kskpieta}.  A
 narrow peak due to $\jpsi\to\KS\Kpm\pimp\eta$ decays is visible.  The \qqbar
 background is shown as the cross-hatched histrogram.  We subtract it
 and then estimate the remaining background, assumed to arise from ISR
 events, as described above.  
 We take the shape of the ISR background $\chi^2$
 distribution directly from the data, as the difference between experimental $\chi^2_{\KS\Kpm\pimp\eta}$ distribution with 
\qqbar background
subtracted and that of the normalized signal simulation (points and open histogram in Fig.~\ref{fig:chi2_kskpieta_exp_mc}).

The total number of signal events obtained in this way is $358\pm24$
 (stat.)  We define the systematic uncertainty in two \Ecm regions to be half
 the number of background events, resulting in a
 relative uncertainty in the signal event yields of 11\% for
 $\Ecm<3$~\gev and 18\% for $\Ecm>3$~\gev.

\subsection{Detection efficiency}

\begin{figure}[t]
\includegraphics[width=0.8\columnwidth]{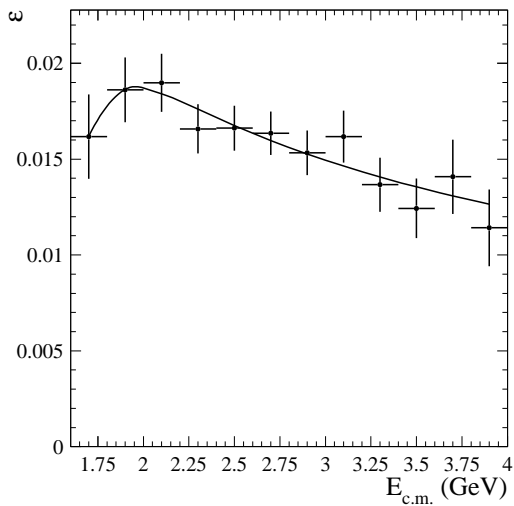}
\caption{Detection efficiency for $\epem\to\KS\Kpm\pimp\eta$ events as a function of
the hadronic invariant mass \Ecm = $m(\KS\Kpm\pimp\eta)$.
The solid curve shows the fitted parametrization.}
\label{fig:effreg_kskpieta}
\end{figure}

The total reconstruction and selection efficiency from the signal
 simulation is shown as a function of \Ecm in Fig.~\ref{fig:effreg_kskpieta}, and is
 parametrized by a smooth function, shown as the solid line.  We apply
 the same corrections for charged-track finding, \KS reconstruction,
 and \Kpm and \pipm identification efficiencies as in Sec.~\ref{sec:efficiency}, and
 evaluate a correction for the shape of the $\chi^2$ distribution in the
 same way.  We do not have a dedicated study of $\eta$ reconstruction
 efficiency, so we assume a correction equal to that on the \piz
 efficiency, but with the uncertainty doubled.

 The momentum and polar angle distributions of the \KS, \Kpm,
 \pipm, and $\eta$ candidates in the data are well described by the
 signal simulation.  To study the effects of resonant substructure, we use
 fast simulations of signal and the ISR $\kstpm\Kmp\eta$ and
 $\kstz\KS\eta$ processes.  Their efficiencies 
 are consistent and we take the largest difference,
 which is 2.5\%, as the systematic uncertainty at all \Ecm,
 to account for potential differences between data and
 simulation for the \Ecm dependence of the efficiency and
 for the resonant structure.
These
 corrections and their uncertainties are listed in Table~\ref{tab:kskpieta_syst}.  The
 total correction is $+0.6\pm5.5$\%.

\subsection{Cross section for $\epem\to\KS\Kpm\pimp\eta$}

The $\epem\to\KS\Kpm\pimp\eta$ cross section is obtained from the analog of
 Eq.~(\ref{eq:xsec}), with the \piz replaced by an $\eta$.  The differential luminosity is
 the same as for the $\KS\Kpm\pimp\piz$ cross section, and the radiative correction is evaluated
 in the analagous way to be R=1.0022$\pm$0.0016, independent of \Ecm.

\begin{figure}[tbh]
\includegraphics[width=0.8\columnwidth]{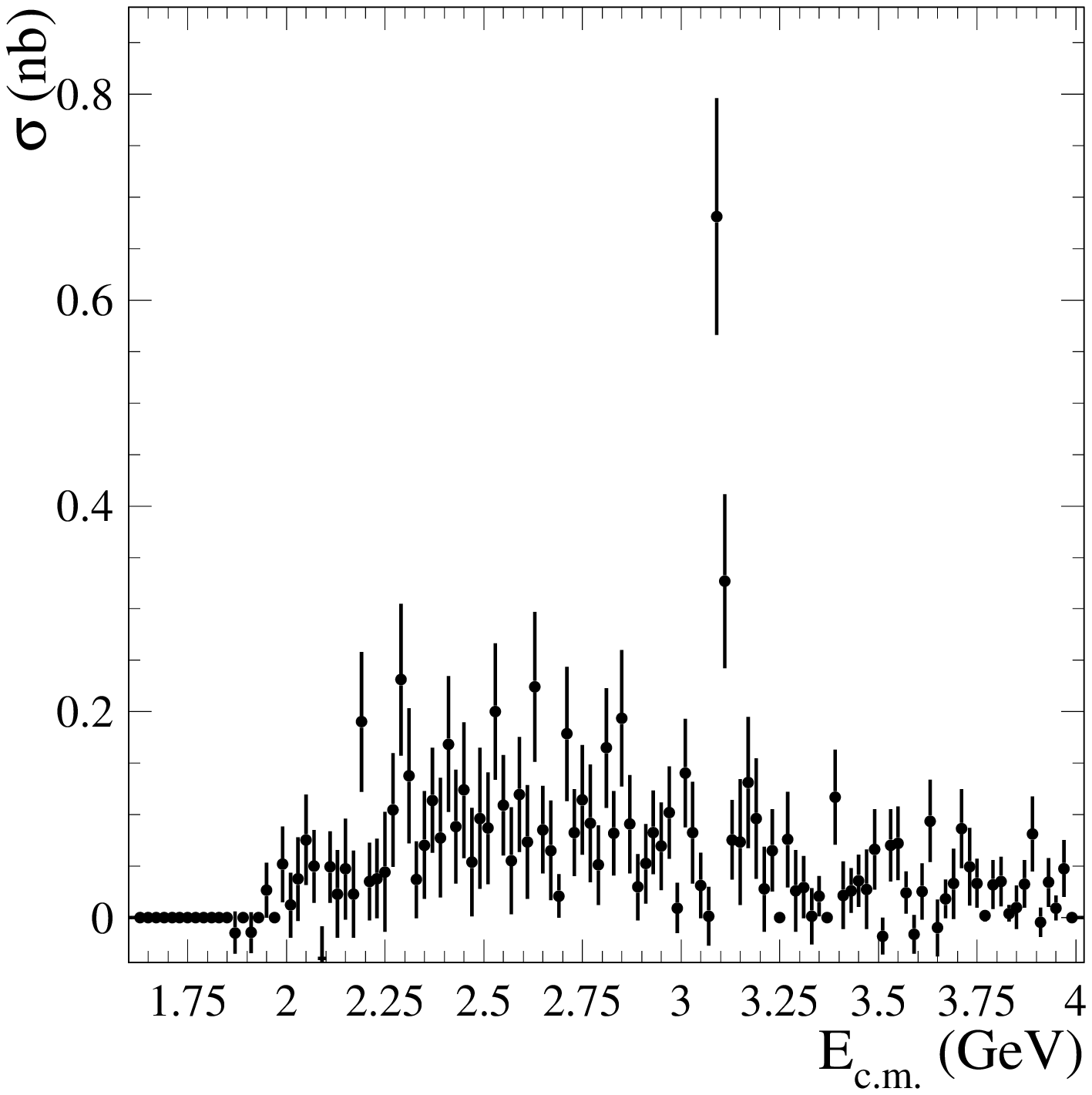}
\caption{Cross section for the process $\epem\to\KS\Kpm\pimp\eta$. Uncertainties are statistical.}
\label{fig:cs_kskpieta}
\end{figure}

The fully corrected cross section is shown in Fig.~\ref{fig:cs_kskpieta} and listed in
 Table~\ref{tab:cs_kskpieta}, with statistical uncertainties only.  The relative systematic
 uncertainties are summarized in Table~\ref{tab:kskpieta_syst}, yielding a total systematic uncertainty 
 of 12.0\%  for $\Ecm<3$~\gev and 19\% for $\Ecm>3$~\gev.

\begin{table}[tbh]
\caption{Summary of the corrections to, and systematic uncertainties in, the $\epem\to\KS\Kpm\pimp\eta$ cross section.}
\label{tab:kskpieta_syst}
\begin{ruledtabular}
\begin{tabular}{l c l}
Source                                    &  Correction    & Systematic           \\
                                          &  (\%)          & uncertainty (\%)     \\                              
\hline
$\eta$ efficiency                         &     +2.0       &  2.0          \\
$\Kpm$, $\pipm$  reconstruction           &     +1.6       &  2.0          \\
$\KS$ reconstruction                      &     +1.1       &  1.0          \\
PID efficiency                            &    \,0.0       &  2.0          \\
\chisq selection                          &     -4.0       &  4.6          \\
Background subtraction                    &     ---        &  11.0, $<$3.0~\gev          \\
                                          &                &  18.0, $>$3.0~\gev         \\
Model acceptance                          &     ---        &  2.5          \\

Luminosity and Rad.Corr.                  &     ---        &  1.4          \\
\hline
 Total                                    &     +0.6       & 12.8, $<$3.0~\gev          \\
                                          &                & 19.1, $>$3.0~\gev         \\
\end{tabular}
\end{ruledtabular}
\end{table}

%\begin{linenomath}
\begin{table*}[tbh]
\caption{Measurement of $e^+e^-\to \KS K^{\pm}\pi^{\mp}\eta$ cross section. The uncertainties are statistical only; 
systematic uncertainties are given in Table~\ref{tab:kskpieta_syst}.}
\label{tab:cs_kskpieta}
\begin{ruledtabular}
\begin{tabular}{c c c c c c c c c c}
 $E_{\rm c.m.}$        & $\sigma$       &  $E_{c.m.}$        & $\sigma$       &  $E_{c.m.}$        & $\sigma$       &  $E_{c.m.}$        & $\sigma$ \\
 (GeV)                 &   (nb)         & (GeV)              &  (nb)          & (GeV)              &  (nb)          &  (GeV)             & (nb) \\
\hline
 2.01 & 0.01 $\pm$ 0.03 & 2.51 & 0.09 $\pm$ 0.05 & 3.01 & 0.14 $\pm$ 0.05 & 3.51 & -0.02 $\pm$ 0.02 \\
 2.03 & 0.04 $\pm$ 0.04 & 2.53 & 0.20 $\pm$ 0.07 & 3.03 & 0.08 $\pm$ 0.05 & 3.53 & 0.07 $\pm$ 0.04 \\
 2.05 & 0.08 $\pm$ 0.04 & 2.55 & 0.11 $\pm$ 0.05 & 3.05 & 0.03 $\pm$ 0.03 & 3.55 & 0.07 $\pm$ 0.04 \\
 2.07 & 0.05 $\pm$ 0.04 & 2.57 & 0.06 $\pm$ 0.05 & 3.07 & 0.00 $\pm$ 0.03 & 3.57 & 0.02 $\pm$ 0.02 \\
 2.09 & -0.04 $\pm$ 0.03 & 2.59 & 0.12 $\pm$ 0.06 & 3.09 & 0.68 $\pm$ 0.11 & 3.59 & -0.02 $\pm$ 0.02 \\
 2.11 & 0.05 $\pm$ 0.03 & 2.61 & 0.07 $\pm$ 0.06 & 3.11 & 0.33 $\pm$ 0.08 & 3.61 & 0.03 $\pm$ 0.03 \\
 2.13 & 0.02 $\pm$ 0.04 & 2.63 & 0.22 $\pm$ 0.07 & 3.13 & 0.08 $\pm$ 0.04 & 3.63 & 0.09 $\pm$ 0.04 \\
 2.15 & 0.05 $\pm$ 0.05 & 2.65 & 0.09 $\pm$ 0.04 & 3.15 & 0.07 $\pm$ 0.06 & 3.65 & -0.01 $\pm$ 0.03 \\
 2.17 & 0.02 $\pm$ 0.04 & 2.67 & 0.07 $\pm$ 0.05 & 3.17 & 0.13 $\pm$ 0.06 & 3.67 & 0.02 $\pm$ 0.02 \\
 2.19 & 0.19 $\pm$ 0.07 & 2.69 & 0.02 $\pm$ 0.02 & 3.19 & 0.10 $\pm$ 0.06 & 3.69 & 0.03 $\pm$ 0.03 \\
 2.21 & 0.04 $\pm$ 0.04 & 2.71 & 0.18 $\pm$ 0.07 & 3.21 & 0.03 $\pm$ 0.04 & 3.71 & 0.09 $\pm$ 0.04 \\
 2.23 & 0.04 $\pm$ 0.04 & 2.73 & 0.08 $\pm$ 0.04 & 3.23 & 0.07 $\pm$ 0.04 & 3.73 & 0.05 $\pm$ 0.04 \\
 2.25 & 0.04 $\pm$ 0.06 & 2.75 & 0.11 $\pm$ 0.05 & 3.25 & 0.00 $\pm$ 0.00 & 3.75 & 0.03 $\pm$ 0.02 \\
 2.27 & 0.10 $\pm$ 0.06 & 2.77 & 0.09 $\pm$ 0.06 & 3.27 & 0.08 $\pm$ 0.05 & 3.77 & 0.00 $\pm$ 0.01 \\
 2.29 & 0.23 $\pm$ 0.07 & 2.79 & 0.05 $\pm$ 0.04 & 3.29 & 0.03 $\pm$ 0.04 & 3.79 & 0.03 $\pm$ 0.02 \\
 2.31 & 0.14 $\pm$ 0.07 & 2.81 & 0.16 $\pm$ 0.06 & 3.31 & 0.03 $\pm$ 0.03 & 3.81 & 0.04 $\pm$ 0.02 \\
 2.33 & 0.04 $\pm$ 0.04 & 2.83 & 0.08 $\pm$ 0.04 & 3.33 & 0.00 $\pm$ 0.03 & 3.83 & 0.00 $\pm$ 0.01 \\
 2.35 & 0.07 $\pm$ 0.05 & 2.85 & 0.19 $\pm$ 0.07 & 3.35 & 0.02 $\pm$ 0.02 & 3.85 & 0.01 $\pm$ 0.02 \\
 2.37 & 0.11 $\pm$ 0.05 & 2.87 & 0.09 $\pm$ 0.05 & 3.37 & 0.00 $\pm$ 0.00 & 3.87 & 0.03 $\pm$ 0.02 \\
 2.39 & 0.08 $\pm$ 0.06 & 2.89 & 0.03 $\pm$ 0.03 & 3.39 & 0.12 $\pm$ 0.05 & 3.89 & 0.08 $\pm$ 0.04 \\
 2.41 & 0.17 $\pm$ 0.07 & 2.91 & 0.05 $\pm$ 0.04 & 3.41 & 0.02 $\pm$ 0.03 & 3.91 & 0.00 $\pm$ 0.01 \\
 2.43 & 0.09 $\pm$ 0.06 & 2.93 & 0.08 $\pm$ 0.04 & 3.43 & 0.03 $\pm$ 0.02 & 3.93 & 0.03 $\pm$ 0.02 \\
 2.45 & 0.12 $\pm$ 0.07 & 2.95 & 0.07 $\pm$ 0.04 & 3.45 & 0.04 $\pm$ 0.03 & 3.95 & 0.01 $\pm$ 0.01 \\
 2.47 & 0.05 $\pm$ 0.05 & 2.97 & 0.10 $\pm$ 0.05 & 3.47 & 0.03 $\pm$ 0.04 & 3.97 & 0.05 $\pm$ 0.03 \\
 2.49 & 0.10 $\pm$ 0.07 & 2.99 & 0.01 $\pm$ 0.02 & 3.49 & 0.07 $\pm$ 0.04 & 3.99 & 0.00 $\pm$ 0.00 \\
\end{tabular}
\end{ruledtabular}
\end{table*}
%\end{linenomath}

\subsection{Substructure in $\KS\Kpm\pimp\eta$}

\begin{figure}
\includegraphics[width=0.48\columnwidth]{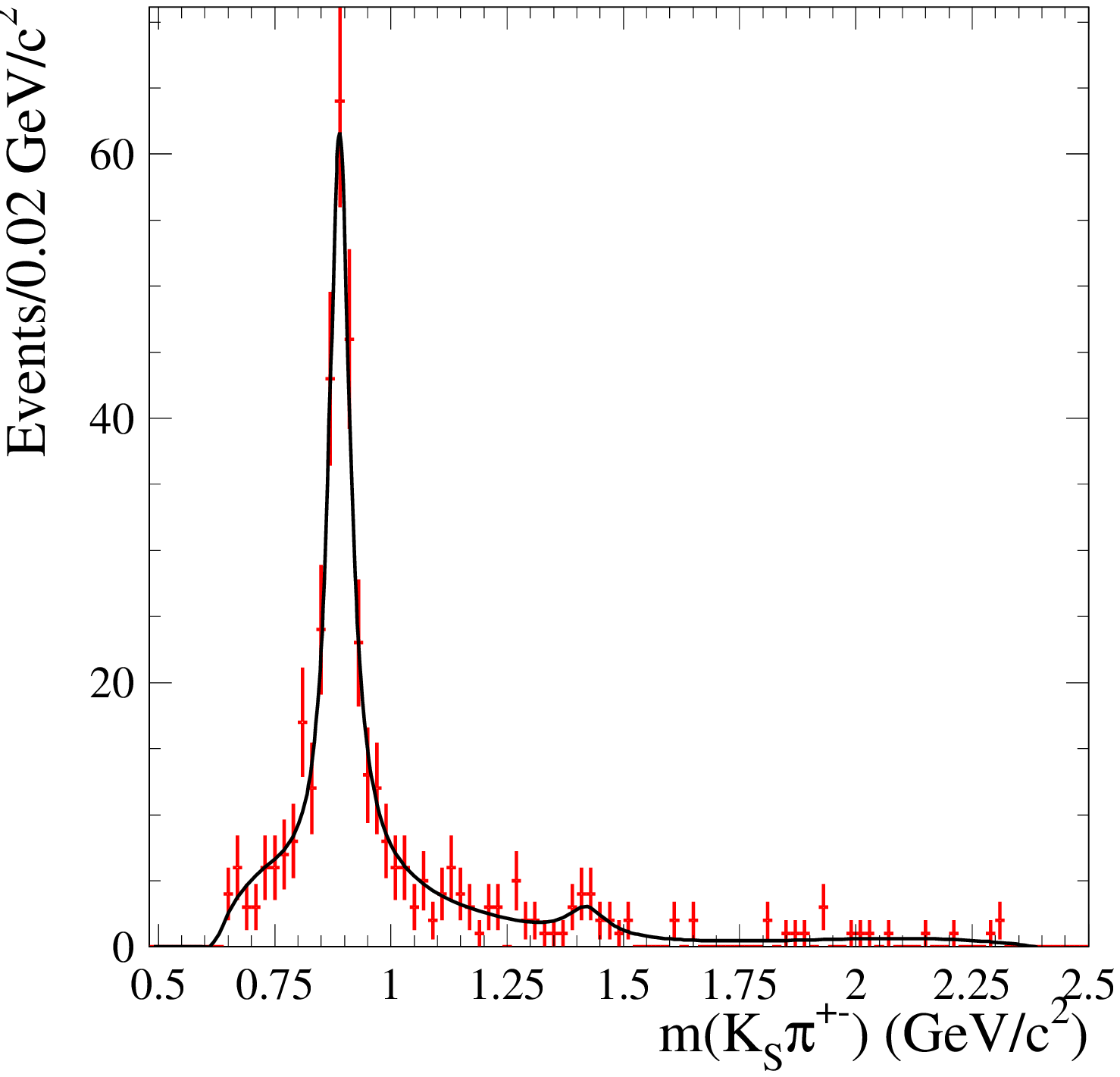}
\put(-25,105){(a)}
\includegraphics[width=0.48\columnwidth]{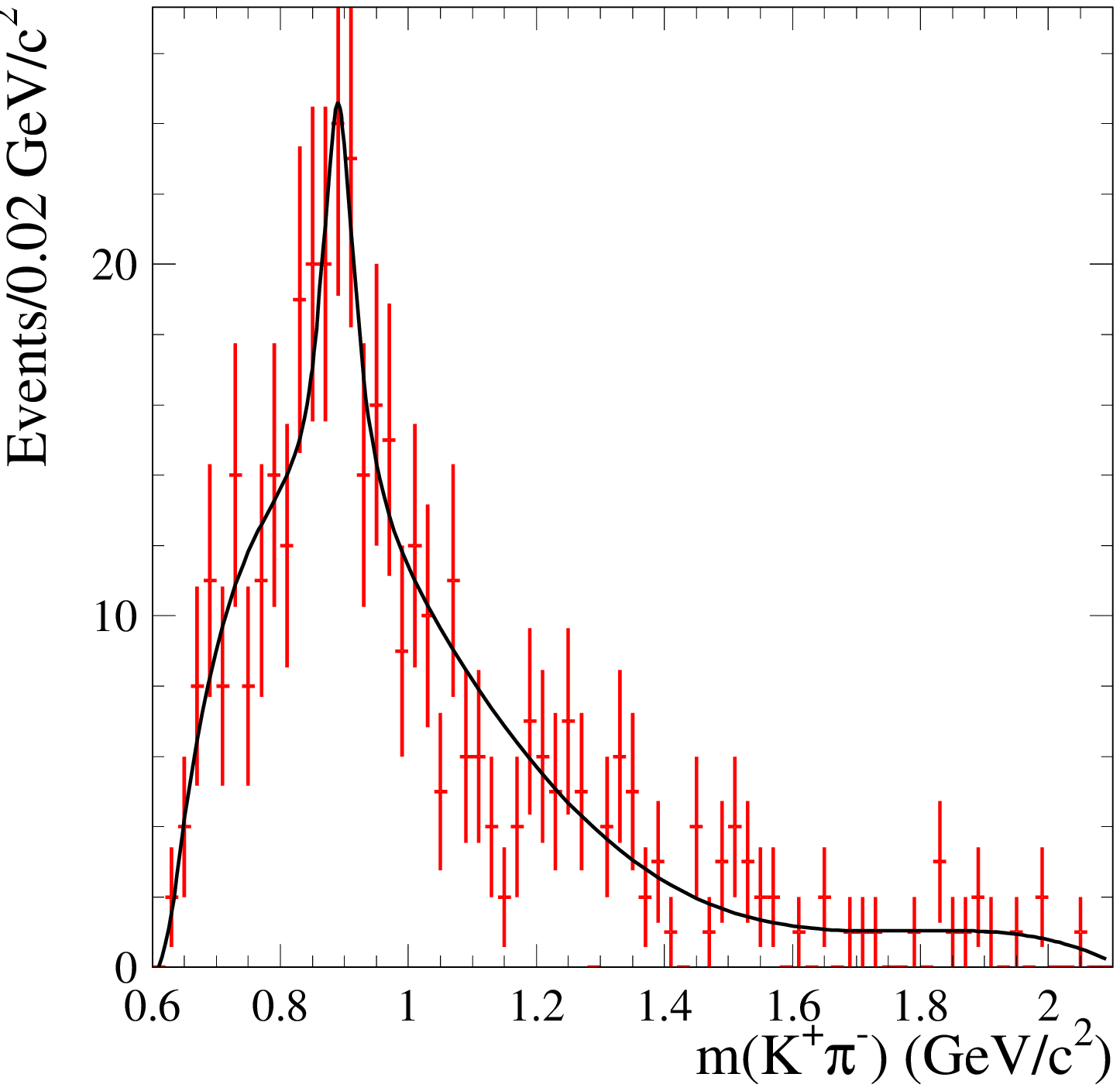}
\put(-25,105){(b)}
\caption{The (a) $\KS\pipm$ and (b) $\Kpm\pimp$ invariant-mass distributions
 in $\epem\to\KS\Kpm\pimp\eta$ events.  The lines represent the results of the
 fits described in the text.}
\label{fig:mkspi_eta}
\end{figure}

We study substructure in the $\KS\Kpm\pimp\eta$ mode in the same way as for the
 $\KS\Kpm\pimp\piz$ mode, using background-subtracted data and excluding the \jpsi
 region $3.0<\Ecm<3.2$~\gev.  Here, we expect far less structure, and
 indeed we see no significant structure in the $\Kpm\KS$, $\Kpm\eta$,
 or $\KS\eta$ invariant-mass distributions.  Figure~\ref{fig:mkspi_eta} shows the
 $\KS\pipm$ and $\Kpm\pimp$ invariant-mass distributions.  The former 
 shows a dominant \kstpm peak, as well as structure near 1.43~\gevcc,
 whereas the latter shows only a modest \kstz peak over a large, broad
 distribution.

We fit the $m(\KS\pipm)$ distribution with a sum of incoherent
 $K^*(892)$ and $K^*_2(1430)$ resonances and a NR component of the
 same form as in Sec.~\ref{sec:kskpipi0_substr}.  The result of the fit is shown in Fig.~\ref{fig:mkspi_eta}(a)
 as the solid line, yielding 242$\pm$21 $\epem\to\kstpm\Kmp\eta$
 events and 10$\pm$5 $\epem\to\ktwo^{\pm}\Kmp\eta$ events, where the uncertainties
 are statistical only.  There is no hint of a $\ktwo^0$ signal in the
 $m(\Kpm\pimp)$ distribution, and we show the result of a
 single-resonance$+$NR fit in Fig.~\ref{fig:mkspi_eta}(b), which yields 123$\pm$36(stat.) 
 $\epem\to\kstz\KS\eta$ events.

We estimate systematic uncertainties due to the fitting procedure as
 above, and summarize these results in Table~\ref{tab:kskpieta_dyn_sum}.  The sum of
 these three resonant yields is consistent with the total number of
 $\KS\Kpm\pimp\eta$ events, and the suppression of neutral with respect to
 charged \Kst production is similar to that seen above in the $\KS\Kpm\pimp\piz$
 final state, and in our previous study of the $\Kp\Km\pip\pim$ final
 state~\cite{kkpipi}.

\begin{figure}
\includegraphics[width=0.48\columnwidth]{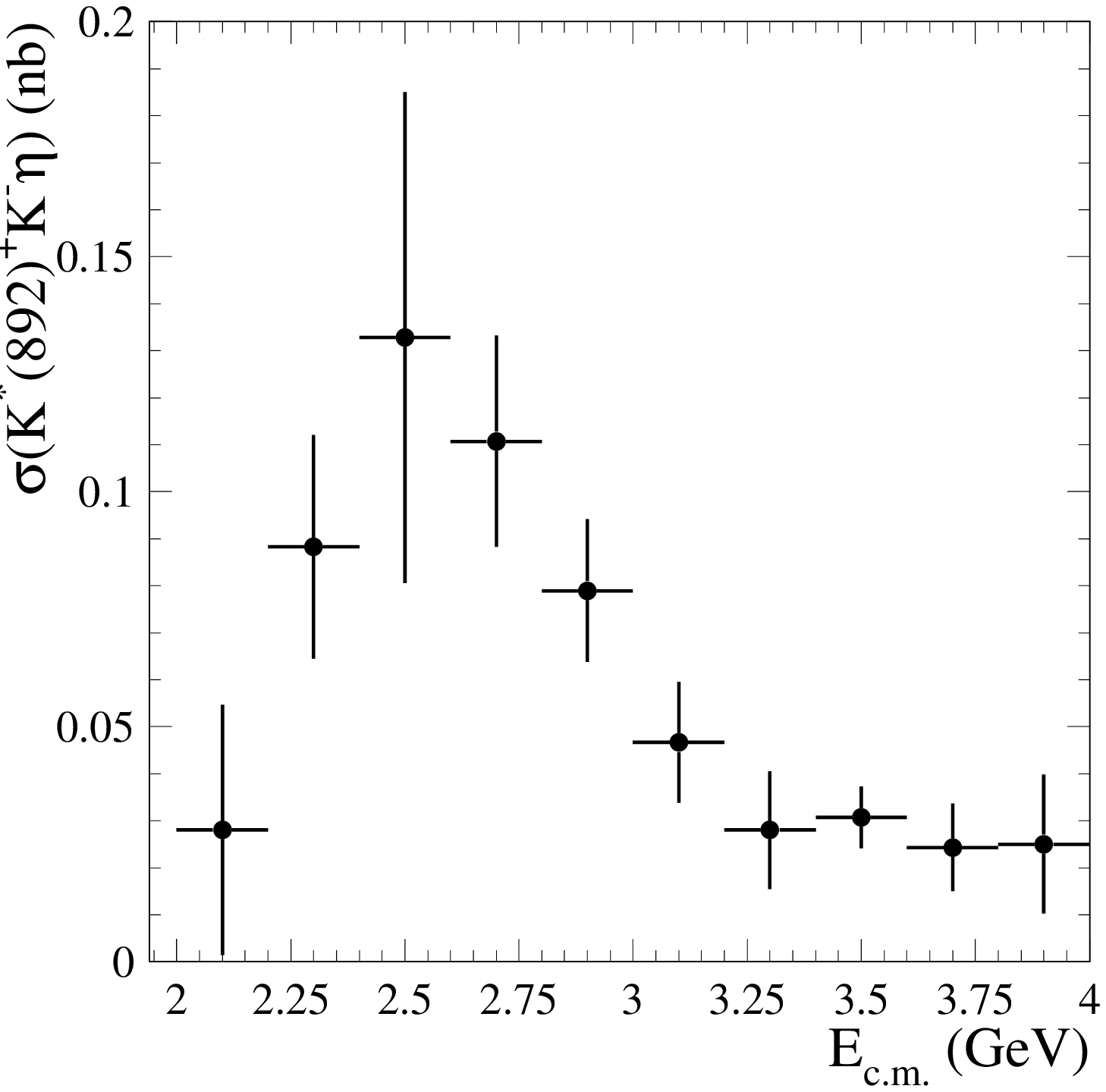}
\put(-25,105){(a)}
\includegraphics[width=0.48\columnwidth]{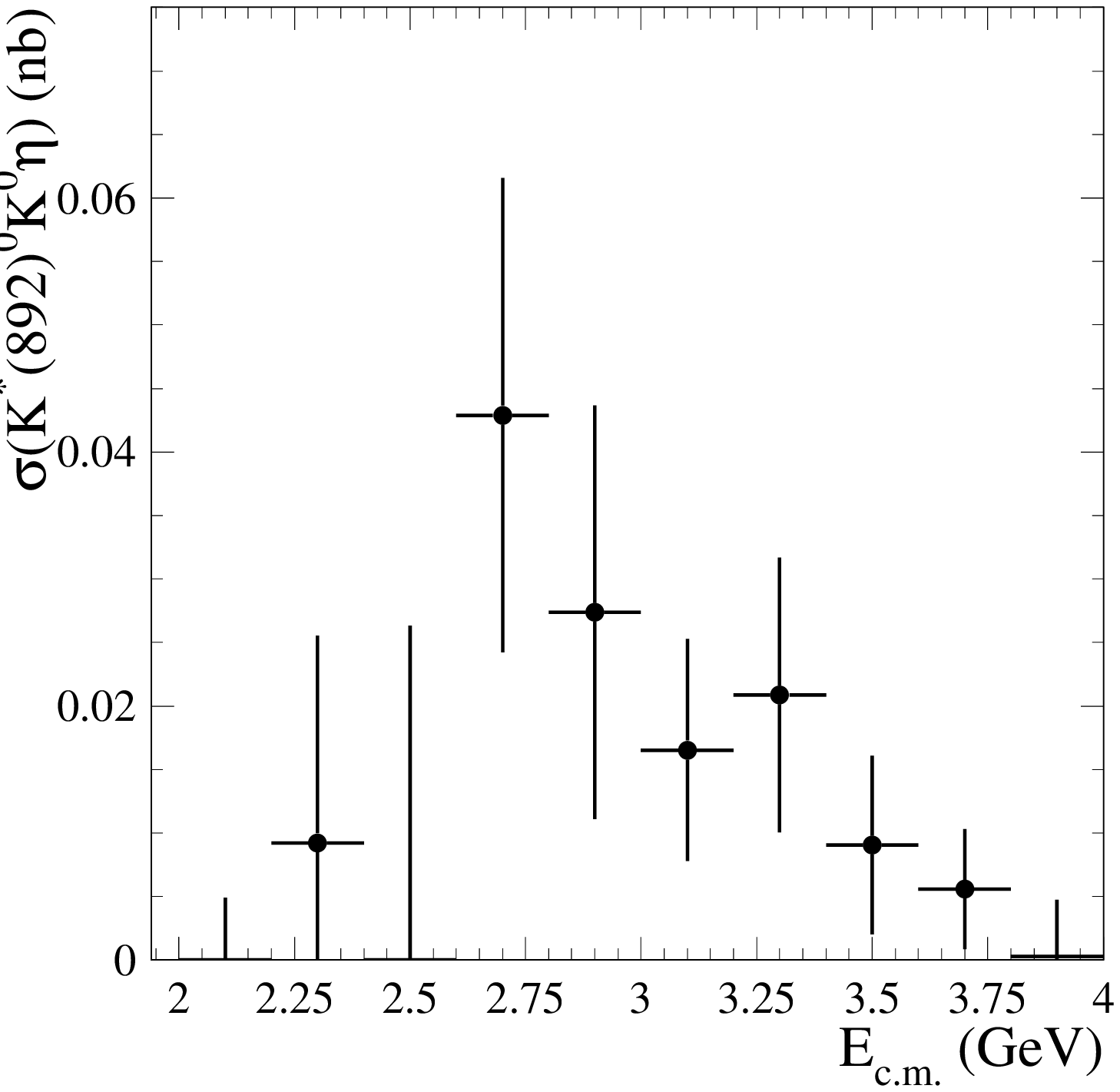}
\put(-25,105){(b)}
\caption{Cross sections for the processes (a) $\epem\to\kstpm\Kmp\eta$ and (b) 
 $\epem\to\kstz\KS\eta$.  The uncertainties are statistical only.}
\label{fig:cs_kstketa}
\end{figure}

Repeating these fits in 0.2~\gev bins of \Ecm, and using Eq.~(\ref{eq:xsec}), we
 extract cross sections for the processes $\epem\to\kstpm\Kmp\eta$
 with $\kstpm\to\KS\pipm$, and $\epem\to\kstz\KS\eta$ with
 $\kstz\to\Kpm\pimp$.  These are shown in Fig.~\ref{fig:cs_kstketa} with statistical
 uncertainties.  A systematic uncertainty of 16\% (21\%) is applicable
 for \Ecm below  (above) 3~\gev.  These are the first measurements
 of these cross sections.  Well above threshold, they become
 consistent with the corresponding $K^*(892)\Kbar\piz$ cross sections.

\begin{table}
\caption{Summary of intermediate processes contributing to the $\KS\Kpm\pimp\eta$ final
state.}
\label{tab:kskpieta_dyn_sum}
\begin{ruledtabular}
\begin{tabular}{lrcccl}
Intermediate state &\multicolumn{5}{c}{Number of events} \\
\hline
$\Kstarz\KS\eta$                                           & 123&$\pm$&36&$\pm$&13\\
$\kstpm\Kmp\eta$                                            & 242&$\pm$&21&$\pm$&24 \\
$\ktwo^{\pm}\Kmp\eta$                                      &  10&$\pm$&5&$\pm$&2 \\
\hline
Total  & 375&$\pm$&42&$\pm$&27 \\
\end{tabular}
\end{ruledtabular}
\end{table}

\begin{figure}[tbh]
\includegraphics[width=0.8\columnwidth]{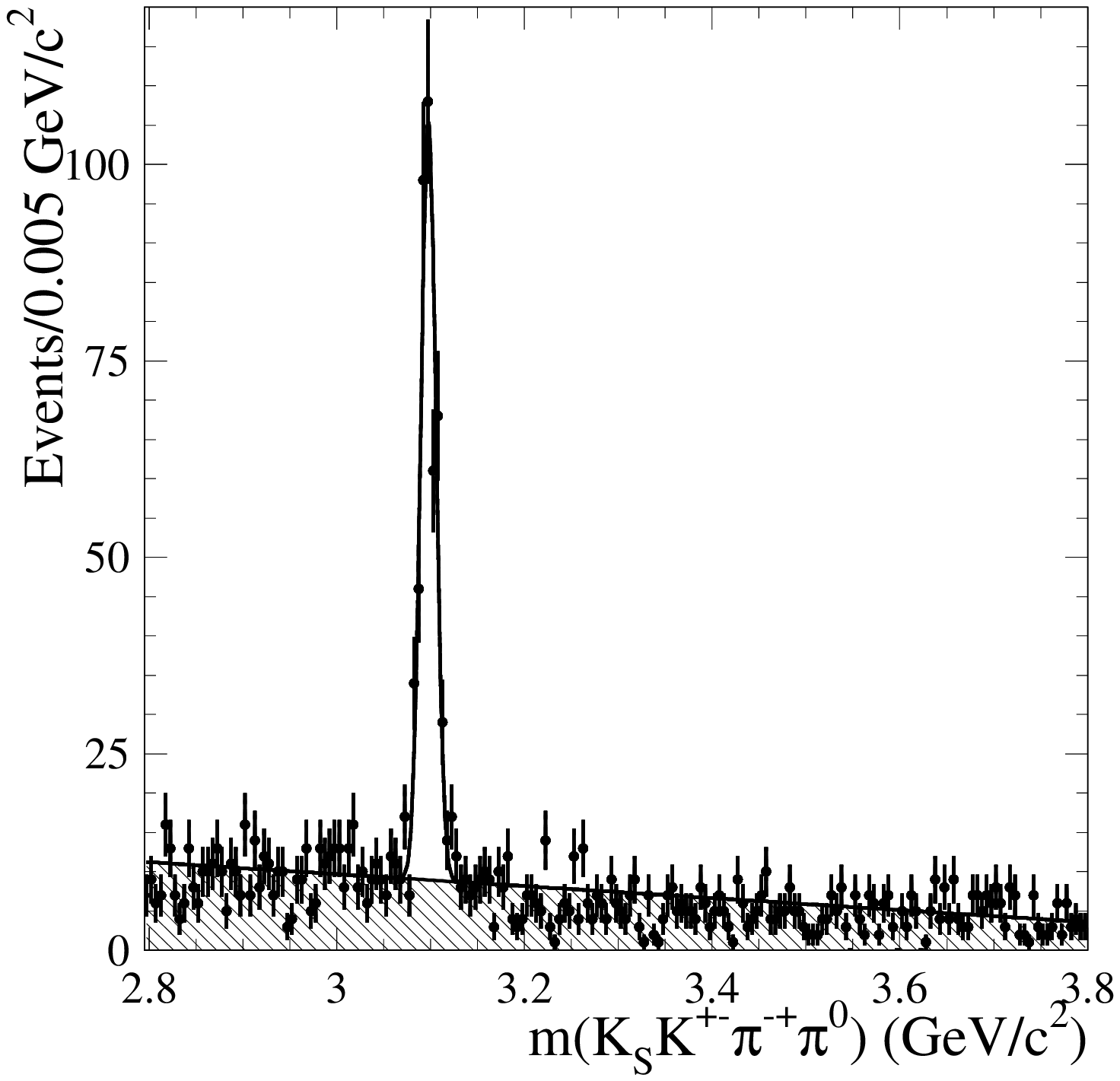}
\caption{The $\KS\Kpm\pimp\piz$ invariant-mass distribution in the \jpsi mass
 region.  The line represents the result of the fit described in the
 text, with the open (hatched) area indicating the (non)resonant
 component.}
\label{fig:jpsi_pi0}
\end{figure}

\begin{figure}[h]
\includegraphics[width=0.8\columnwidth]{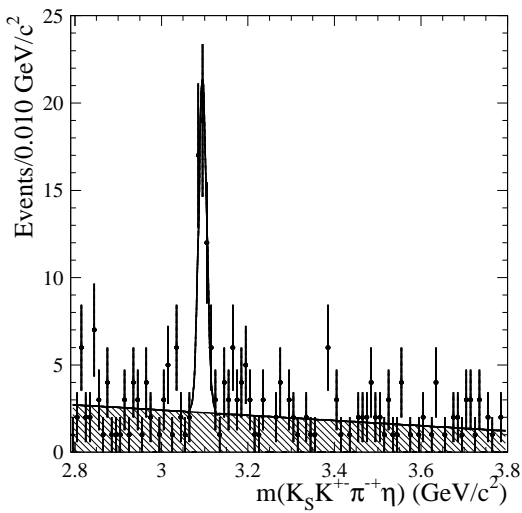}
\caption{The $\KS\Kpm\pimp\eta$ invariant-mass distribution in the \jpsi mass
 region.  The line represents the result of the fit described in the
 text, with the open (hatched) area indicating the (non)resonant
 component.}
\label{fig:jpsi_eta}
\end{figure}

\section{The $\jpsi$ region\label{sec:jpsi}}

Figures~\ref{fig:jpsi_pi0} and~\ref{fig:jpsi_eta} show expanded views of the mass distributions in
 Figs.~\ref{fig:mkskpipi0_sig_bkg} and~\ref{fig:bkg_kskpieta}, respectively, in the 2.8--3.8~\gevcc mass region. 
 They show clear \jpsi signals, and no other significant structure.
 Fitting each of these distributions with
 the sum of a Gaussian describing the \jpsi signal shape and a first-order
 polynomial function yields $393\pm23$ $\jpsi\to\KS\Kpm\pimp\piz$
 decays and $44\pm7$ $\jpsi\to\KS\Kpm\pimp\eta$ decays. In these fits, the Gaussian center is fixed
 to the nominal \jpsi mass~\cite{pdg}, and the fitted widths of
 8--9~\mevcc are consistent with the simulated resolution. The results
 of the fits are shown as solid lines on Figs.~\ref{fig:jpsi_pi0} and~\ref{fig:jpsi_eta}, with the
 hatched areas representing the non-\jpsi components.  

Using the simulated selection efficiencies with all the corrections
 described above and the differential luminosity, and dividing by the
 $\KS\to\pip\pim$ and $\piz/\eta\to\gamma\gamma$ branching fractions~\cite{pdg},
 we calculate the products of the \jpsi electronic width and
 branching fractions to these modes, and list them in Table~\ref{tab:jpsi_decays}.  The
 first uncertainties are statistical, and the second include all the
 systematic uncertainties applied to the cross sections, described
 above.

Using the world-average value of $\Gamma_{ee}^\jpsi = 5.55$~\kev~\cite{pdg},  
 we obtain the corresponding \jpsi branching fractions, also listed in
 Table~\ref{tab:jpsi_decays}.  The results
for $\mathcal{B}^{\jpsi}_{\KS\Kpm\pimp\piz}$ and $\mathcal{B}^{\jpsi}_{\KS\Kpm\pimp\eta}$
include the contributions of both nonresonant and
intermediate resonant states.
The systematic uncertainties now include the uncertainty in
$\Gamma^{\jpsi}_{ee}$.
Our result for $\mathcal{B}^{\jpsi}_{\KS\Kpm\pimp\eta}$ is consistent with,
and more precise than, the world average value~\cite{pdg}.
Our result for $\mathcal{B}^{\jpsi}_{\KS\Kpm\pimp\piz}$ is the first measurement
of this branching fraction. Our result, $\mathcal{B}^{\jpsi}_{\KS\Kpm\pimp\piz} = (5.7\pm 0.3\pm 0.4)x10^{-3}$,
is consistent with our previous measurement of
$\mathcal{B}^{\jpsi}_{\Kp\Km\pip\pim}=(6.84\pm0.28)x10^{-3}$~\cite{kkpipi} within around
two standard deviations, and larger than our
 $\mathcal{B}^\jpsi_{\Kp\Km\piz\piz}=(2.12\pm 0.21)\times10^{-3}$~\cite{kkpipi},
 $\mathcal{B}^\jpsi_{\KS\KL\pip\pim}=(3.7 \pm 0.7)\times10^{-3}$~\cite{ksklpipi}, and 
 $\mathcal{B}^\jpsi_{\KS\KS\pip\pim}=(1.68\pm 0.17)\times10^{-3}$~\cite{ksklpipi}.

\subsection{Substructure in $\jpsi\to\KS\Kpm\pimp\piz$ decays}

\begin{figure}[h] 
\includegraphics[width=0.8\columnwidth]{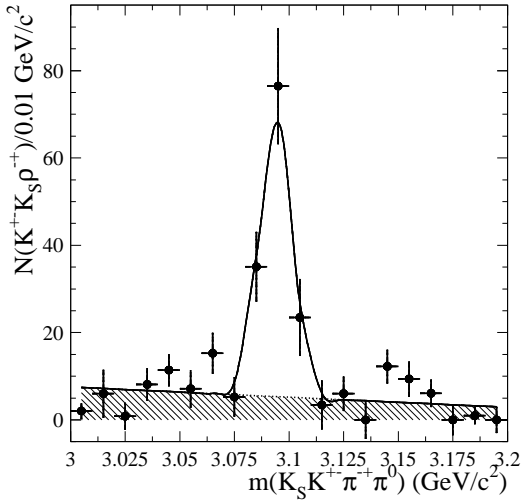}
\caption{The number of $\KS\Kpm\rho^{\mp}$ events  as a function of the
 $\KS\Kpm\pimp\piz$ invariant mass in the \jpsi region.  The line
 represents the result of the fit described in the text, with the open
 (hatched) area indicating the resonant (nonresonant) component."}
\label{fig:jpsi2rho}   
\end{figure}

We study the $\KS\Kpm\rho^{\mp}$ and $K^*K\pi$ contributions to the 
 $\jpsi\to\KS\Kpm\pimp\piz$ decay in a manner similar to that
 described in Sec.~\ref{sec:kskpipi0_substr}.  Fitting the $\pipm\piz$ invariant mass
 distribution (see Fig.~\ref{fig:rho}(a)) in 10~\mevcc bins of the
 $\KS\Kpm\pimp\piz$ invariant mass yields the numbers of
 $\KS\Kpm\rho^{\mp}$ events per bin shown in Fig.~\ref{fig:jpsi2rho}.  A fit to a Gaussian
 plus first-order polynomial (line and hatched area, respectively, in
 Fig.~\ref{fig:jpsi2rho}) yields $130\pm 12\pm 19$ $\jpsi\to\KS\Kpm\rho^{\mp}$ decays, where the
 first uncertainty is statistical and the second is the systematic uncertainty associated with
 the fit to the $\pimp\piz$ invariant-mass distribution, described
 above.  We correct for efficiency, and calculate the product 
 $\Gamma^\jpsi_{ee}\mathcal{B}^\jpsi_{\KS\Kpm\rho^{\mp}}$ from which we determine the
 branching fraction.  The results, listed in Table~\ref{tab:jpsi_decays}, represent the
 first measurement of this $\jpsi$\ decay mode.

\begin{figure}
\begin{tabular}{cc}
\includegraphics[width=0.48\columnwidth]{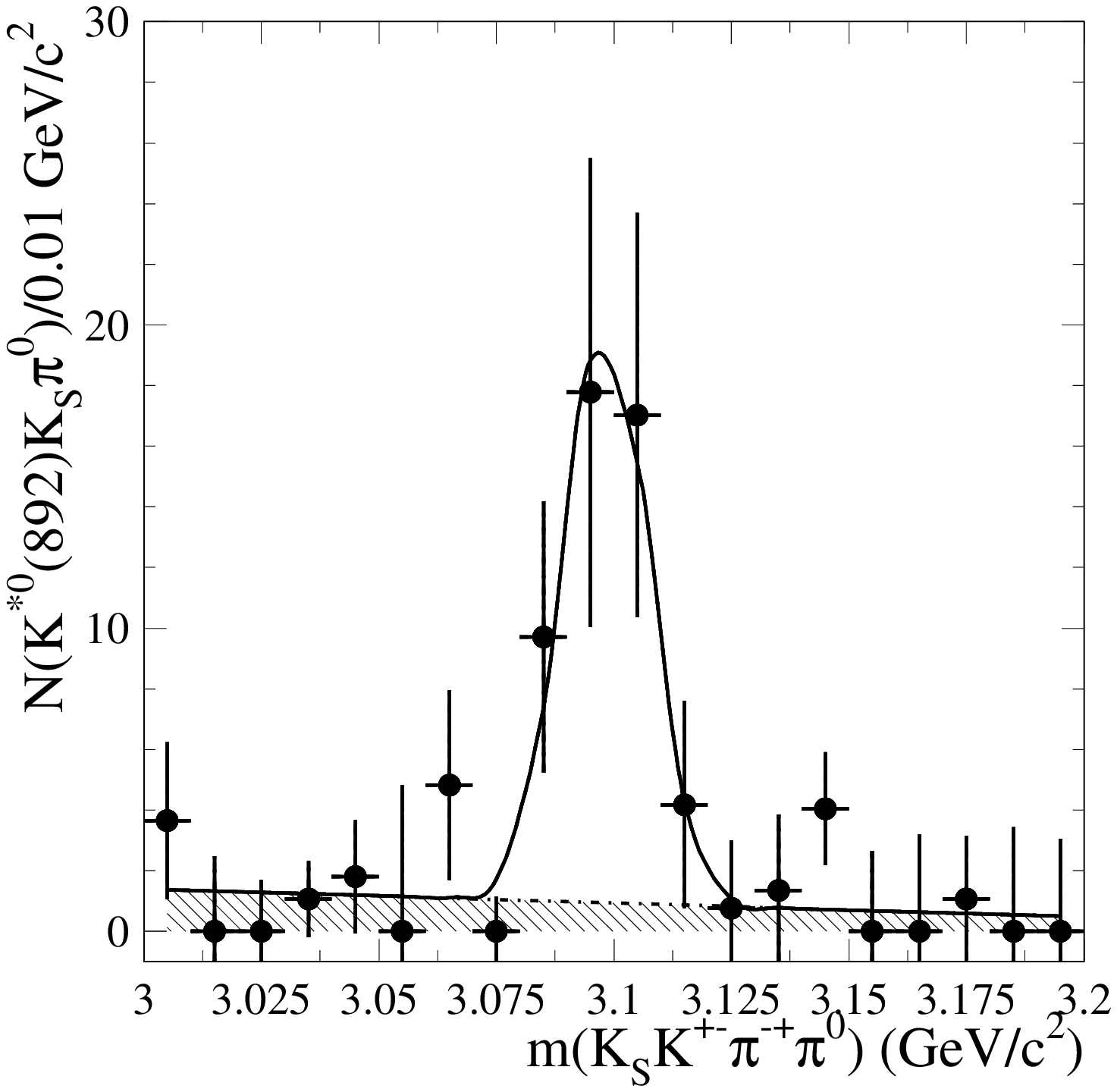}
\put(-25,105){(a)} &
\includegraphics[width=0.48\columnwidth]{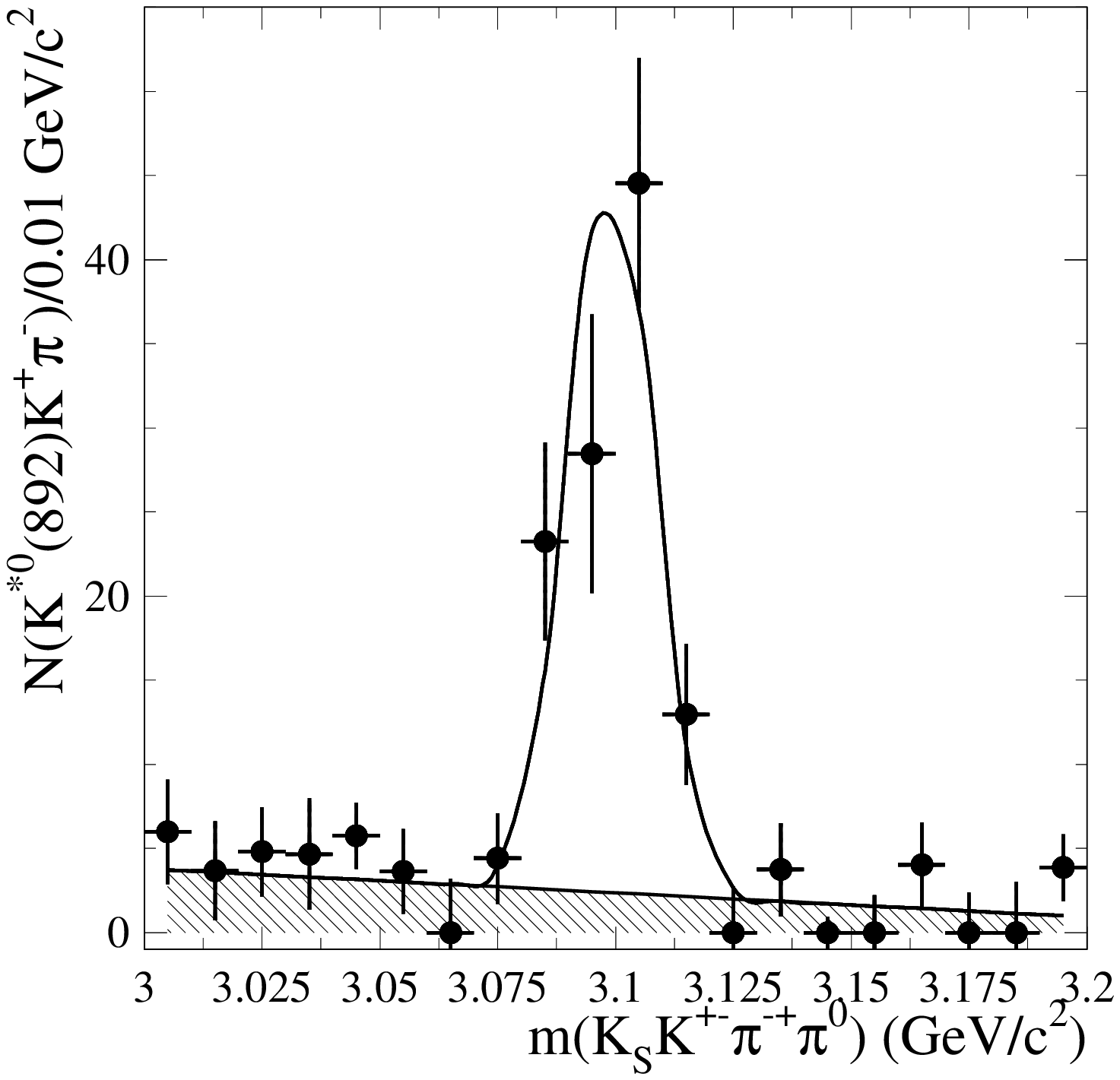}
\put(-25,105){(b)}\\
\includegraphics[width=0.48\columnwidth]{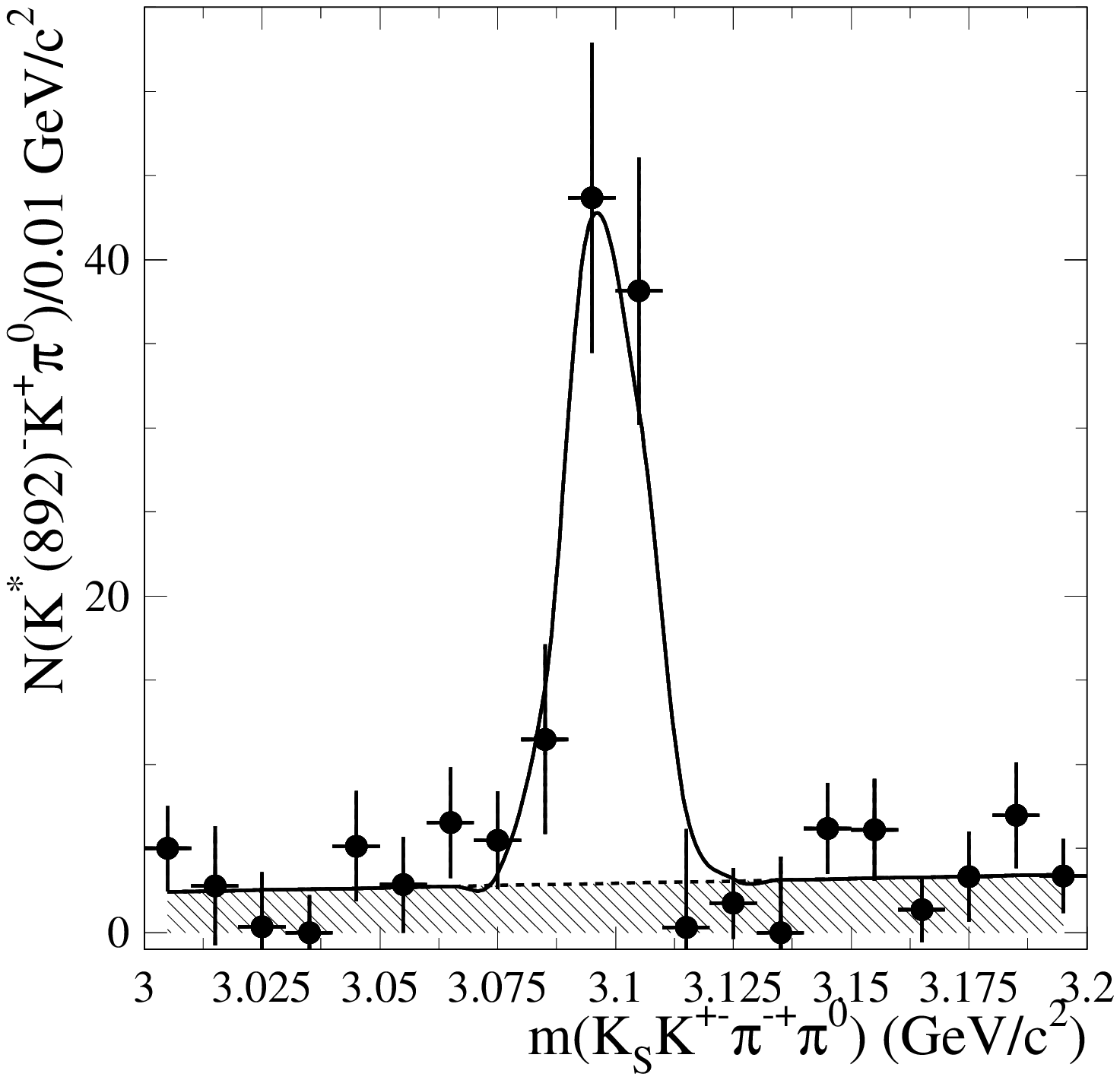} &
\put(-25,105){(c)}
\includegraphics[width=0.48\columnwidth]{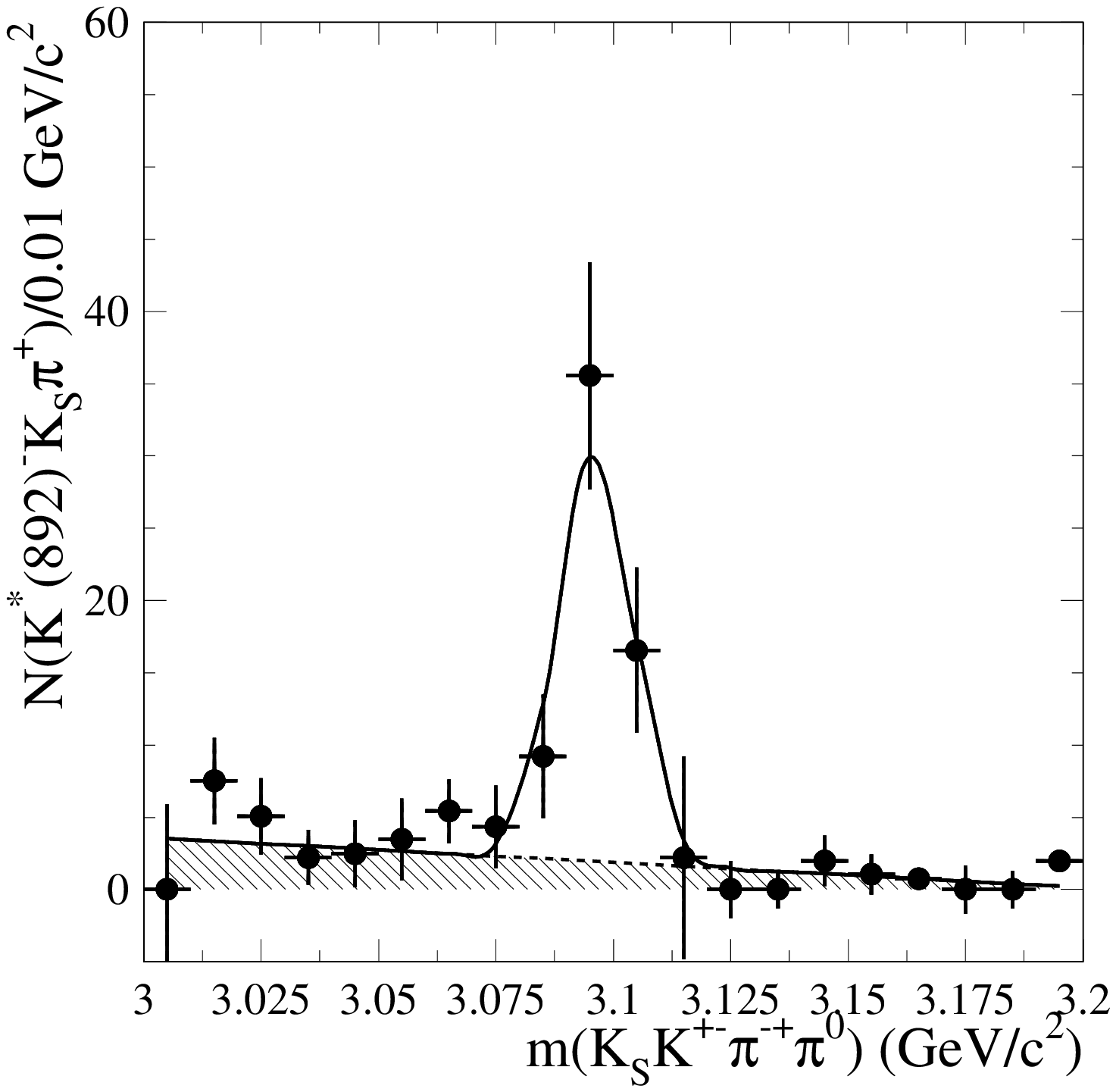}
\put(-25,105){(d)}
\end{tabular}
\caption{Event yields of (a) $\Kstarz\KS\piz$, (b) $\Kstarz\Kpm\pimp$, 
 (c) $\Kstarpm\Kmp\piz$, and (d) $\Kstarpm\KS\pimp$ final states as functions of
 the $\KS\Kpm\pimp\piz$ invariant mass in the \jpsi region.  The lines
 represent the results of the fits described in the text, with the
 open (hatched) areas indicating the resonant  (nonresonant) components.}
\label{fig:jpsi2kstkpi}
\end{figure}

We perform fits in bins of \Ecm between 3.0 and 3.2 GeV, 
analogous to those shown in Figs.~\ref{fig:kst0kst0} and~\ref{fig:nkst0corr},
of the $\KS\piz$, $\Kpm\pimp$, $\KS\pipm$ and $\Kpm\piz$ invariant-mass distributions,
to determine the number of respective $\jpsi\to K\pi$ decays.
Systematic uncertainties for these results are determined
as described in Sec.~\ref{sec:kskpipi0_substr}.
We fit each of the four distributions in Fig.~\ref{fig:jpsi2kstkpi} with a Gaussian 
plus first-order polynomial function to obtain
 $34\pm 6\pm 22$ $\jpsi\to\Kstarz\KS\piz$ decays,
 $99\pm10\pm 17$ $\jpsi\to\Kstarz\Kpm\pimp$ decays,
 $80\pm10\pm 24$ $\jpsi\to\kstpm\Kmp\piz$ decays, and
 $64\pm 9\pm 22$ $\jpsi\to\kstpm\KS\pimp$ decays.
 Here, the first uncertainties are statistical and the second
 systematic, where these latter terms result from the fit procedure.  We 
correct for efficiency
 and calculate the products 
 $\Gamma^\jpsi_{ee}\mathcal{B}^\jpsi_{K^*(892)\Kbar\pi}\mathcal{B}^{K^*(892)}_{K\pi}$,
 and then the products of the \jpsi and $K^*(892)$ branching fractions, and
 list them in Table~\ref{tab:jpsi_decays}.  With the current data samples, we are not able
 to study $\jpsi\to\Kstar\Kstarb$ decays.

 There are no previous measurements of these decay chains.  The measurement 
$\mathcal{B}^\jpsi_{\kstpm\KS\pimp}\mathcal{B}^{\kstpm}_{\KS\pipm}=(2.6\pm 0.9)\times 10^{-3}$~\cite{ksklpipi}
is about half as large as our result for $\mathcal{B}^\jpsi_{\KS\Kpm\pimp\piz}$; this difference is consistent with 
expectations for isospin conservation. In Ref.~\cite{kkpipi} it was found that the $\Kp\Km\pip\pim$ mode is dominated by the $\kstz\Kpm\pimp$ 
channel,  which originates predominantly from the decay of $\Kstarz(892)\Kbar^*_{(0,2)}(1430)$ apart from a small contribution from
$\Kstarz(892)\Kstarzb(892)$.  
Our results are consistent with this pattern, and the
 world-average $\mathcal{B}^\jpsi_{\Kstarz\Kstarzb}=(0.23\pm0.07)\times10^{-3}$~\cite{pdg} is
 well below our values for $\mathcal{B}^\jpsi_{\Kstarz(892) K\pi}$.  On the other
 hand, the sum of our $\mathcal{B}^\jpsi_{\kstpm K\pi}$ modes is only about
 twice the world-average $\mathcal{B}^\jpsi_{\kstpm\kstmp}=1.00^{+0.22}_{-0.40}\times10^{-3}$~\cite{pdg}.
\begin{table*}[tbh]
\caption{Summary of $\jpsi$ decay measurements from this analysis. Here, $\mathcal{B}_f$ represents the $\jpsi$ branching 
fraction to the indicated final state and $\Gamma^{\jpsi}_{ee}$ the
partial width for $\jpsi$ decay to $ee$.}
%Here, $\mathcal{B}_f$ represents the product of the \jpsi
% branching fraction to the given mode and any $K^*$ branching fraction required to reach the $\KS\Kpm\pimp\piz$ final state.}
\label{tab:jpsi_decays}
\begin{ruledtabular}
\begin{tabular}{c c c c }
 & \multicolumn{2}{c}{This work} & PDG(2014) \\
final state & $\mathcal{B}_{f}\cdot\Gamma^{\jpsi}_{ee}$ (eV) & $\mathcal{B}_f$ (10$^{-3}$) &  $\mathcal{B}_f$ (10$^{-3}$) \\
\hline
\KS\Kpm\pimp\piz            & 31.7$\pm$1.9$\pm$1.8 & 5.7$\pm$0.3$\pm$0.4    &                    ---                  \\
$\KS\Kpm\pimp\eta$          & 7.3$\pm$1.4$\pm$0.4  & 1.30$\pm$0.25$\pm$0.07 &                 2.2$\pm$0.4             \\
                            &                      &                        &                                         \\
$\KS\Kpm\rho(770)^{\mp}$    & 10.4$\pm$1.0$\pm$1.9 & 1.87$\pm$0.18$\pm$0.34 &                    ---                  \\
\kstz\Km\pip + c.c.         & 7.1$\pm$0.8$\pm$1.2  & 1.3$\pm$0.1$\pm$0.2    &                    ---                  \\
\kstz\KS\piz + c.c.         & 2.4$\pm$0.5$\pm$1.5  & 0.43$\pm$0.01$\pm$0.27 &                    ---                  \\
$\kstpm\Kmp\piz$            & 5.7$\pm$0.7$\pm$1.7  & 1.0$\pm$0.1$\pm$0.3    &                    ---                  \\ 
$\kstpm\KS\pimp$            & 4.6$\pm$0.6$\pm$1.6  & 0.8$\pm$0.1$\pm$0.3    &                    ---                  \\
\end{tabular}
\end{ruledtabular}
\end{table*}

\section{Summary}

We have presented the first measurements of the $\epem\to\KS\Kpm\pimp\piz$ and
$\epem\to\KS\Kpm\pimp\eta$ cross sections. The measurements are performed  over the c.m. energy ranges from
their respective threshold to 4~\gev.  The total uncertainty in the $\KS\Kpm\pimp\piz$ cross
section ranges from 6.3\% at low masses, to 11.5\% at
3~\gev, increasing with higher masses.  That on the $\KS\Kpm\pimp\eta$ cross
section is 12.8\% (19.1\%) below (above) 3~\gev.  These results are
useful inputs into the total hadronic cross section, and the
theoretical calculation of $(g-2)_{\mu}$.

The \KS\Kpm\pimp\piz cross section exhibits  a slow rise from threshold, then 
a steep rise from 1.6~\gev to a peak
value of about 2~nb near 1.9~\gev, followed by a slow decrease with
increasing mass.  There is a clear \jpsi signal, but no other
significant structure.  The cross section is about half that of 
$\epem\to\Kp\Km\pip\pim$~\cite{kkpipi}, and about twice that of
$\epem\to\KS\KL\pip\pim$~\cite{ksklpipi} or $\Kp\Km\piz\piz$~\cite{kkpipi}.

The $\KS\Kpm\pimp\eta$ cross section is much smaller, and consistent with zero between
threshold and 2~\gev.  It then demonstrates a slow rise
to a value of about 0.1~nb over a wide range around
2.5~\gev, followed by a slow decrease with increasing mass.  There is
a clear \jpsi signal and no other significant structure.

Several intermediate resonant states are evident in the $\KS\Kpm\pimp\piz$ data,
and we have measured cross sections into this final state via $\epem\to\Kstarz\Kstarzb$,
$\Kstarz\KS\piz$ + c.c, $\Kstarz\Km\pip$ + c.c., $\kstpm\kstmp$, $\kstpm\KS\pimp$,
\kstpm\Kmp\piz, and $\KS\Kpm\rho^{\mp}$.  There are also signals for the
production of at least one $\Kstar(1430)$ state, and at least three $K_1$
states.  Together, these channels dominate $\KS\Kpm\pimp\piz$ production, and the
$\kstp\kstm$ channel dominates both \kstpm\KS\pimp and
$\kstpm\Kmp\piz$ production.  The cross sections are consistent with
previous results in other final states.

The $\KS\Kpm\pimp\eta$ final state includes contributions from $\Kstarz\KS\eta+c.c.$,
$\kstpm\Kmp\eta$, and $\ktwo^{\pm}\Kmp\eta$, and no other significant
substructure.  We have obtained the first measurements of the
$\epem\to\Kstarz\KS\eta+c.c.$ and $\kstpm\Kmp\eta$ cross sections, and these
channels dominate the overall $\KS\Kpm\pimp\eta$ production. 

With the results of this analysis, \babar\ has now provided the cross section measurements for the complete set of allowed
$\epem\to K\Kbar\pi$ and $K\Kbar\pi\pi$ processes except for those containing a $\KL\KL$ pair. 
Since the latter modes are expected to be the same as the corresponding
modes with a $\KS\KS$ pair,
the $K\Kbar\pi$ and $K\Kbar\pi\pi$ contributions to $g_\mu-2$ can be
calculated using this set of exclusive cross section measurements,
with no assumptions or isospin relations.
We expect a reduction in the total uncertainties of these contributions by a factor of five to eight
compared with current estimates~\cite{g-2}.

%Signals for $\jpsi$ production are evident in both final states and several of
%the submodes.  We have measured the first $\jpsi$ branching fractions to
%inclusive $\KS\Kpm\pimp\piz$ and $\KS\Kpm\pimp\eta$, 
We have measured the $\jpsi$ branching fraction
to $\KS\Kpm\pimp\eta$, and presented the first $\jpsi$ branching fraction
measurement to $\KS\Kpm\pimp\piz$
as well as the branching fractions to the $\KS\Kpm\pimp\piz$ final state via
$\Kstarz\Kpm\pimp$+c.c., $\Kstarz\KS\piz$+c.c., $\kstp\Km\piz$+c.c.,
$\kstpm\KS\pimp$, and $\KS\Kpm\rho^{\mp}$.  
%We make the first measurements of the branching
%fractions of the $\jpsi$ to inclusive $\KS\Kpm\pimp\piz$ and $\KS\Kpm\pimp\eta$, 
%also the branching fractions to the $\KS\Kpm\pimp\piz$ final state via
%$\Kstarz\Kpm\pimp$+c.c., $\Kstarz\KS\piz$+c.c., $\kstp\Km\piz$+c.c.,
%$\kstpm\KS\pimp$, and $\KS\Kpm\rho^{\mp}$.  
We cannot extract branching
fractions for $\Kstarz\Kstarzb$ or $\kstp\kstm$, but our results for 
$\kstp\Km\piz$+c.c. and $\kstpm\KS\pimp$ are both consistent with the
world-average value for $\kstp\kstm$, indicating the same dominance of
$\kstp\kstm$ as in non-$\jpsi$ data. Our results for
$\Kstarz\Kp\pim$ + c.c. and $\Kstarz\KS\piz$ + c.c. respectively are about five and two times
larger than the world-average value for $\Kstarz\Kstarzb$.

\section*{Acknowledgements}
We are grateful for the 
extraordinary contributions of our \pep2\ colleagues in
achieving the excellent luminosity and machine conditions
that have made this work possible.
The success of this project also relies critically on the 
expertise and dedication of the computing organizations that 
support \babar.
The collaborating institutions wish to thank 
SLAC for its support and the kind hospitality extended to them. 
This work is supported by the
US Department of Energy
and National Science Foundation, the
Natural Sciences and Engineering Research Council (Canada),
the Commissariat \`a l'Energie Atomique and
Institut National de Physique Nucl\'eaire et de Physique des Particules
(France), the
Bundesministerium f\"ur Bildung und Forschung and
Deutsche Forschungsgemeinschaft
(Germany), the
Istituto Nazionale di Fisica Nucleare (Italy),
the Foundation for Fundamental Research on Matter (The Netherlands),
the Research Council of Norway, the
Ministry of Education and Science of the Russian Federation, 
Ministerio de Econom\'{\i}a y Competitividad (Spain), the
Science and Technology Facilities Council (United Kingdom),
and the Binational Science Foundation (U.S.-Israel).
Individuals have received support from 
the Marie-Curie IEF program (European Union) and the A. P. Sloan Foundation (USA). 

% NOTES:
% add "and the Binational Science Foundation (U.S.-Israel)"  07-Oct-2013 Bill Gary (Abi Soffer request)

\end{document}